\documentclass[]{elsarticle}

\usepackage{lineno,hyperref}
\modulolinenumbers[5]
\usepackage{subfigure}
\usepackage{subscript}
\usepackage{amsmath}
\usepackage{url}

\journal{Journal of XXXX}

\graphicspath{{figs/}}

\begin{document}

\begin{frontmatter}

\title{Language of fungi\\  derived from electrical spiking activity}

\author{Andrew Adamatzky}

\address{Unconventional Computing Laboratory, UWE, Bristol, UK}

\begin{abstract}
Fungi exhibit oscillations of extracellular electrical potential recorded via differential electrodes inserted into a substrate colonised by mycelium or directly into sporocarps. We analysed electrical activity of 
ghost fungi (\emph{Omphalotus nidiformis}),
Enoki fungi (\emph{Flammulina velutipes}), split gill fungi (\emph{Schizophyllum commune}) and  caterpillar fungi (\emph{Cordyceps militari}).  The spiking characteristics are species specific: a spike duration varies from one to 21 hours and an amplitude from 0.03~mV to 2.1mV. We found that spikes are often clustered into trains.  Assuming that spikes of electrical activity are used by fungi to communicate and process information in mycelium networks, we group spikes into words and provide a  linguistic and information complexity analysis of the fungal spiking activity. We demonstrate that distributions of fungal word lengths match that of human languages. We also construct algorithmic and Liz-Zempel complexity hierarchies of fungal sentences and show that species  \emph{S. commune} generate most complex sentences.
\end{abstract}


\end{frontmatter}

\section{Introduction}

Spikes of electrical potential are typically considered to be key attributes of neurons and neuronal spiking activity is interpreted as a language of a nervous system~\cite{baslow2009languages,andres2015language, pruszynski2019language}. However, almost all creatures without nervous system produce spikes of electrical potential --- Protozoa~\cite{eckert1972sensory,bingley1966membrane,ooyama2011hierarchical}, Hyrdoroza~\cite{hanson2021spontaneous}, slime moulds~\cite{iwamura1949correlations,kamiya1950bioelectric} and plants~\cite{trebacz2006electrical,fromm2007electrical,zimmermann2013electrical}. Fungi also exhibit trains of action-potential like spikes, detectable by intra- and extra-cellular recordings~\cite{slayman1976action,olsson1995action,adamatzky2018spiking}. In experiments with recording of electrical potential of  oyster fungi \emph{Pleurotus djamor} we discovered two types of spiking activity:  high-frequency (period 2.6~min) and low-freq (period 14~min)~\cite{adamatzky2018spiking}. While studying other species of fungus,  \emph{Ganoderma resinaceum}, we found that most common width of an electrical potential spike is 5-8~min~\cite{AdamatzkyGanoderma}. In both species of fungi we observed bursts of spiking in the trains of the spike similar to that observed in central nervous system~\cite{cocatre1992identification,legendy1985bursts}. Whilst the similarly could be just phenomenological this indicates a possibility that mycelium networks transform information via interaction of spikes and trains of spikes in manner homologous to neurons. First evidence has been obtained that indeed fungi respond to mechanical, chemical and optical stimulation by changing pattern of its electrically activity and, in many cases, modifying characteristics of their spike trains~\cite{adamatzky2021fungal,adamatzky2021reactive}. There is also evidence of electrical current participation in the interactions between mycelium and plant roots during formation of mycorrhiza~\cite{berbara1995electrical}. In~\cite{dehshibi2021electrical} we compared complexity measures of the fungal spiking train and sample text in European languages and found that the 'fungal language' exceeds the European languages in morphological complexity. In our venture to decode the language of fungi a first step would be to uncover if all species of fungi exhibit similar characteristics of electrical spiking activity.  We recorded and analysed, as detailed in Sect.~\ref{methods}, electrical activity of ghost fungi (\emph{Omphalotus nidiformis}),
Enoki fungi (\emph{Flammulina velutipes}), split gill fungi (\emph{Schizophyllum commune}) and  caterpillar fungi (\emph{Cordyceps militari}). The phenomenological characteristic of the spiking behaviour discovered are presented in Sect.~\ref{results}. Linguistic analysis and information and algorithmic complexity estimates of the spiking patterns are given in Sect.~\ref{language}.

\section{Experimental laboratory methods and analysis}
\label{methods}

\begin{figure}[!tbp]
    \centering
\subfigure[]{\includegraphics[width=0.49\textwidth]{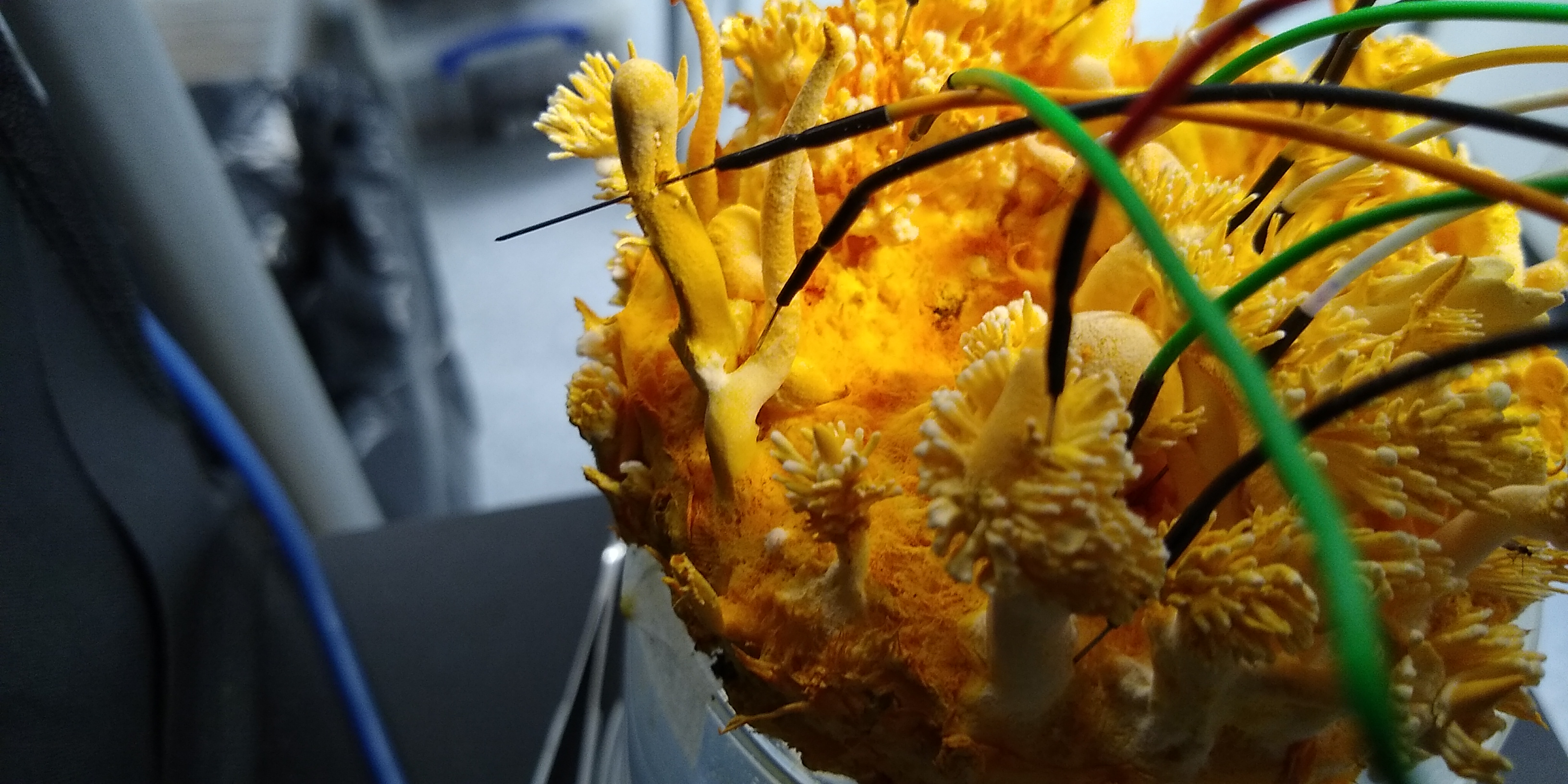}\label{corcydepssetup}}
\subfigure[]{\includegraphics[width=0.49\textwidth]{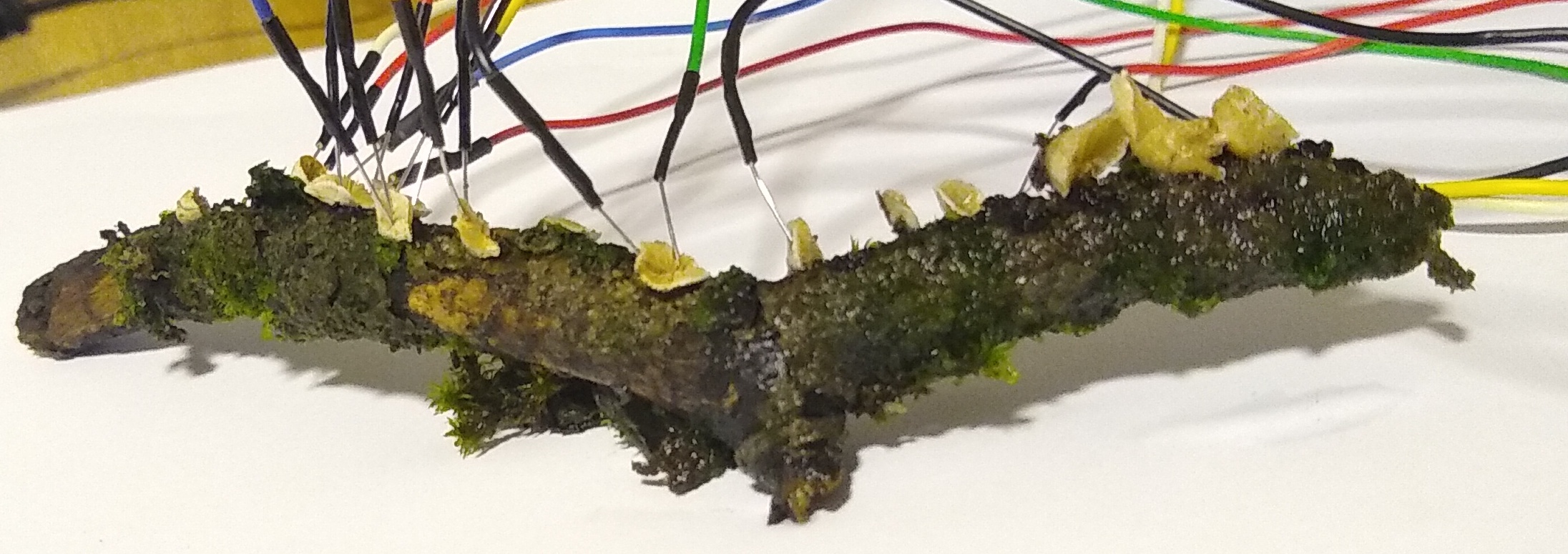}\label{Schizophyllum_communesetup}}
\subfigure[]{\includegraphics[width=0.49\textwidth]{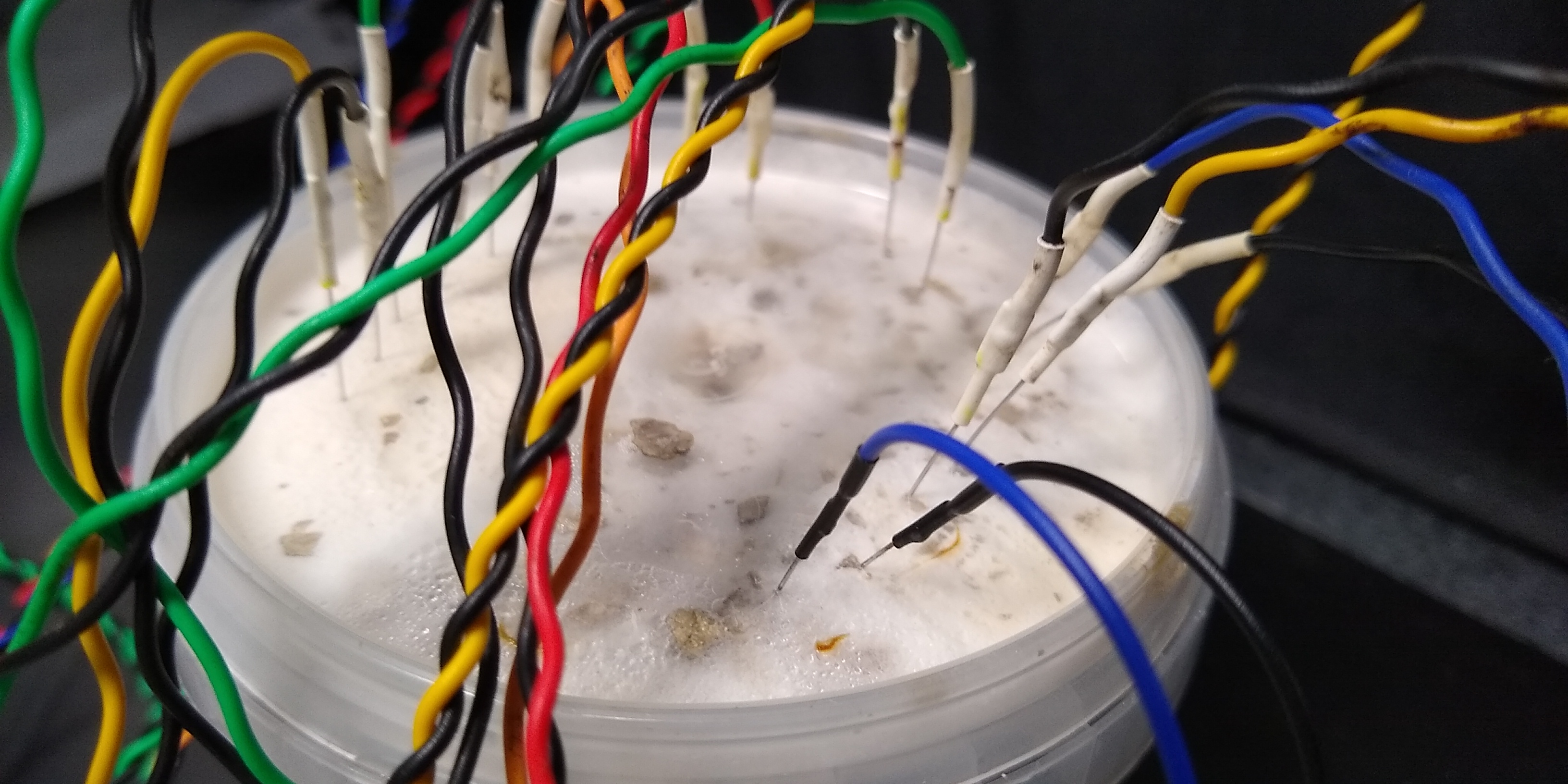}\label{ghostsetup}}
    \caption{Photographs of pairs of differential electrodes inserted in 
    (a)~\emph{C. militaris}, the block of a substrate colonised by the fungi was removed from the plastic container to make a photo after the experiments, 
    (b)~\emph{S. commune}, the twig with the fungi was removed from the humid plastic container to make a photo after the experiment, 
    (c)~\emph{F. velutipes}, the container was kept sealed and electrodes pierced through the lid.}
    \label{fig:electrodes}
\end{figure}

Four species of fungi have been used in experiments: 
\emph{Omphalotus nidiformis} and
\emph{Flammulina velutipes}, supplied by Mycelia NV, Belgium (\url{mycelium.be}),  \emph{Schizophyllum commune}, collected near Chew Valley lake, Somerset, UK (approximate coordinates 51.34949164156282, -2.622511962302647), 
\emph{Cordyceps militaris}, supplied by Kaizen Cordyceps, UK (\url{kaizencordyceps.co.uk}).

Electrical activity of the fungi was recorded using pairs of iridium-coated stainless steel sub-dermal needle electrodes (Spes Medica S.r.l., Italy), with twisted cables and  ADC-24 (Pico Technology, UK) high-resolution data logger with a 24-bit A/D converter, galvanic isolation and software-selectable sample rates all contribute to a superior noise-free resolution.  Each pair of electrodes  reported a potential difference between the electrodes. The pairs of electrodes were pierced into the substrates colonised by fungi or, as in case of \emph{S. commune}, in the sporocarps are shown in Fig.~\ref{fig:electrodes}.  Distance between electrodes was 1-2~cm.  We recorded electrical activity one sample per second. We recorded 8 electrode pairs simultaneously. During the recording, the logger has been doing as many measurements as possible (typically up to 600 per second) and saving the average value. The acquisition voltage range was 78~mV. \emph{S. commune} has been recorded for 1.5 days, other species for c. 5 days.

\begin{figure}[!tbp]
    \centering
    \includegraphics[width=0.7\textwidth]{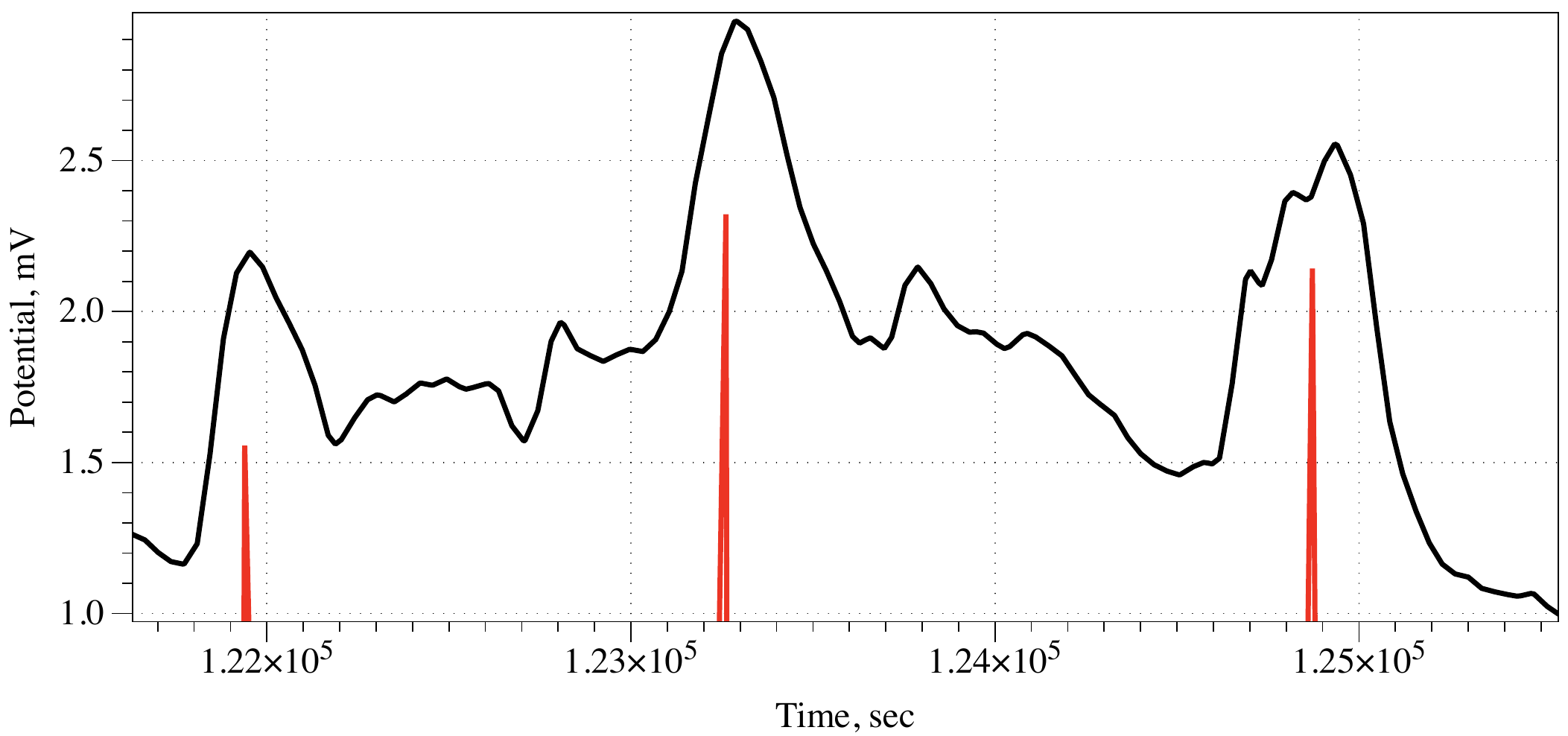}
    \caption{Example of spike detection.}
    \label{fig:spikingdetection}
\end{figure}

Spikes of electrical potential have been detected in a semi-automatic mode as follows. For each sample measurement $x_i$ we calculated average value of its neighbourhood as 
$a_i=(4 \cdot w)^{-1} \cdot \sum_{i-2\cdot w \leq j \leq i+2\cdot w} x_j$. The index $i$ is considered a peak of the local spike if $|x_i|-|a_i|>\delta$. The list of spikes were further filtered by removing false spikes located at a distance $d$ from a given spike. Parameters were species specific, for \emph{C. militaris} and \emph{F. velutipes}  $w=200$, $\delta=0.1$, $d=300$; for \emph{S. commune} $w=100$, $\delta=0.005$, $d=100$;
for \emph{O. nidiformis} $w=50$, $\delta=0.003$, $d=100$. An example of the spikes detected is shown in Fig.~\ref{fig:spikingdetection}. Over 80\% of spikes have been detected by such a technique.

\section{Characterisation of the electrical spiking of fungi}
\label{results}

\begin{figure}[!tbp]
    \centering
\subfigure[]{\includegraphics[width=0.75\textwidth]{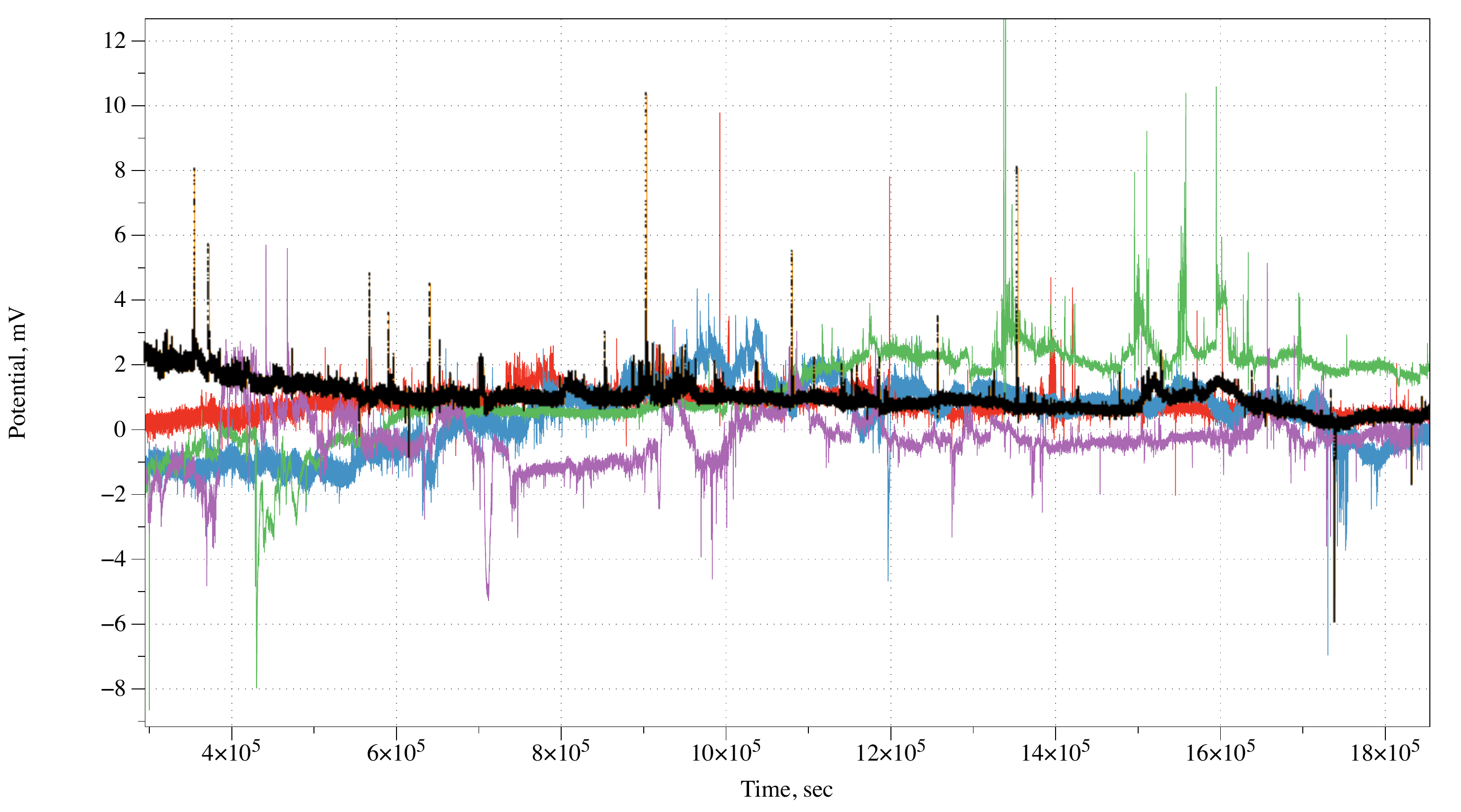}\label{cordyceps}}
\subfigure[]{\includegraphics[width=0.65\textwidth]{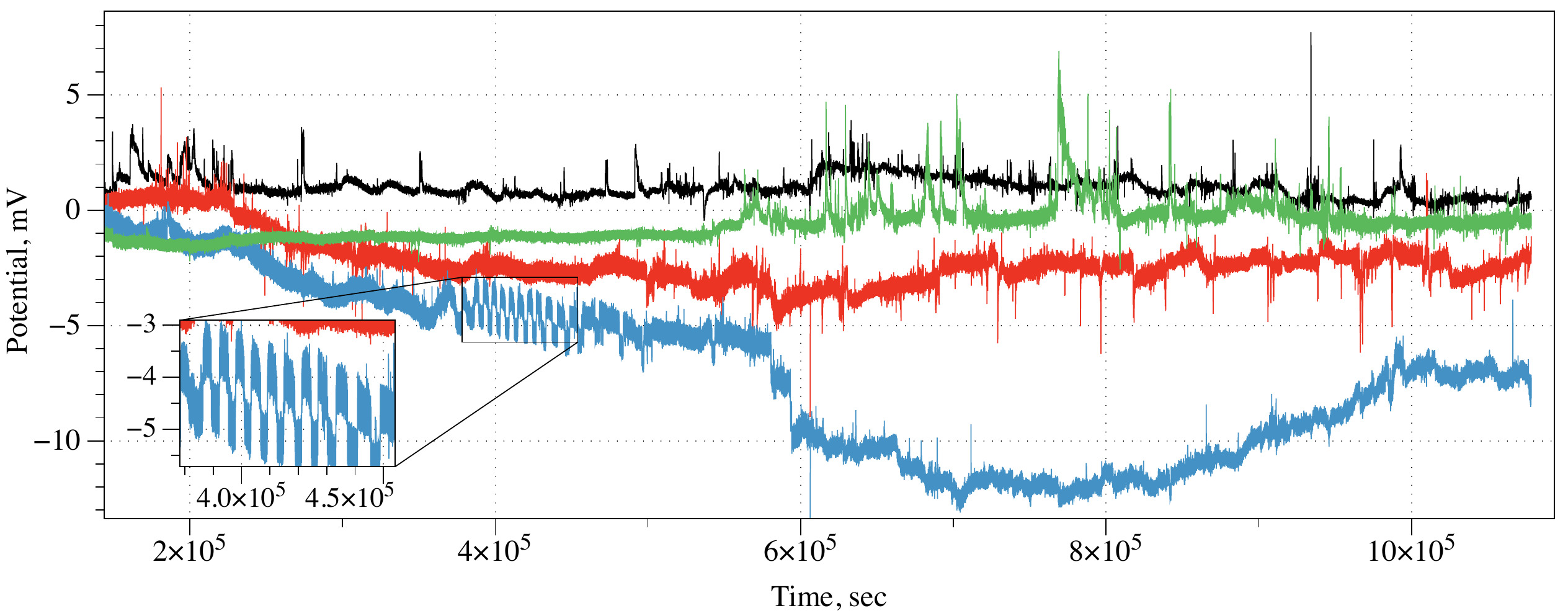}~\label{enokispiking}}
\subfigure[]{\includegraphics[width=0.55\textwidth]{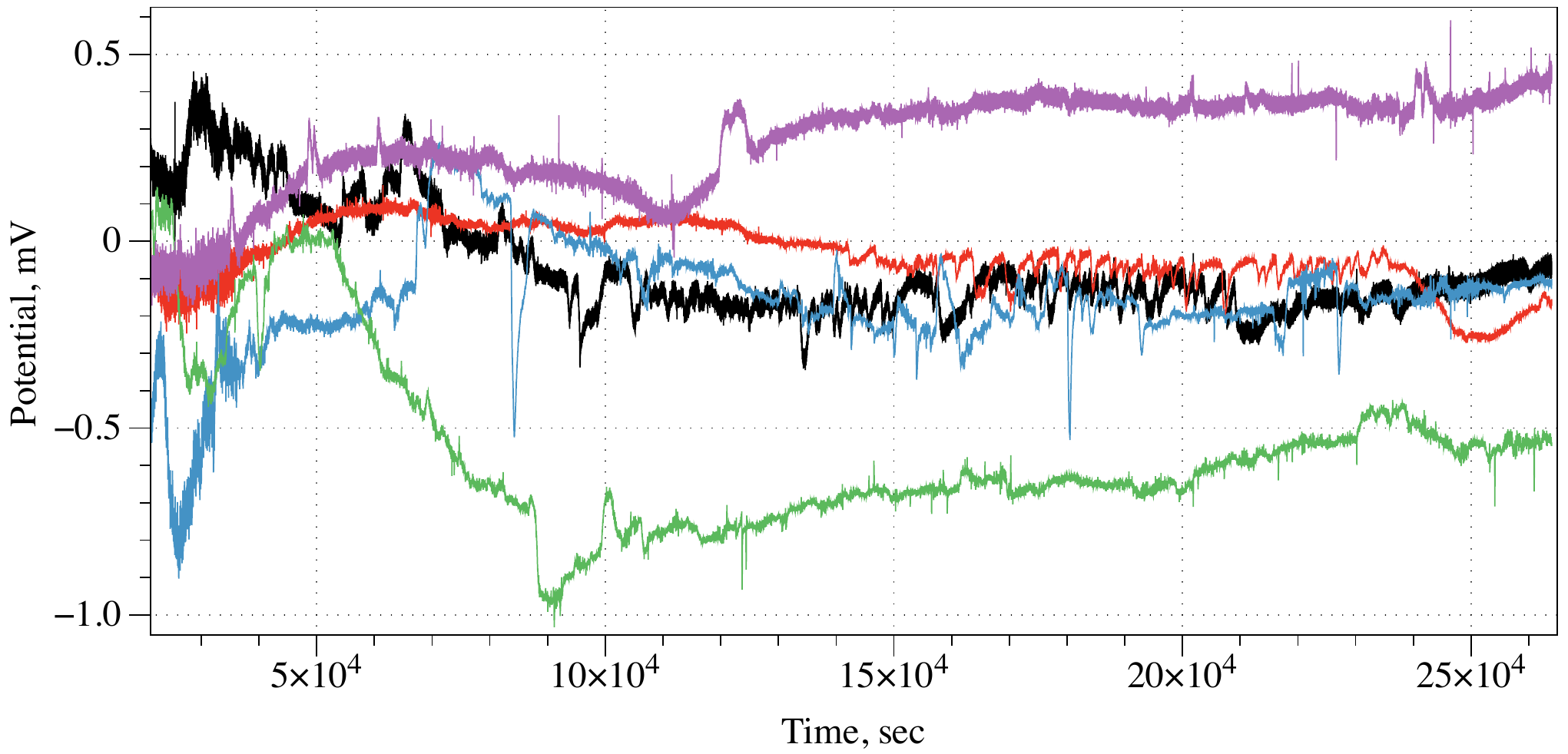}\label{schizophyllumspiking}}
\subfigure[]{\includegraphics[width=0.55\textwidth]{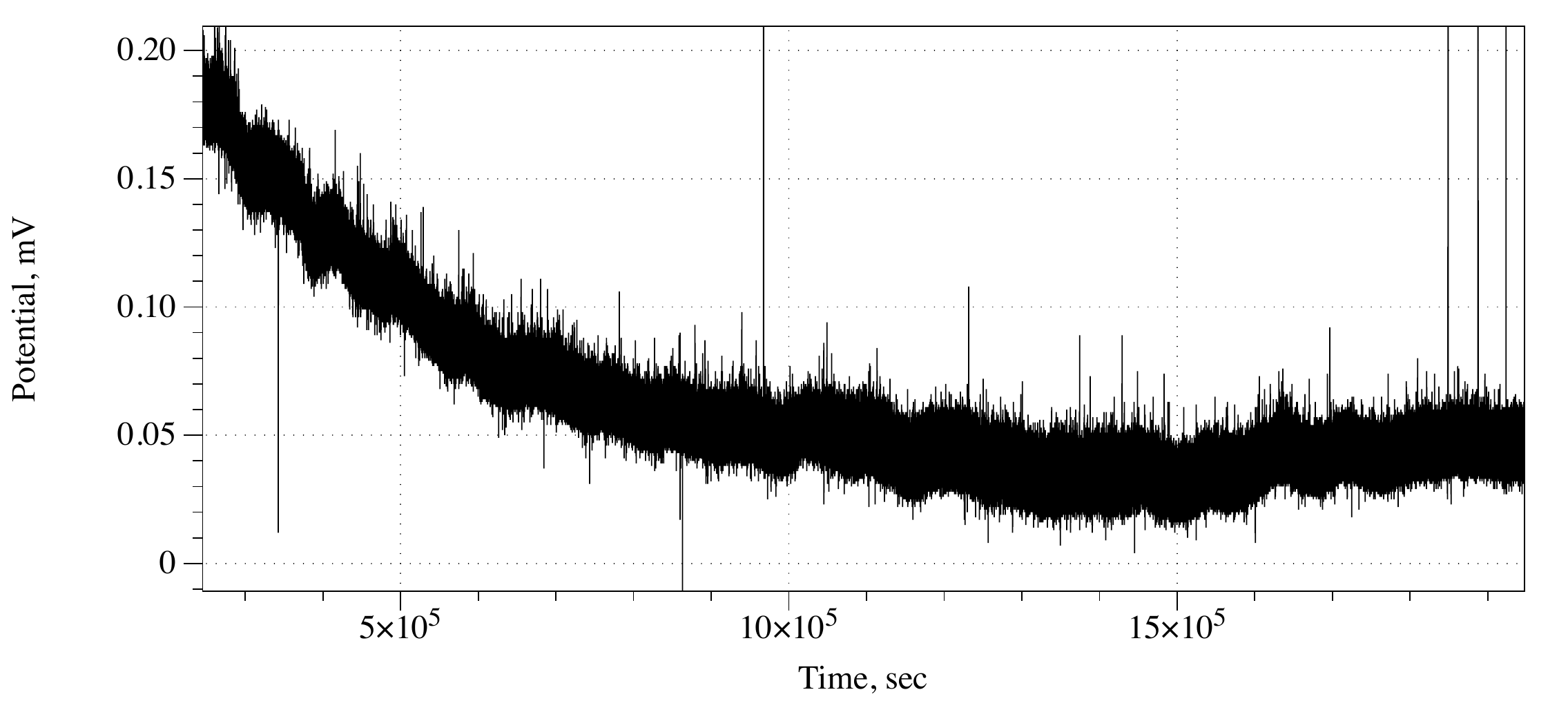}\label{ghostspiking}}
    \caption{Examples of electrical activity of 
    (a)~\emph{C. militaris},
    (b)~\emph{F. velutipes}, insert shows zoomed in burst of high-frequency spiking.
    (d)~\emph{S. commune},
    (c)~\emph{O. nidiformis}. 
        Colours reflect recordings from different channels.}.
    \label{fig:spiking}
\end{figure}

\begin{figure}[!tbp]
    \centering
    \subfigure[]{\includegraphics[width=0.49\textwidth]{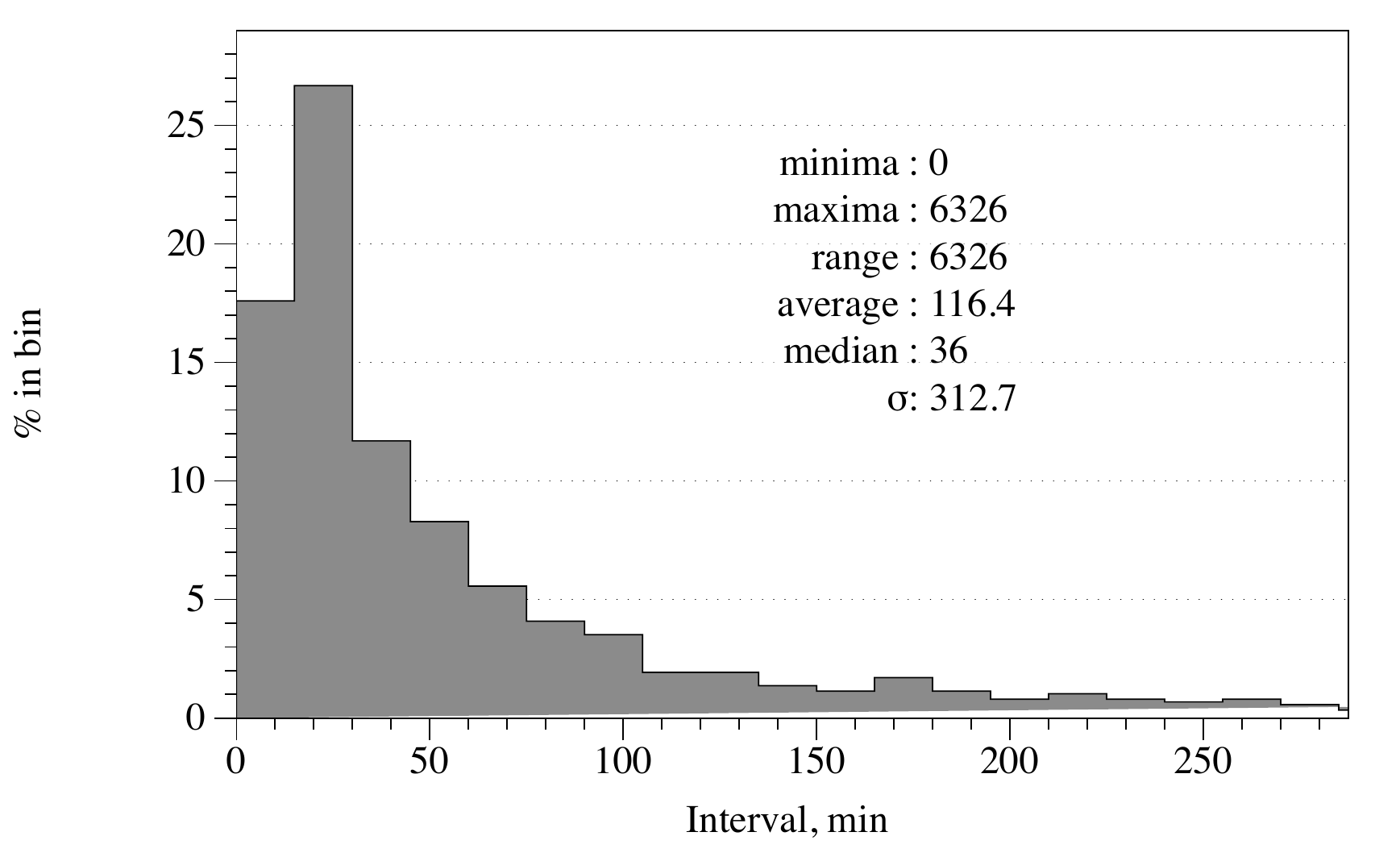}}
    \subfigure[]{\includegraphics[width=0.49\textwidth]{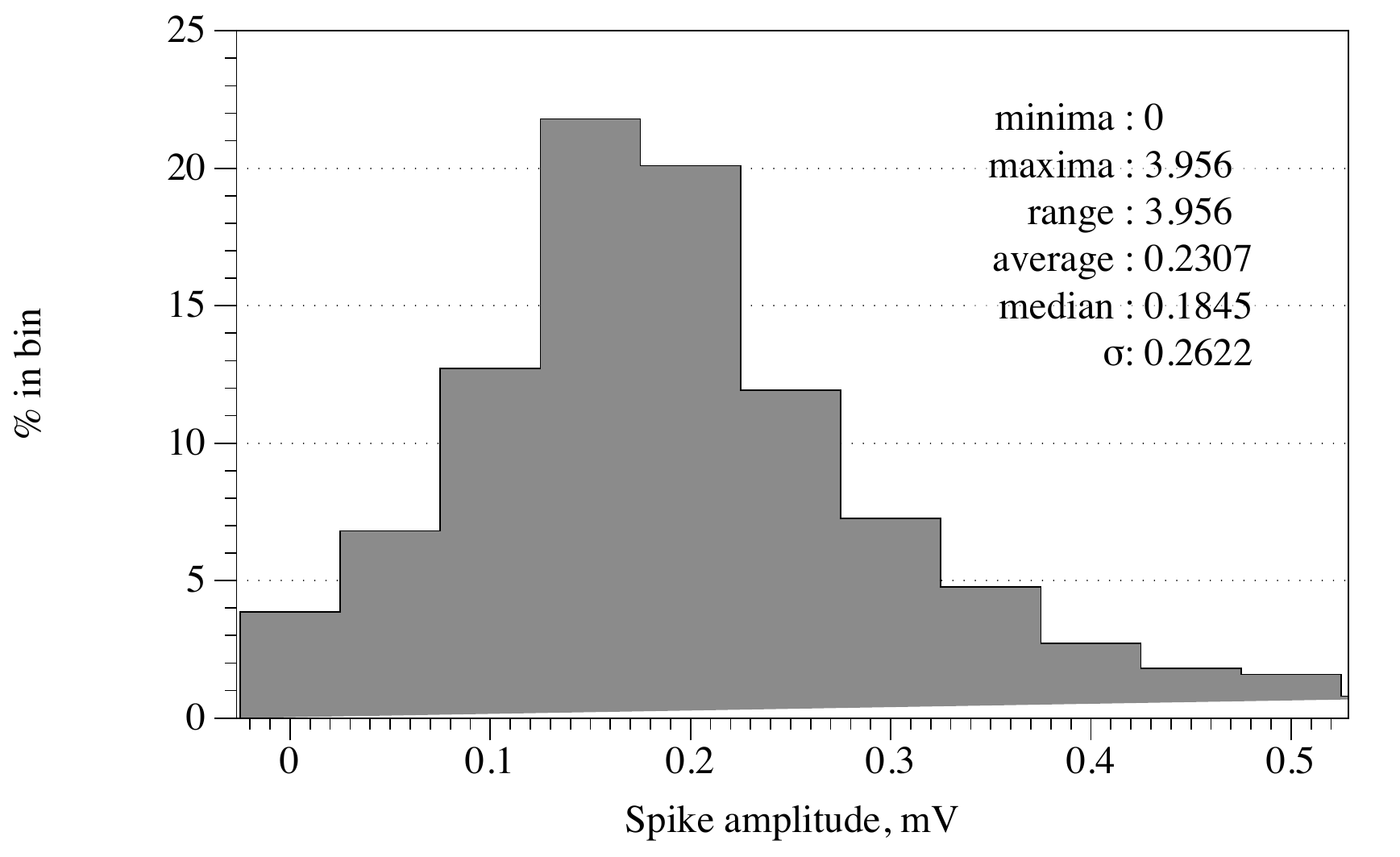}}
     \subfigure[]{\includegraphics[width=0.49\textwidth]{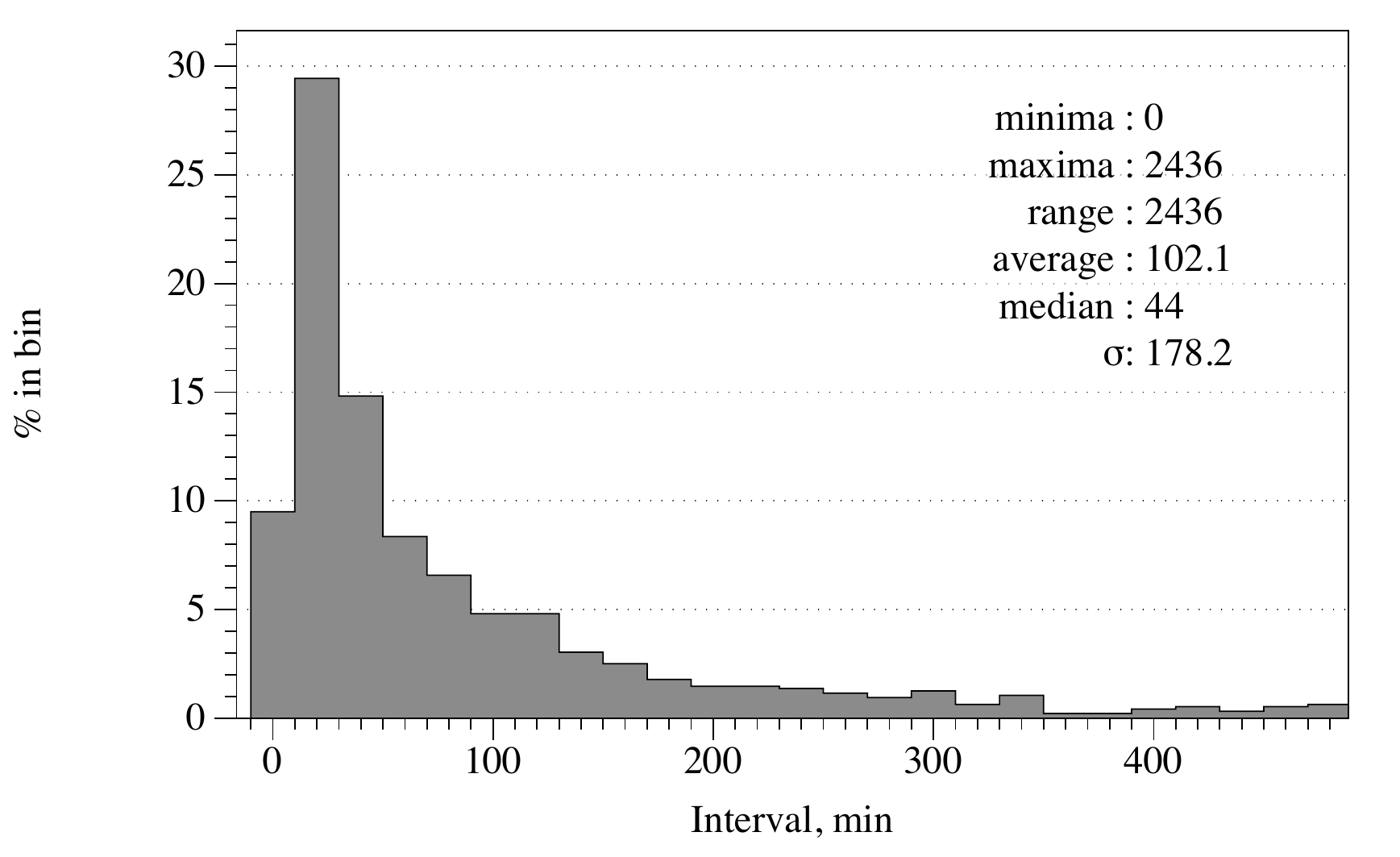}}
    \subfigure[]{\includegraphics[width=0.49\textwidth]{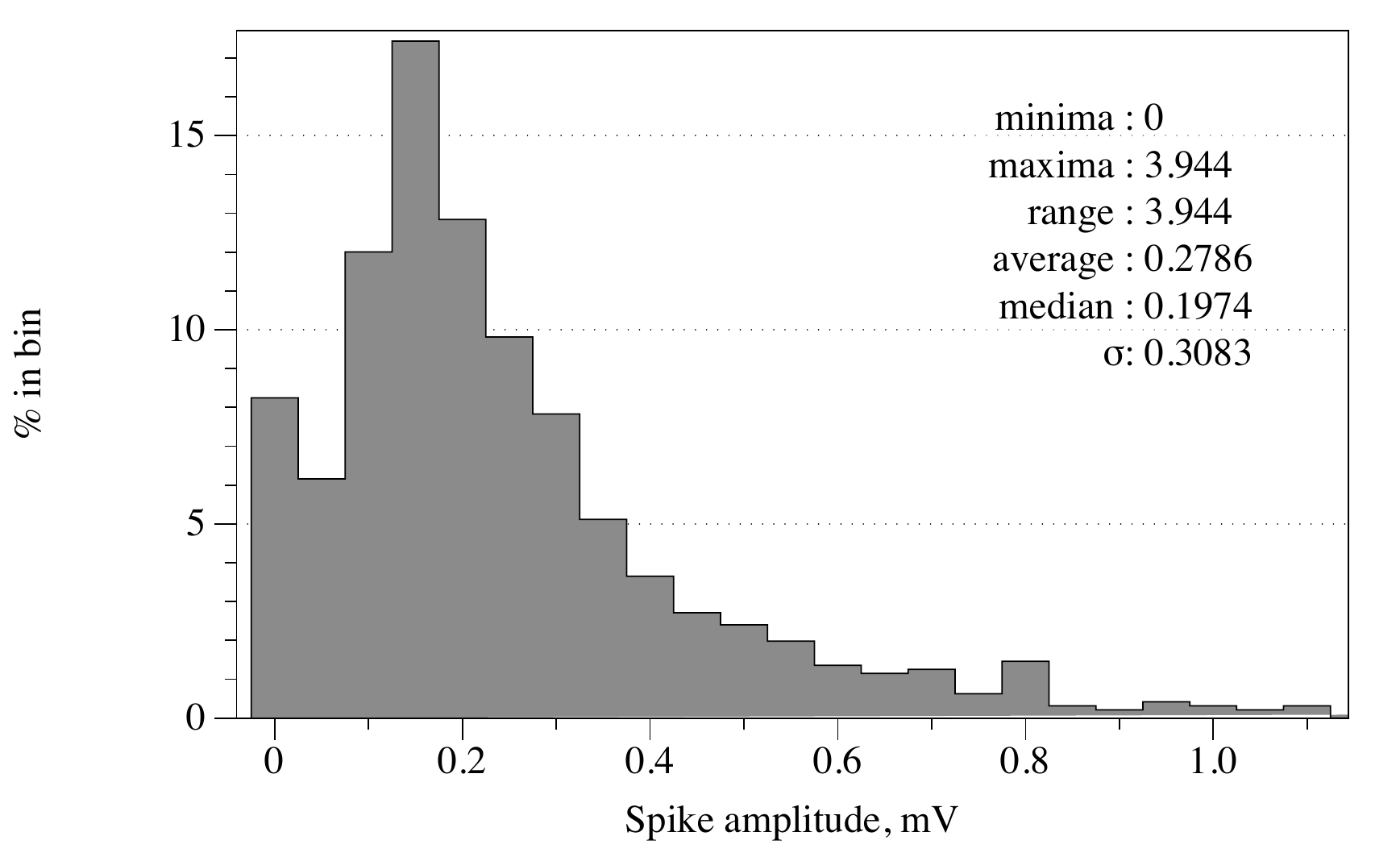}}
     \subfigure[]{\includegraphics[width=0.49\textwidth]{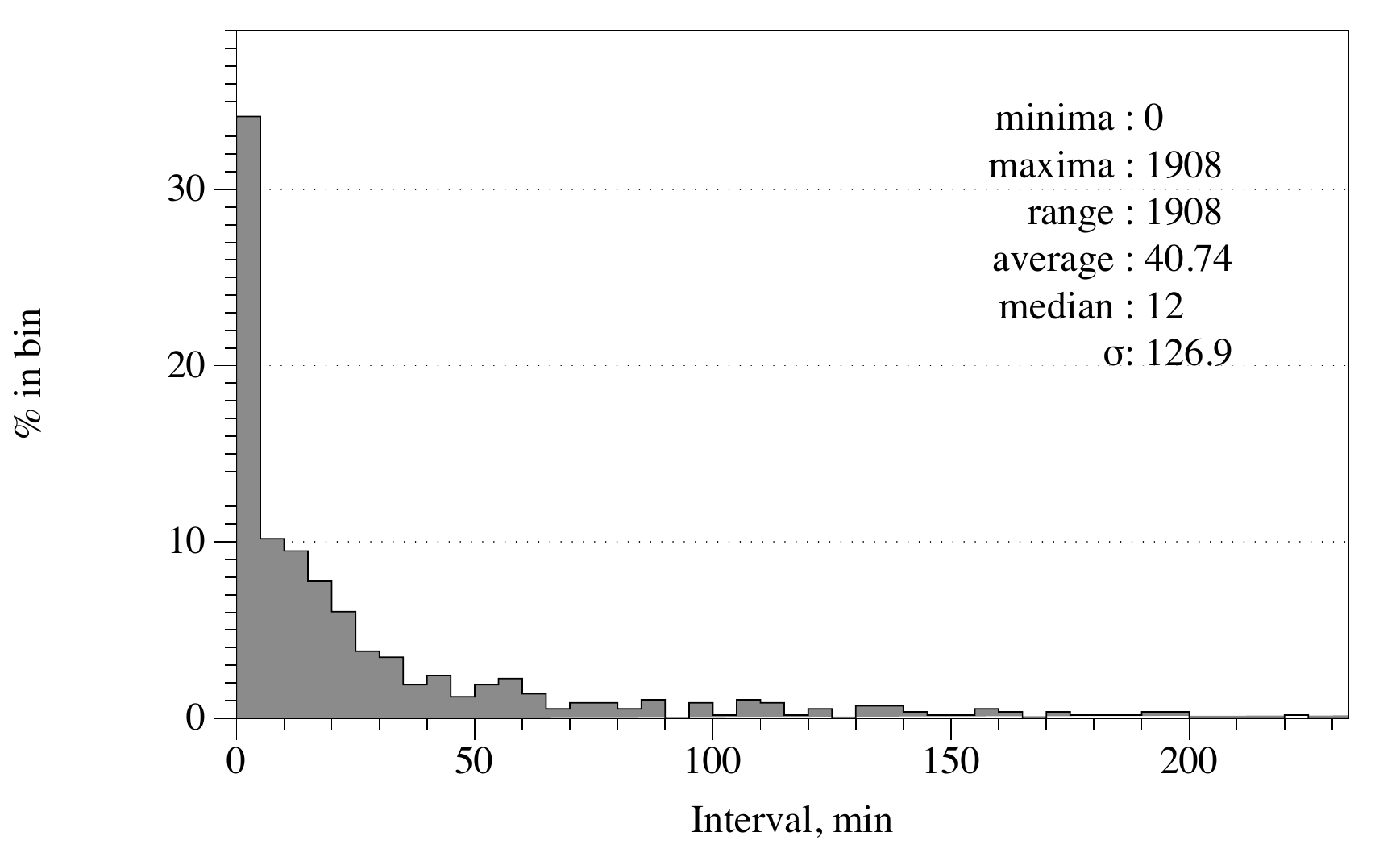}}
    \subfigure[]{\includegraphics[width=0.49\textwidth]{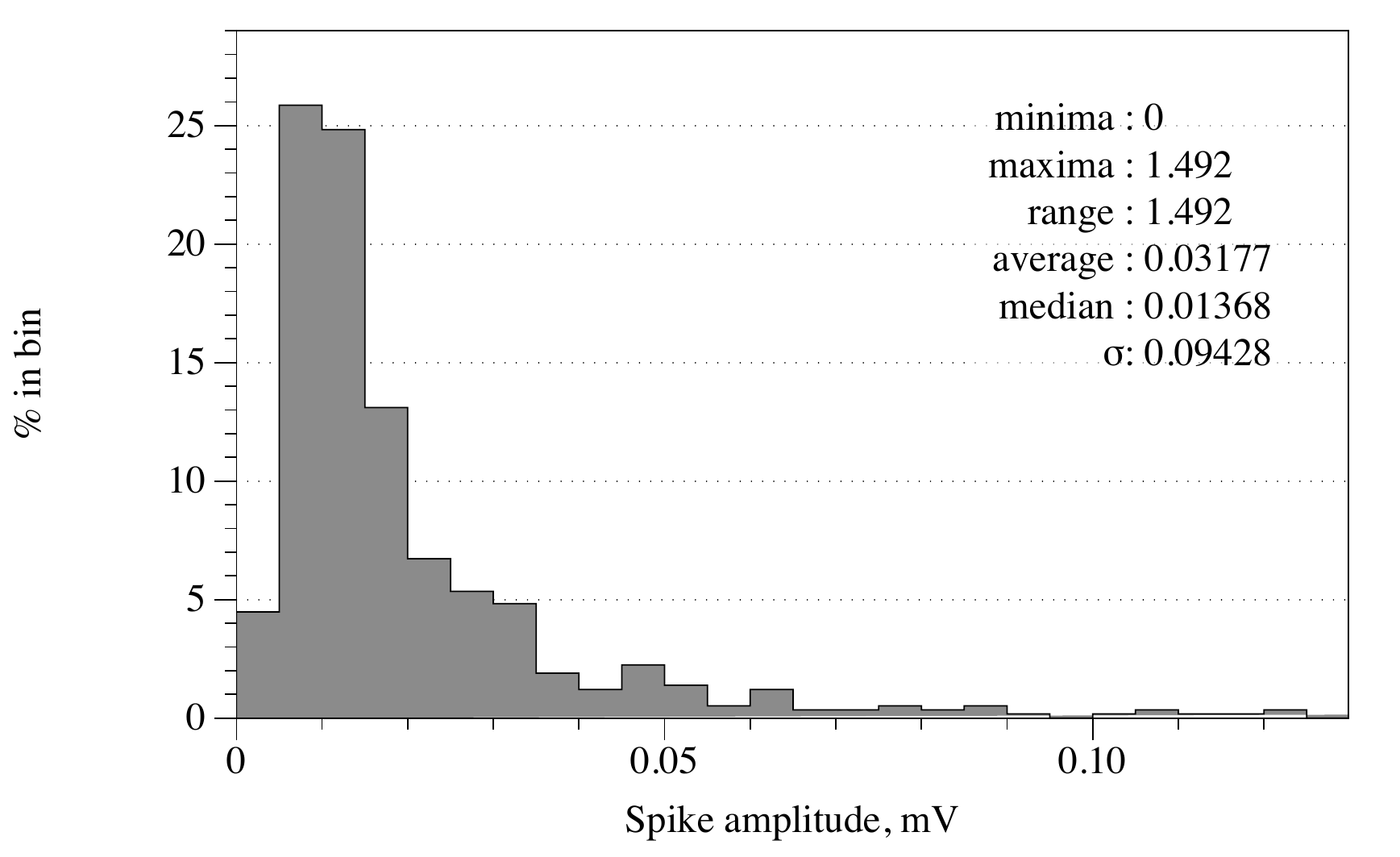}}
     \subfigure[]{\includegraphics[width=0.49\textwidth]{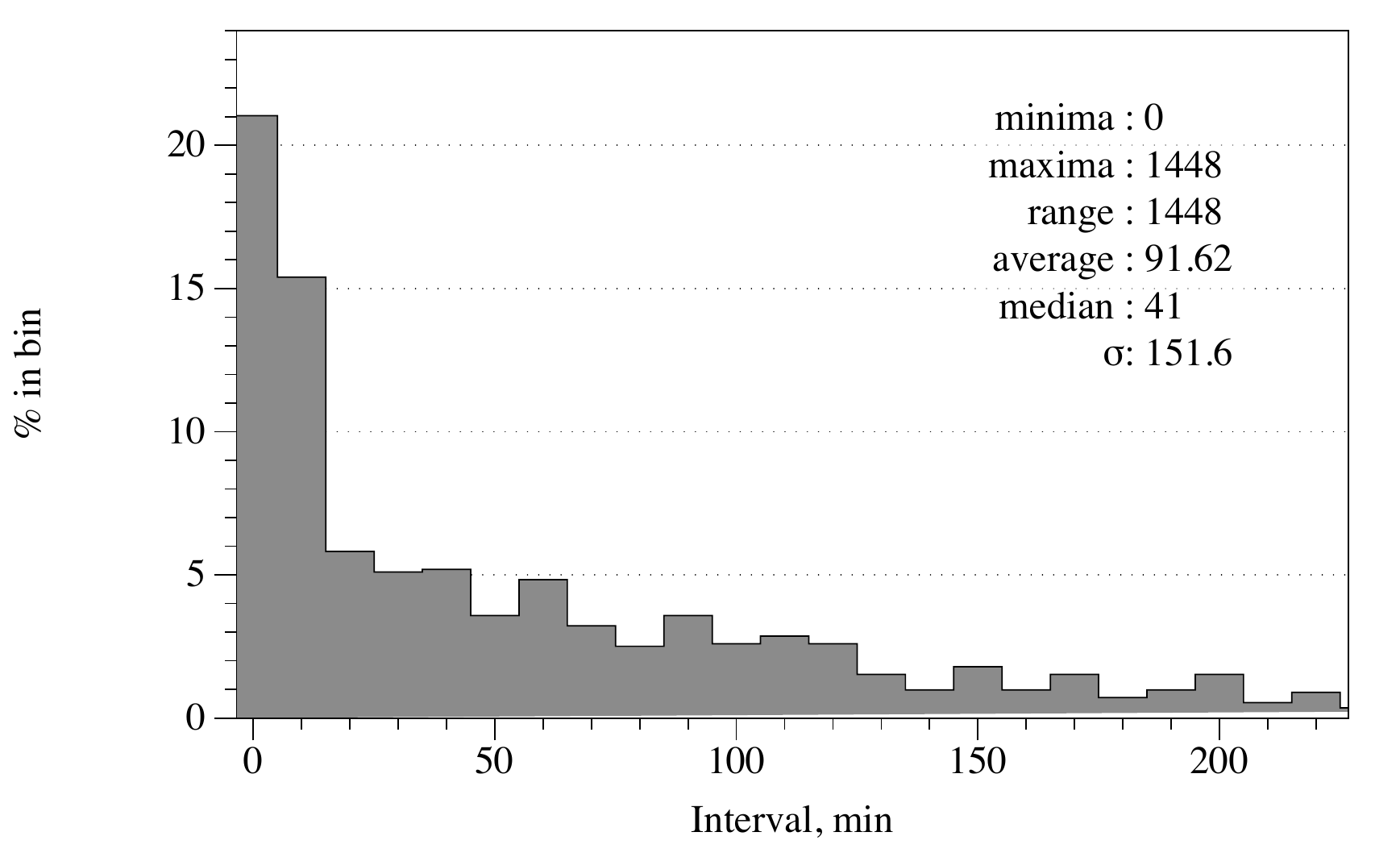}}
    \subfigure[]{\includegraphics[width=0.49\textwidth]{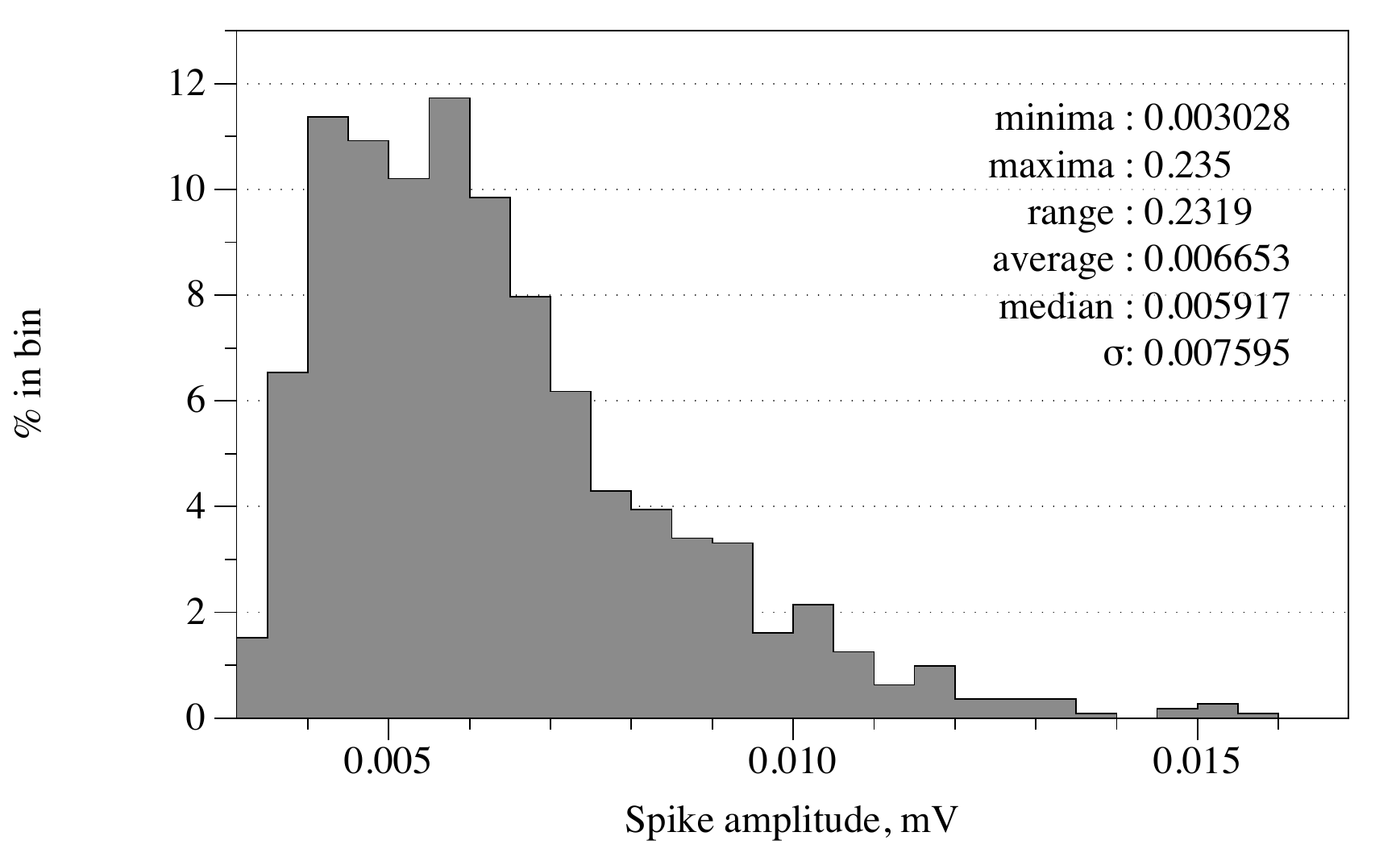}}
    \caption{Distribution of intervals between spikes (aceg) and average spike amplitude (bdfh) of (ab)~\emph{C. militaris}, (cd)~\emph{F. velutipes}, (ef)~\emph{S. commune}, (gh)~\emph{O. nidiformis}.}
    \label{fig:distribution}
\end{figure}

\begin{table}[]
    \centering
    \begin{tabular}{p{3cm}|l|l|l}
Species  & Number of spikes & Interval, min & Amplitude, mV  \\ \hline
\emph{C. militaris} &  881  & 116  & 0.2    \\
\emph{F. velutipes}  & 958 & 102  &  0.3 \\
\emph{S. commune}  &  530 & 41 & 0.03     \\
\emph{O. nidiformis} & 1117 & 92 & 0.007  \\
    \end{tabular}
    \caption{Characteristics of electrical potential spiking: number of spikes recorded, average interval between spikes, and average amplitude of a spike.}
    \label{tab:spiking}
\end{table}

Examples of electrical activity recorded are shown in Fig.~\ref{fig:spiking}.
Intervals between the spikes and amplitudes of spikes are characterised in Fig.~\ref{fig:distribution} and Tab.~\ref{tab:spiking}. 

\emph{C. militaris} shows the lowest average spiking frequency amongst the species recorded (Fig.~\ref{cordyceps} and Fig.~\ref{fig:distribution}a): average interval between spikes is nearly two hours. The diversity of the frequencies recorded is highest amongst the species studied: standard deviation is over five hours. The spikes detected in \emph{C. militaris} and \emph{F. velutipes} have highest amplitudes: 0.2~mV and 0.3~mV, respectively. Variability of the amplitudes in both species is high, standard deviation nearly 0.3. 

Enoki fungi \emph{F. velutipes} show a rich spectrum of diverse patterns of electrical activity which combines low and high frequency oscillations (Fig.~\ref{enokispiking}). Most commonly exhibited patterns are characterised by low frequency irregular oscillations: average amplitude 0.3~mV (Fig.~\ref{fig:distribution}d) and average interval between two spikes is just over 1.5 hr (Fig.~\ref{fig:distribution}c and Tab.~\ref{tab:spiking}). There are also bursts of spiking showing a transition from a low frequency spiking to high frequency and back, see recording in blue in Fig.~\ref{enokispiking}. There are 12 spikes in the train, average amplitude is 2.1~mV, $\sigma=0.1$, average duration of a spike is 64~min, $\sigma=1.7$.

\emph{O. nidiformis} also show low amplitude and low frequency electrical spiking activity with the variability of the characteristics highest amongst species recorded (Fig.~\ref{ghostspiking} and Tab.~\ref{tab:spiking}).  
Average interval between the spikes is just over 1.5 hr with nearly 2.5 hr standard variation (Fig.~\ref{fig:distribution}g).
Average amplitude is 0.007~mV but the variability of the amplitudes is very high:  $\sigma=0.006$ (Fig.~\ref{fig:distribution}h).

\begin{figure}[!tbp]
    \centering
    \subfigure[]{\includegraphics[width=0.49\textwidth]{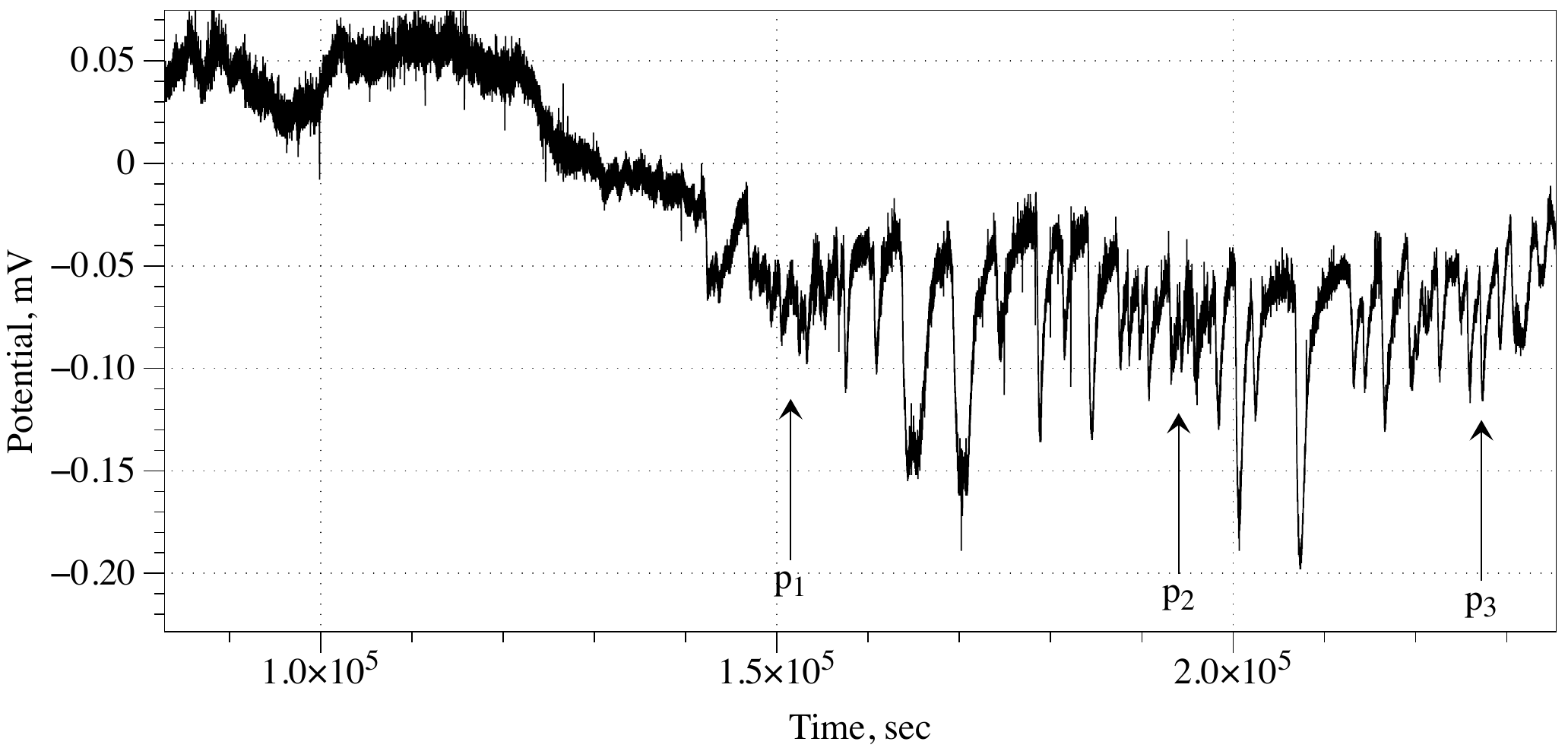}\label{Schizophyllum_Transition2Spiking}}
    \subfigure[]{\includegraphics[width=0.49\textwidth]{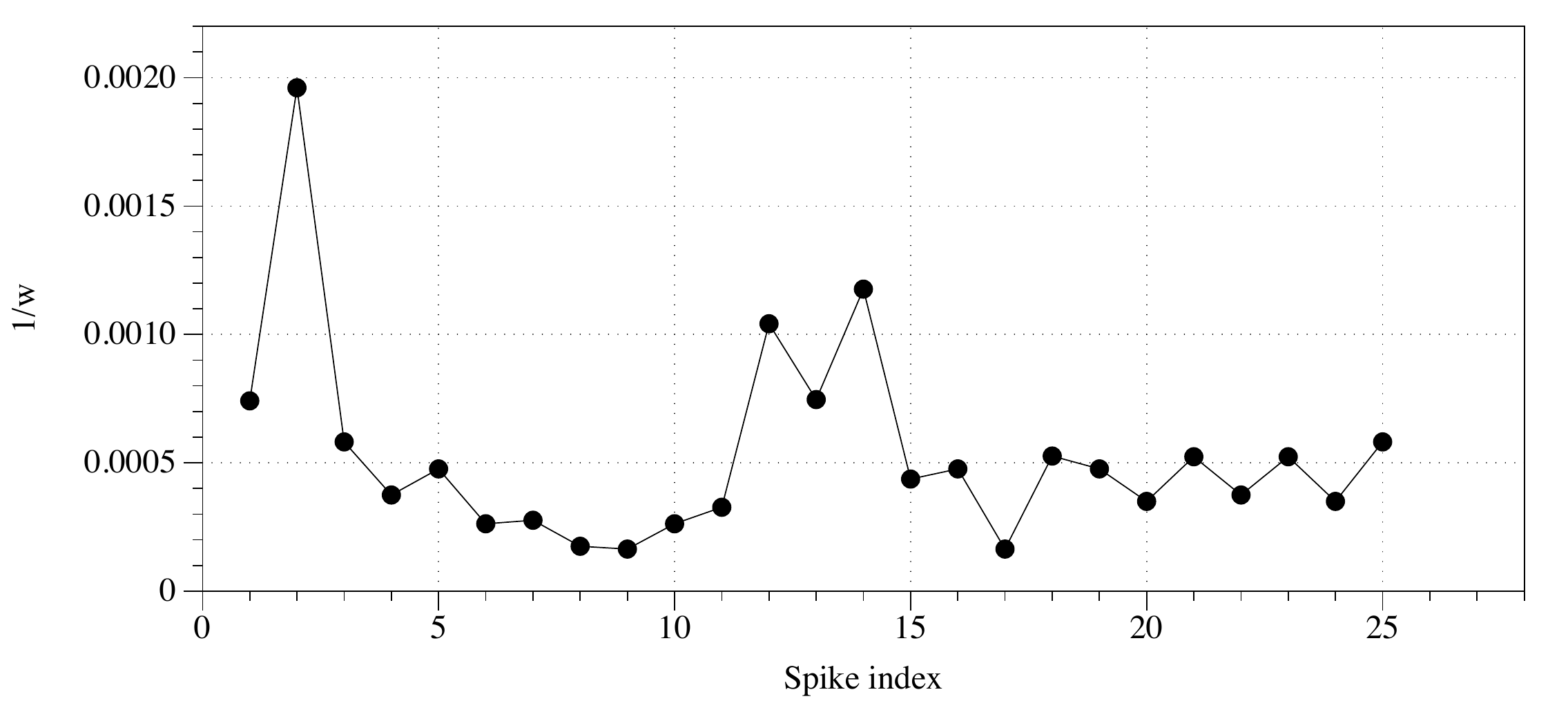}\label{frequyencyOfspikingInTheTransition2Spiking}}
        \subfigure[]{\includegraphics[width=0.49\textwidth]{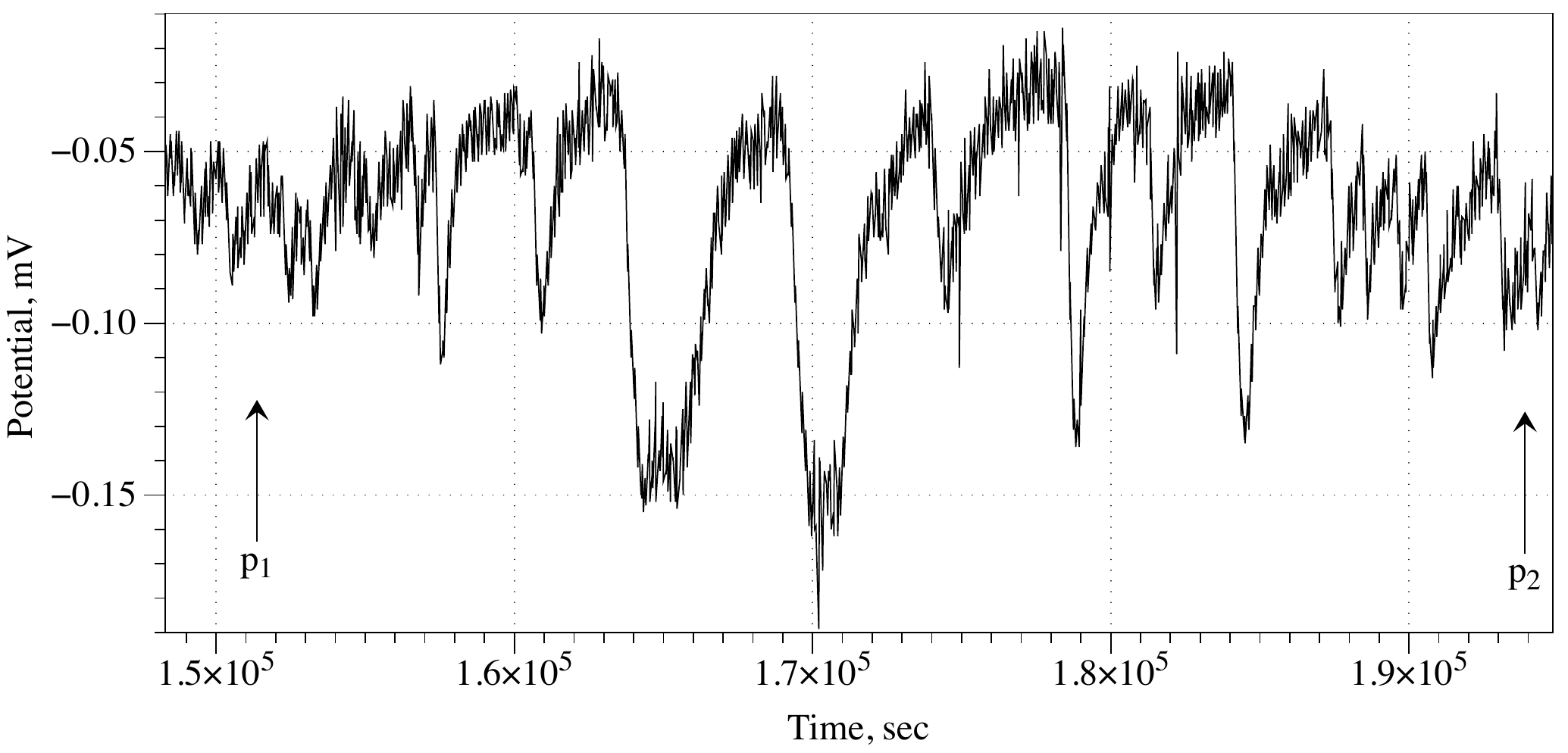}\label{Schizophyllum_One_Wave_Packets}}
    \caption{Transition to spikes outburst in \emph{S. commune}.}
    \label{fig:Schizophyllum_Transition2Spiking}
\end{figure}

\begin{figure}[!tbp]
    \centering
    \subfigure[]{\includegraphics[width=0.49\textwidth]{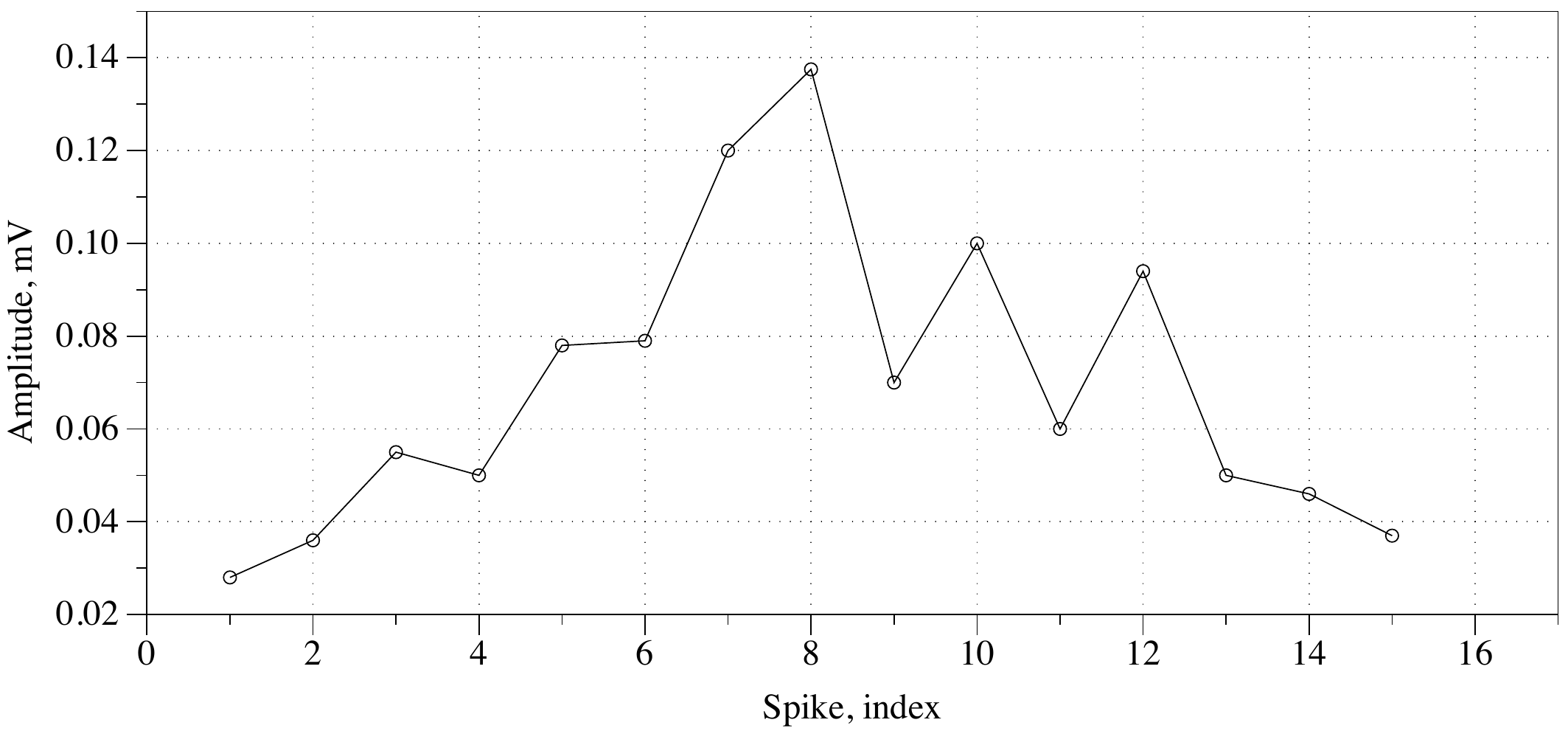}\label{WavePacketSpikeAmplitude}}
    \subfigure[]{\includegraphics[width=0.49\textwidth]{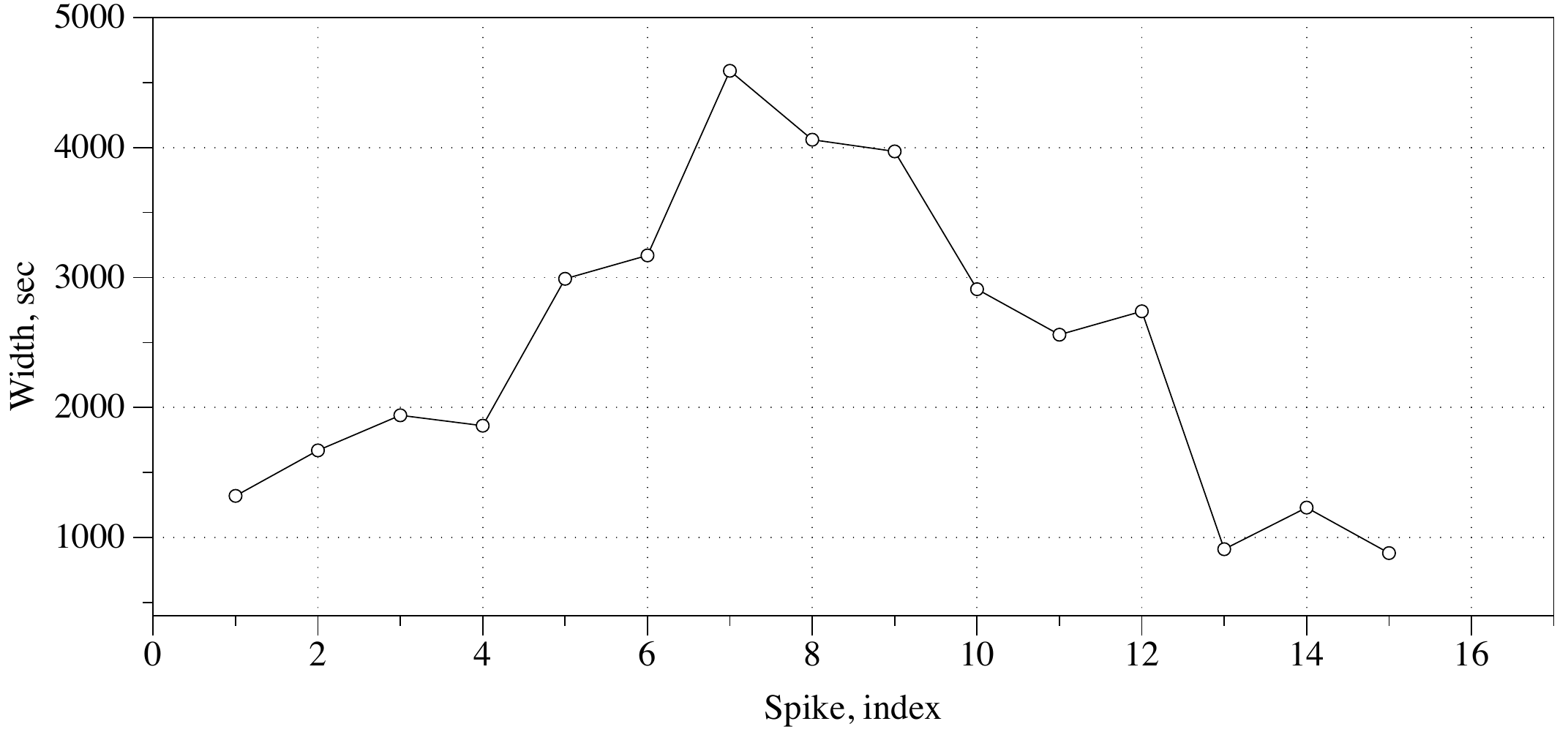}\label{WavePacketSpikeWidth}}
    \caption{Characteristics of an exemplar wave packet of electrical potential oscillation in \emph{S. commune}: (a)~evolution of spike amplitude, (b)~evolution of spike width.}
    \label{fig:wavepacket}
\end{figure}

\emph{S. commune} electrical activity is remarkably diverse (Figs.~\ref{schizophyllumspiking} and \ref{fig:distribution}ef). Typically, there are low amplitude spikes detected (Fig.~\ref{fig:distribution}f), due to the reference electrodes in each differential pair being inserted into the host wood. However, they are the fastest spiking species, with an average interval between spikes is just above half-an-hour (Fig.~\ref{fig:distribution}e).  We observed transitions between different types of spiking activity from low amplitude and very low frequency spikes to high amplitude high frequency spikes (Fig.~\ref{fig:Schizophyllum_Transition2Spiking}). 
A dynamic change in spikes frequency in the transition Fig.~\ref{fig:Schizophyllum_Transition2Spiking} is shown in Fig.~\ref{frequyencyOfspikingInTheTransition2Spiking}. A closer look at the spiking discovers presence of two wave packets labelled $(p_1,p_2)$ and $(p_2,p_3)$ in Fig.~\ref{Schizophyllum_Transition2Spiking}. One of the wave-packets is shown in Fig.~\ref{Schizophyllum_One_Wave_Packets}, and the key characteristics are shown in Fig.~\ref{fig:wavepacket}.

\begin{figure}[!tbp]
    \centering
    \subfigure[]{\includegraphics[width=0.49\textwidth]{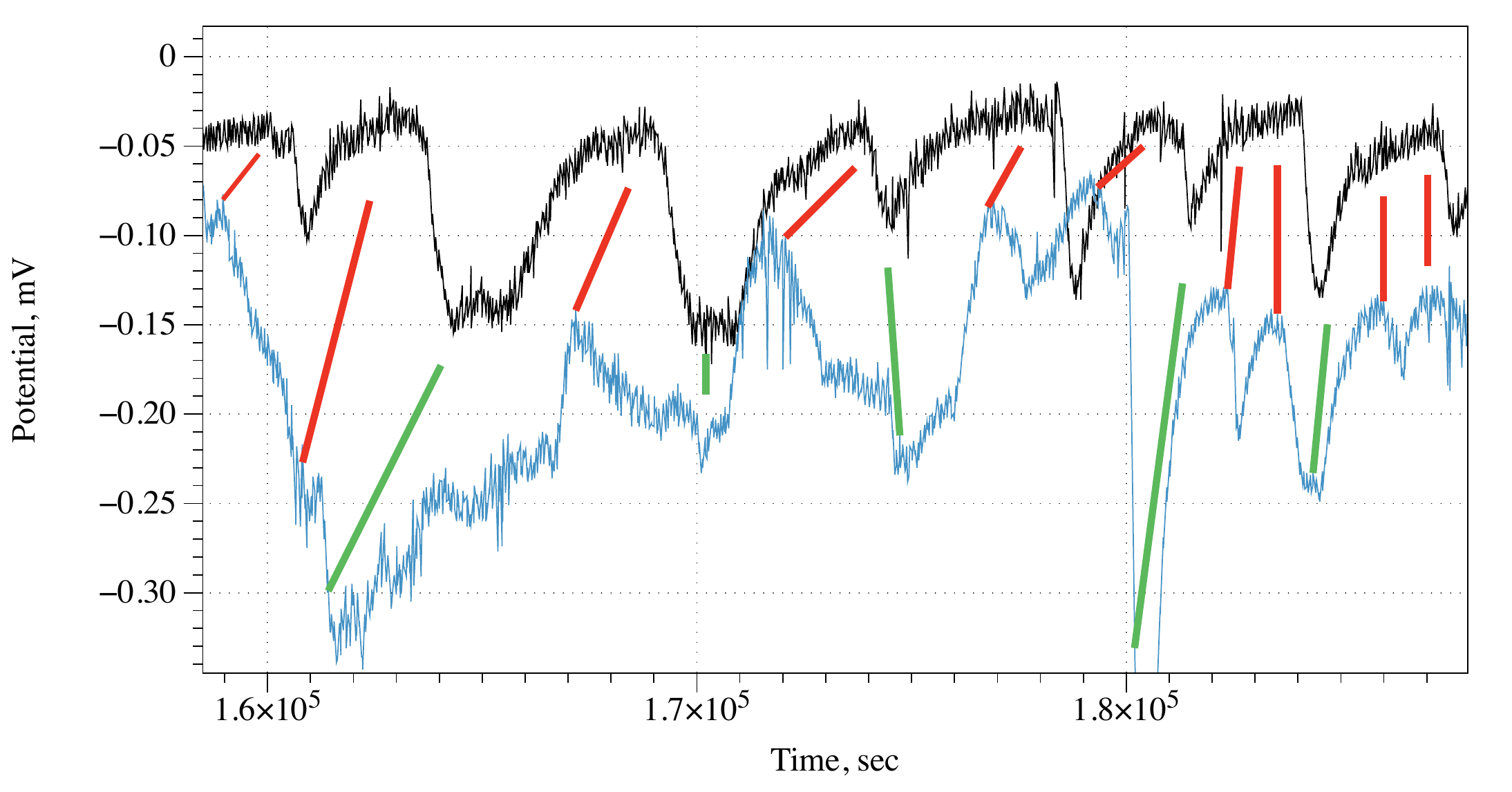}\label{SchizophyllumSynchronisation}}
    \subfigure[]{\includegraphics[width=0.49\textwidth]{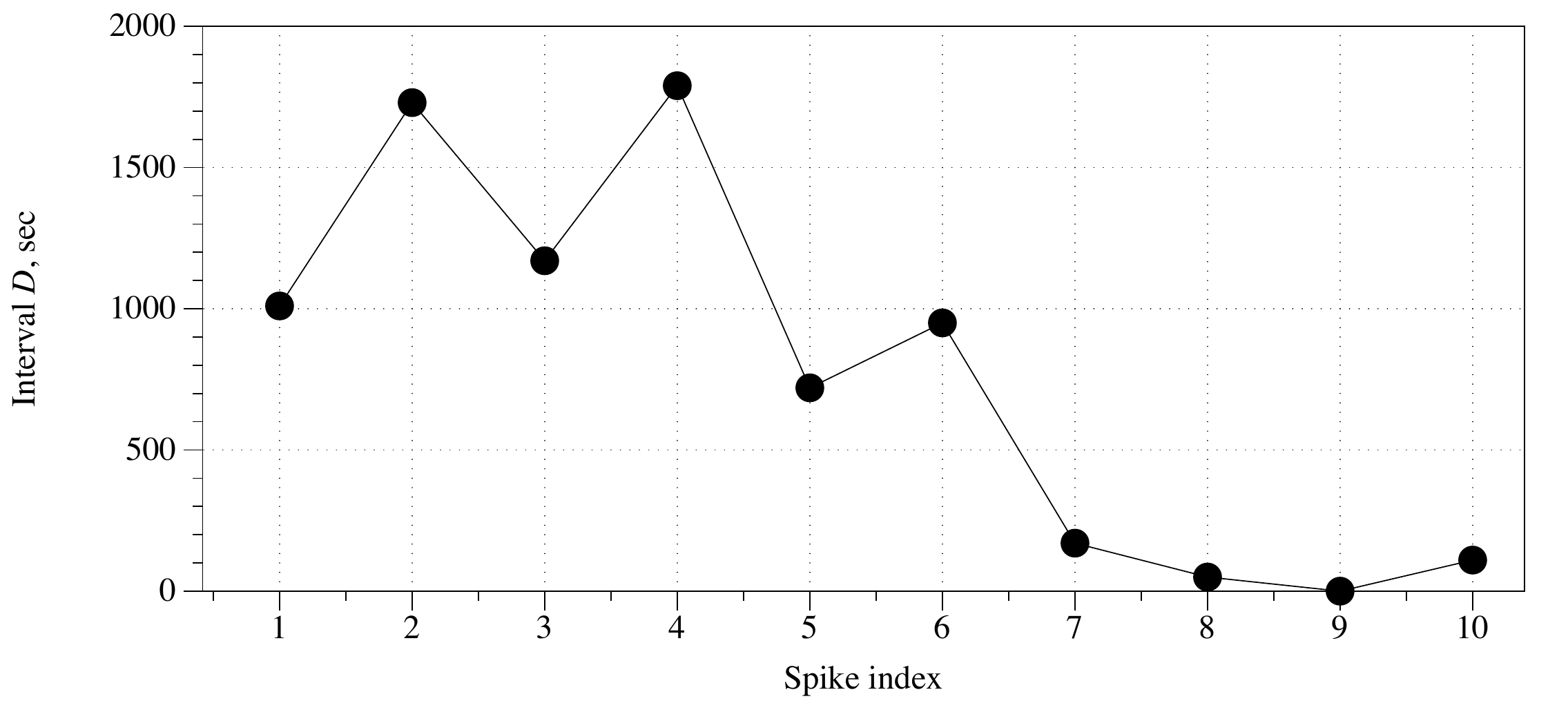}\label{IntervalBetweenSpikesOnNeighbouringChannels}}
    \caption{Exemplar synchronisation of spikes in two neighbouring sporocarps of \emph{S. commune}: channel (3-4), second sporocarp in Fig.~\ref{Schizophyllum_communesetup}, and channel (5-6),  third sporocarp in Fig.~\ref{Schizophyllum_communesetup}. (a)~Spiking activity, corresponding spikes of increased voltage are linked by red lines and decreased voltage by green line.}
    \label{fig:SchizophyllumSynchronisationExample}
\end{figure}

In experiments with  \emph{S. commune} we observed synchronisation of the electrical potential spikes recorded on the neighbouring fruit bodies. This is illustrated in Fig.~\ref{fig:SchizophyllumSynchronisationExample}. The dependencies between the spikes are shown by red (increase of potential spike) and green (decrease of potential spike) lines in Fig.~\ref{SchizophyllumSynchronisation}. Time intervals between peaks of the spikes occurred on neighbouring fruit bodies are illustrated in Fig.~\ref{IntervalBetweenSpikesOnNeighbouringChannels}. Average interval between first four spikes is 1425~sec ($\sigma=393$), next three spikes 870~sec ($\sigma=113$), and last four spikes 82 ($\sigma=73$).

\section{Towards language of fungi}
\label{language}

Are the elaborate patterns of electrical activity used by fungi to communicate states of the mycelium and its environment and to transmit and process information in the mycelium networks? Is there a language of fungi?  When interpreting fungal spiking patterns as a language we should consider a number of linguistic phenomena as have been successfully used to decode pictish symbols revealed as a written language in \cite{lee2010pictish}: (1) type of characters used to code, (2) size of the character lexicon, (3) grammar, (4) syntax (word order), (5) standardised spelling. These phenomena, apart from grammar and spelling, are analysed further.

\begin{figure}[!tbp]
    \centering
    \includegraphics[width=0.99\textwidth]{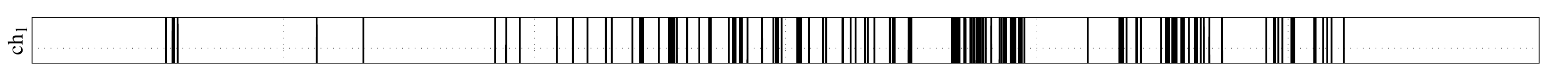}
    \includegraphics[width=0.99\textwidth]{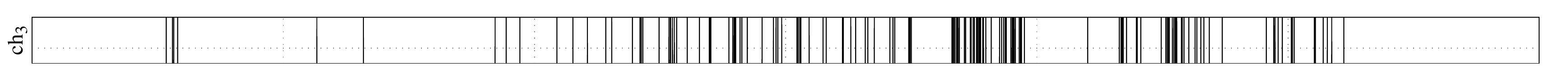}
    \includegraphics[width=0.99\textwidth]{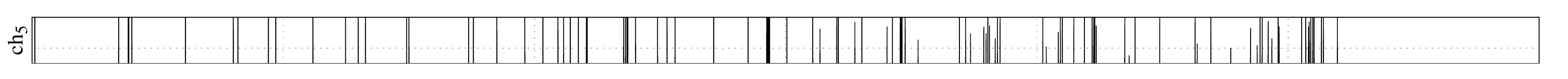}
    \includegraphics[width=0.99\textwidth]{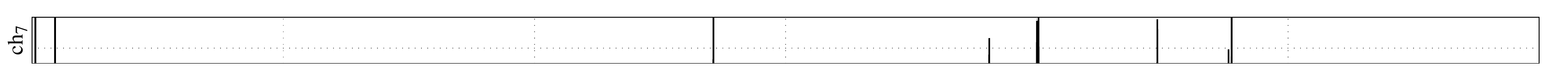}
    \includegraphics[width=0.99\textwidth]{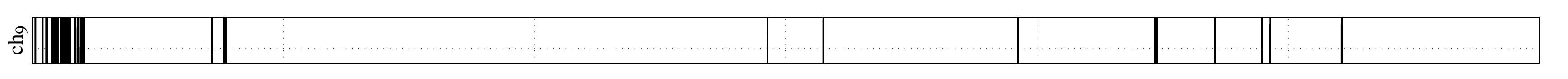}
    \includegraphics[width=0.99\textwidth]{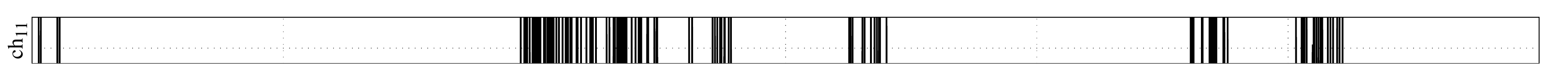}
    \includegraphics[width=0.99\textwidth]{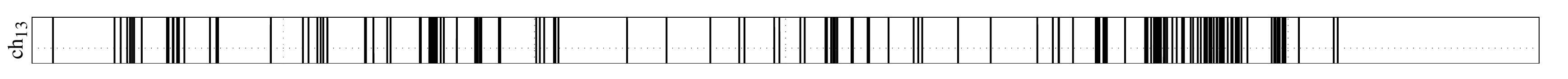}
    (a)
    
\vspace{5mm}

    \includegraphics[width=0.99\textwidth]{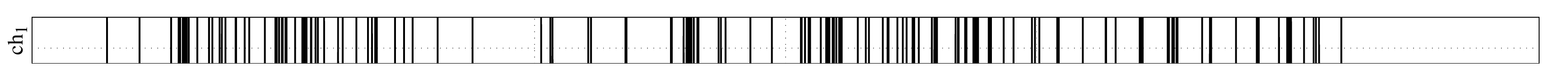}
    \includegraphics[width=0.99\textwidth]{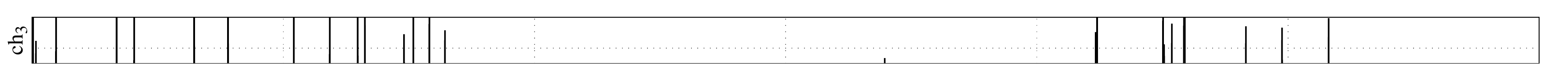}
    \includegraphics[width=0.99\textwidth]{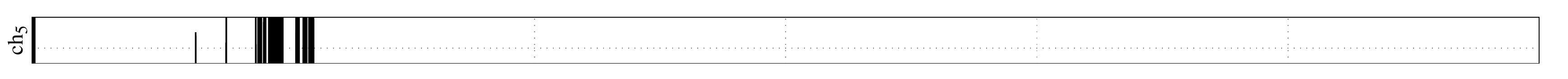}
    \includegraphics[width=0.99\textwidth]{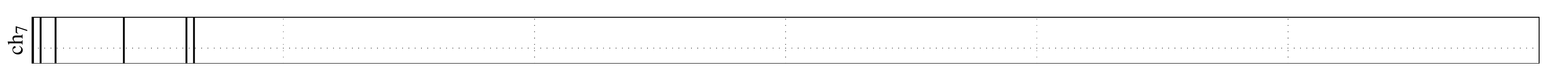}
    \includegraphics[width=0.99\textwidth]{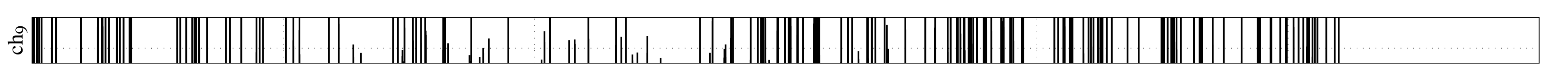}
    \includegraphics[width=0.99\textwidth]{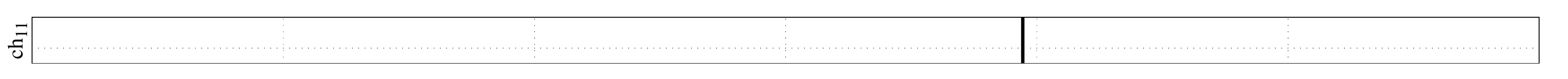}
    \includegraphics[width=0.99\textwidth]{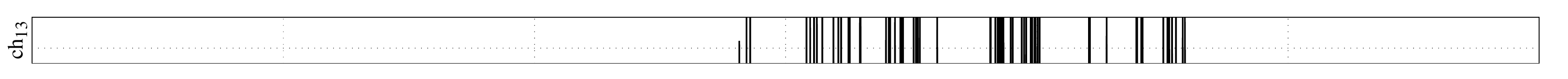}
    (b)
\caption{Bar code like presentation of spikes recorded in 
(a)~\emph{C. milataris}
(b)~\emph{F. velutipes}, 
five days of recording.}
    \label{fig:Barcode}
\end{figure}

To quantify types of characters used and a size of lexicon we convert 
the spikes detected in experimental laboratory recordings to binary strings $s$, where index $i$ is the index of the sample taken at $i$th second of recording and $s_i=1$ if there is a spike's peak at $i$th second and $s_0=0$ otherwise. Examples of the binary strings, in a bar code like forms, extracted from the electrical activity of \emph{C. militaris}  and \emph{F. velutipes} are shown in Fig.~\ref{fig:Barcode}.

\begin{figure}[!tbp]
    \centering
    \subfigure[]{\includegraphics[width=0.49\textwidth]{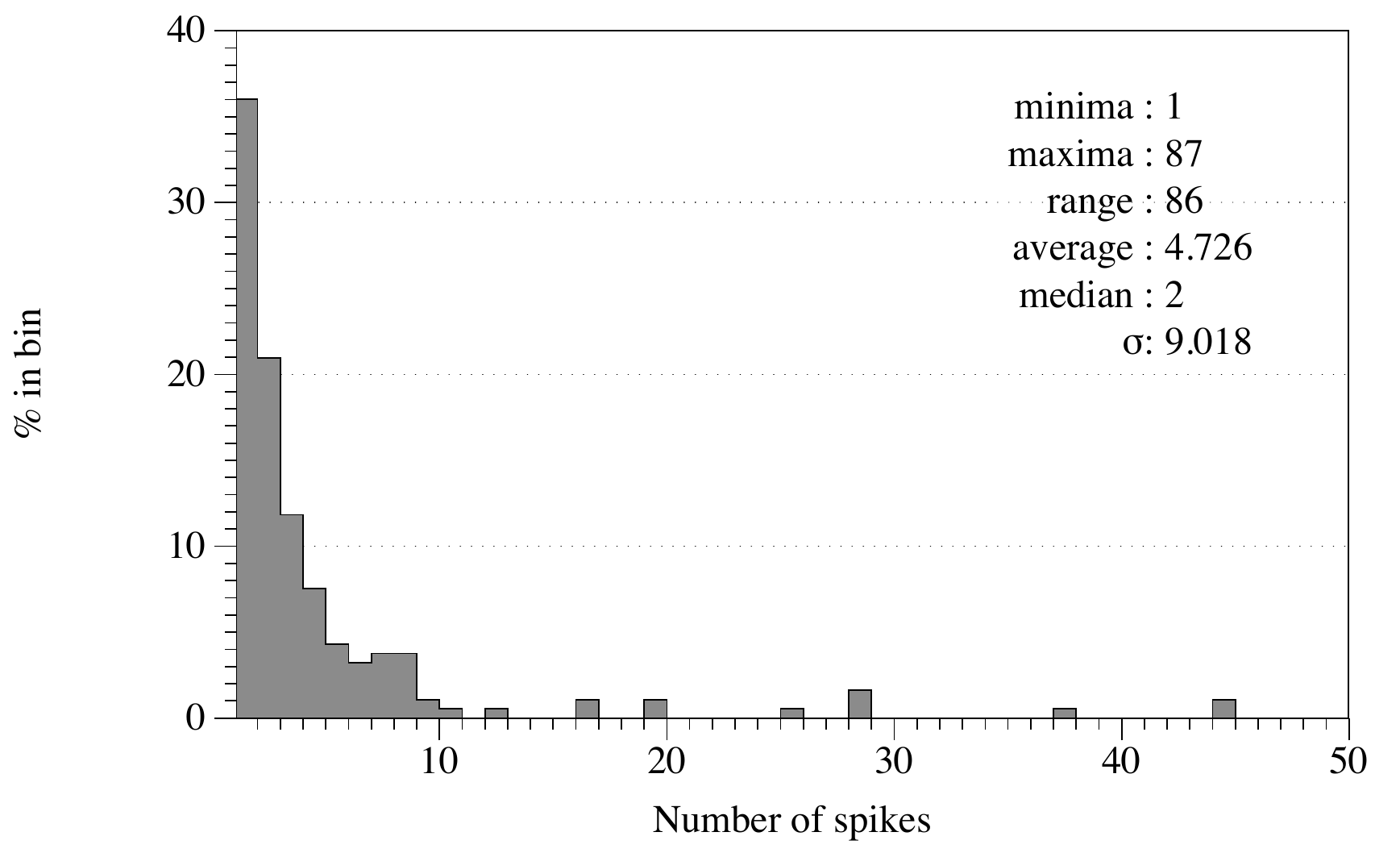}}
        \subfigure[]{\includegraphics[width=0.49\textwidth]{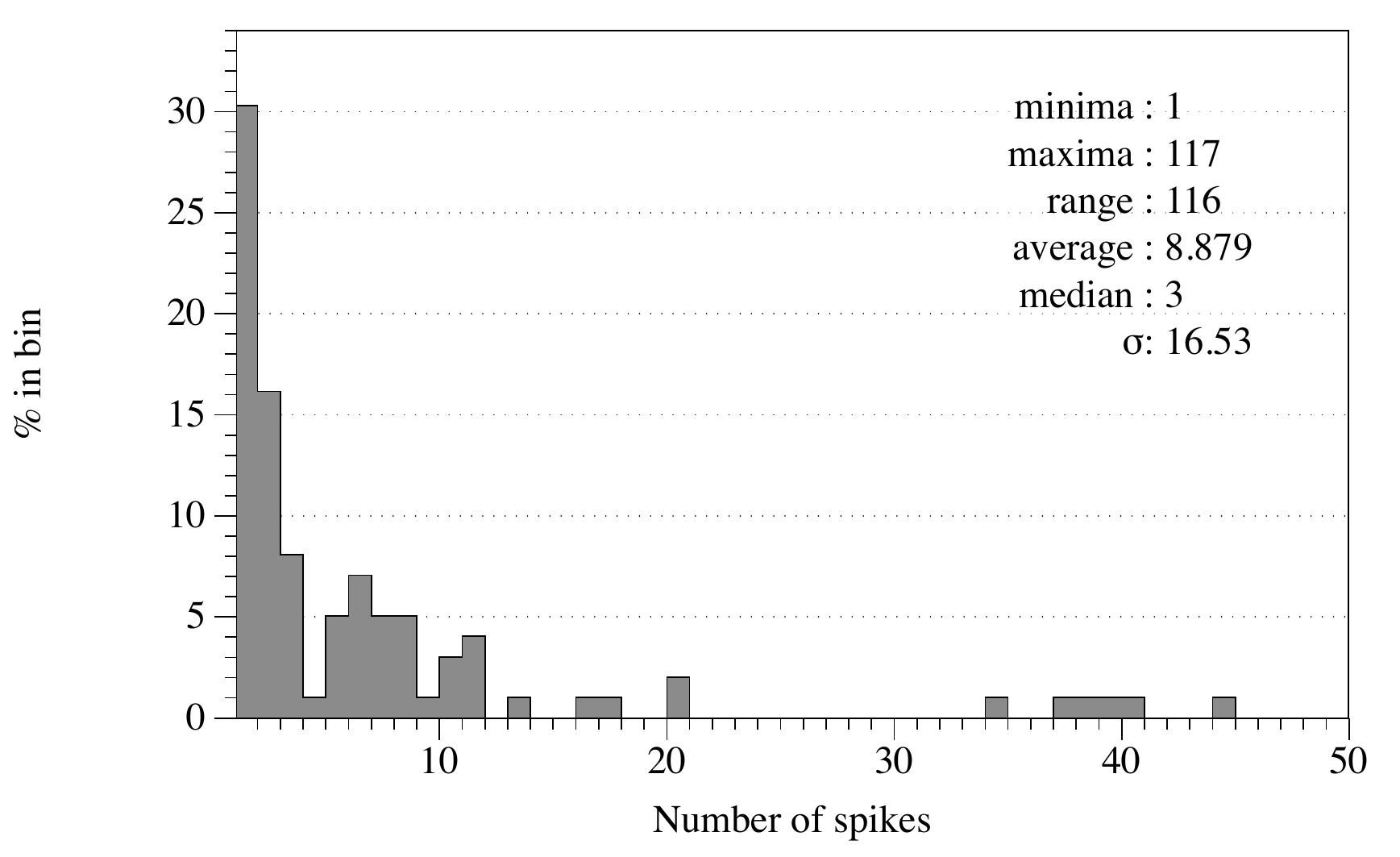}}
     \subfigure[]{\includegraphics[width=0.49\textwidth]{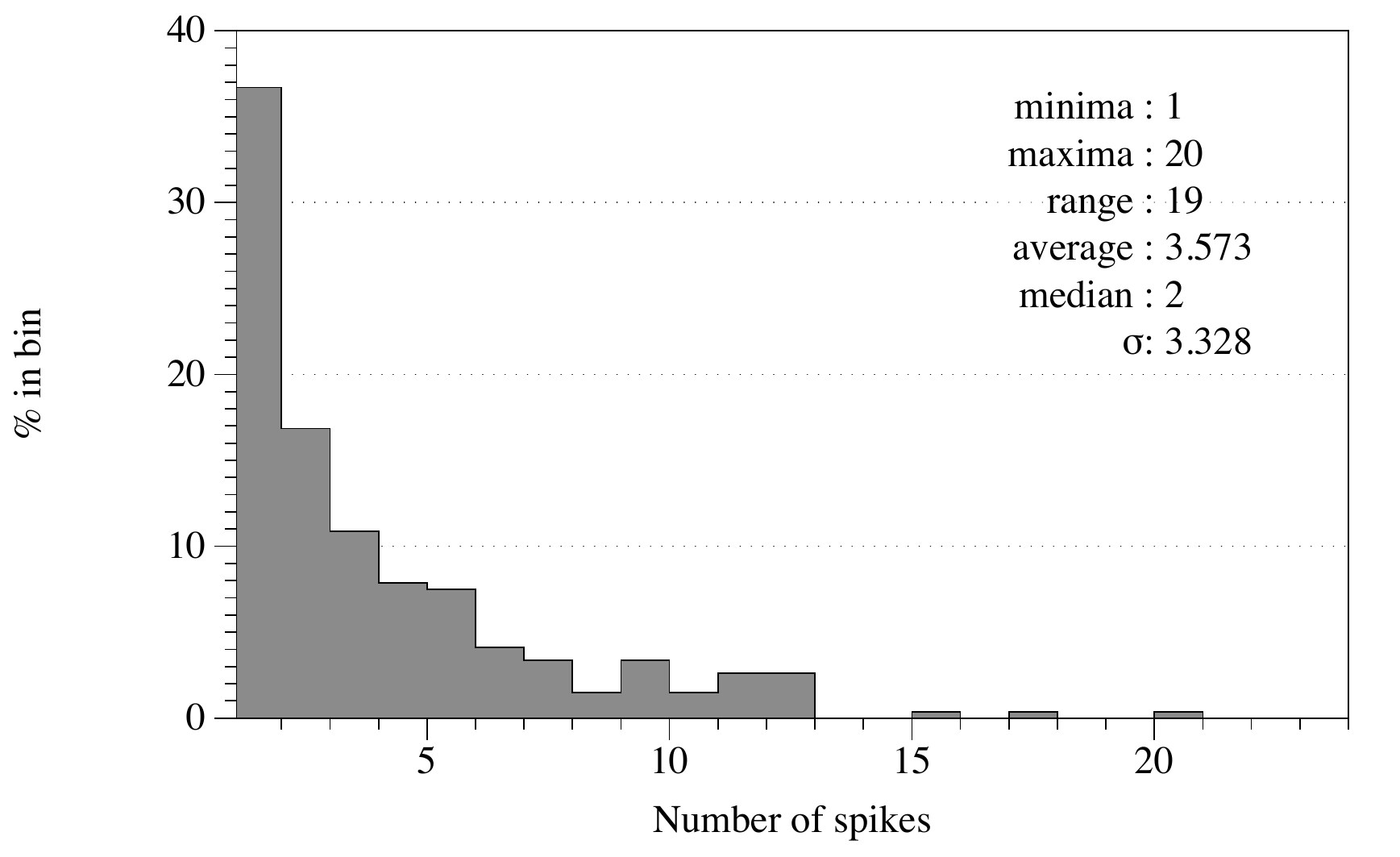}}
     \subfigure[]{\includegraphics[width=0.49\textwidth]{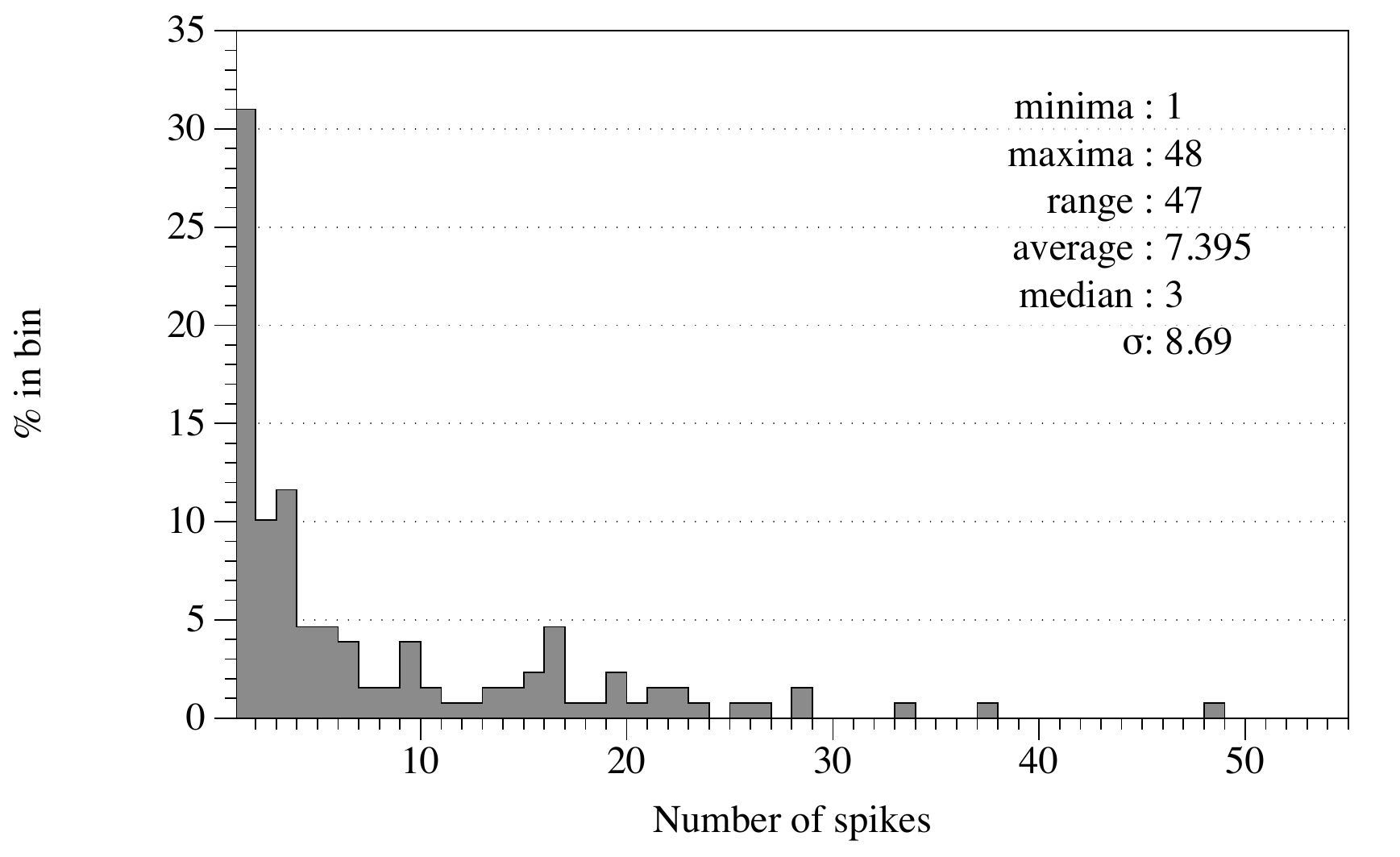}}
     \subfigure[]{\includegraphics[width=0.49\textwidth]{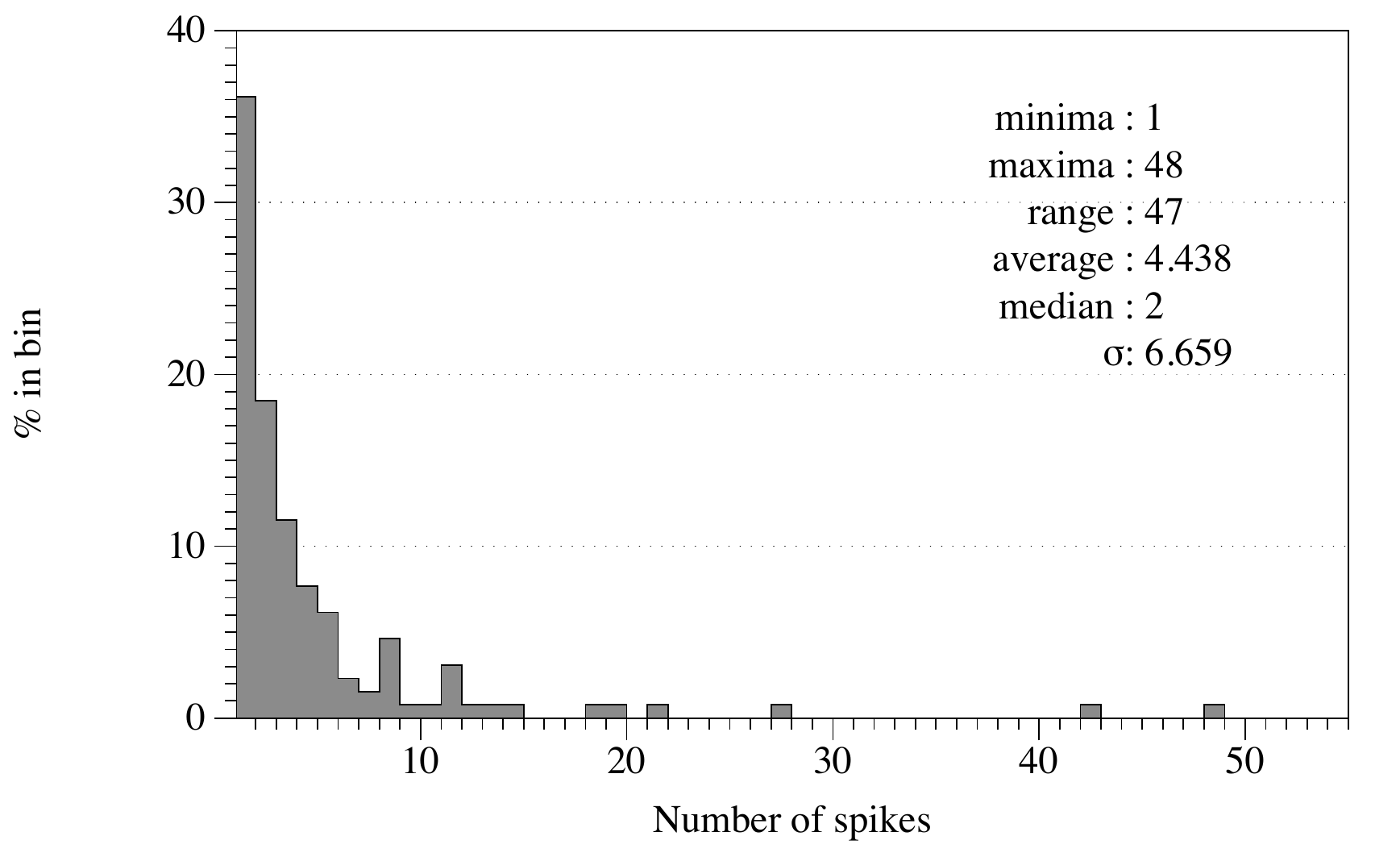}}
     \subfigure[]{\includegraphics[width=0.49\textwidth]{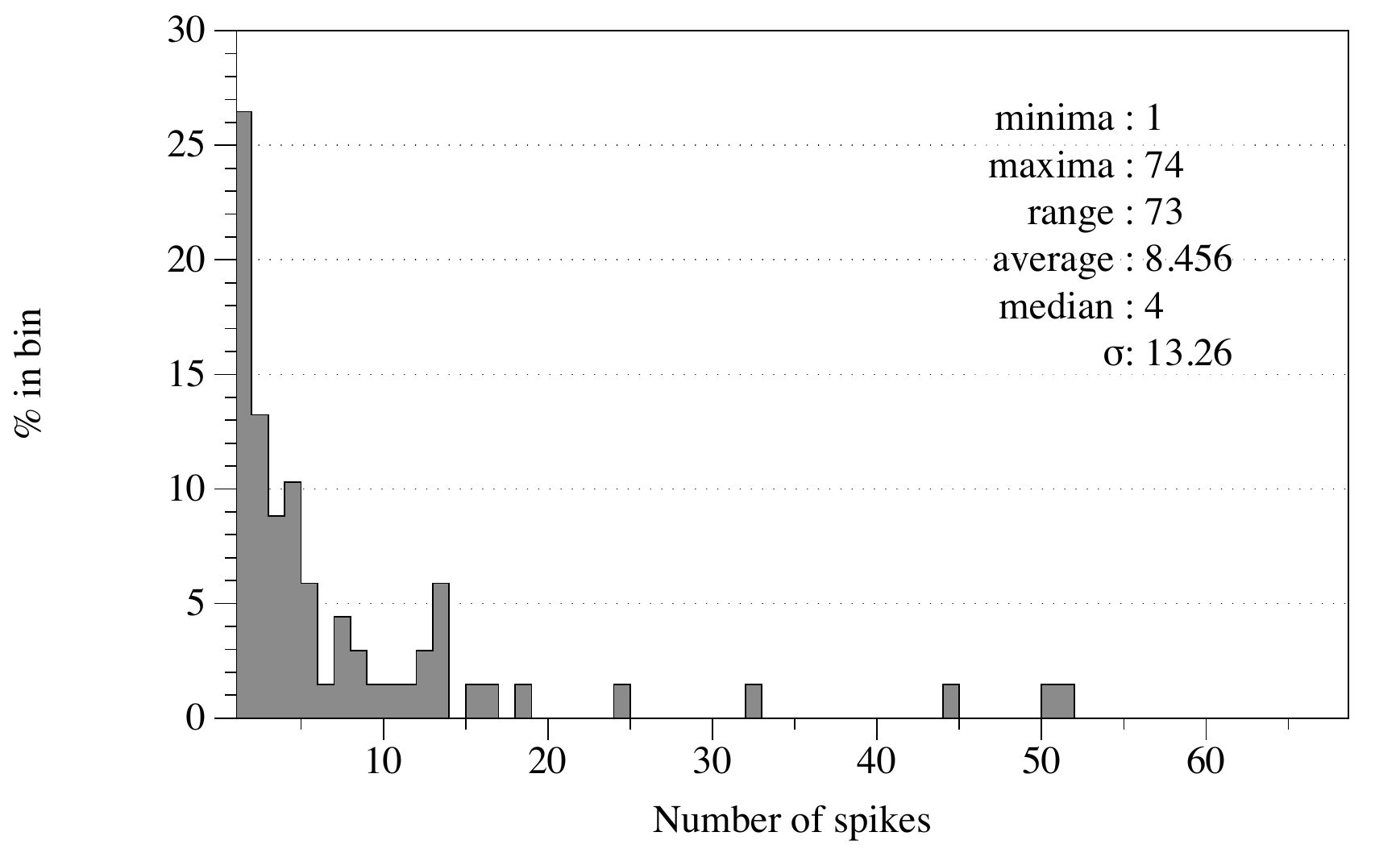}}
     \subfigure[]{\includegraphics[width=0.49\textwidth]{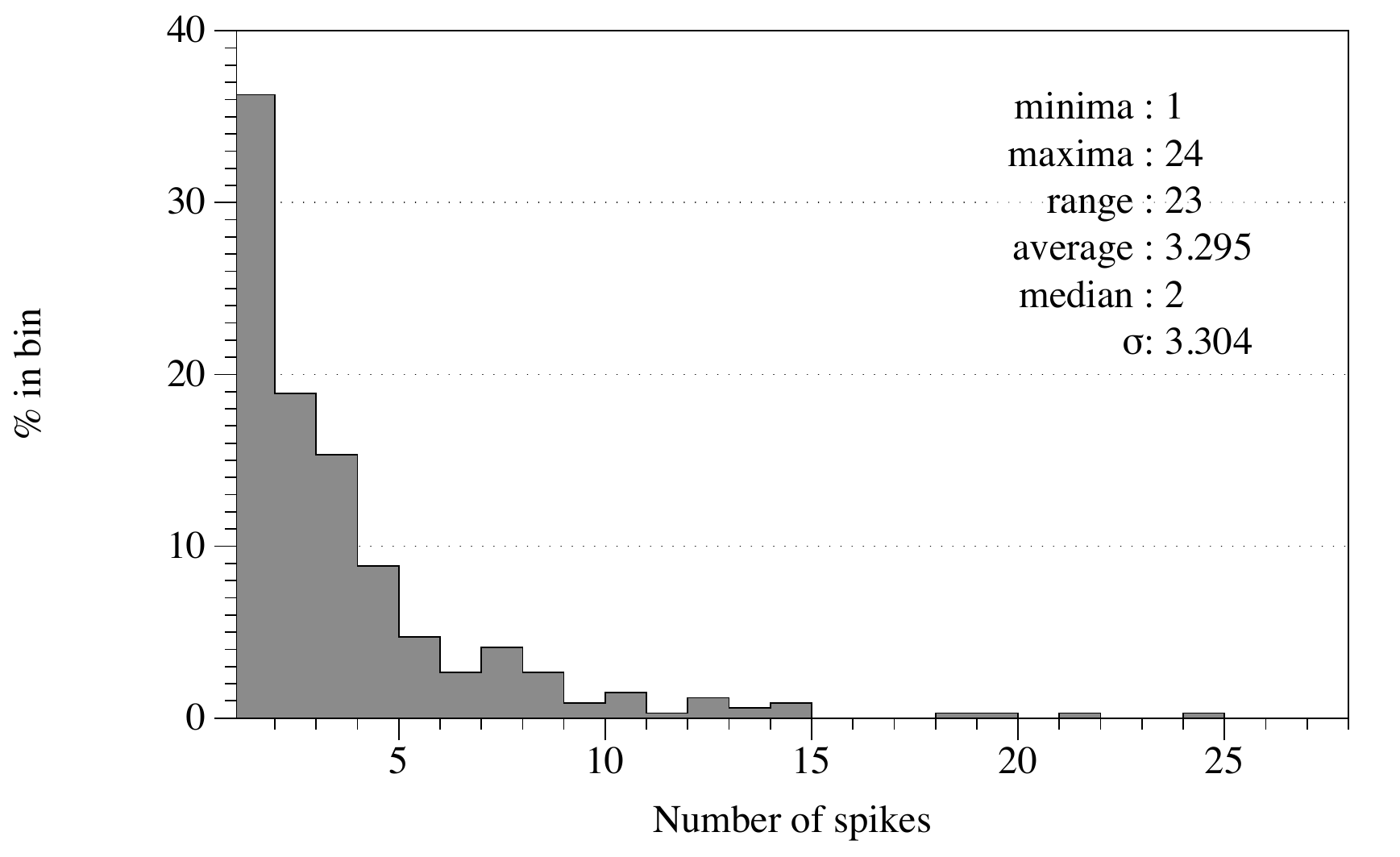}}
     \subfigure[]{\includegraphics[width=0.49\textwidth]{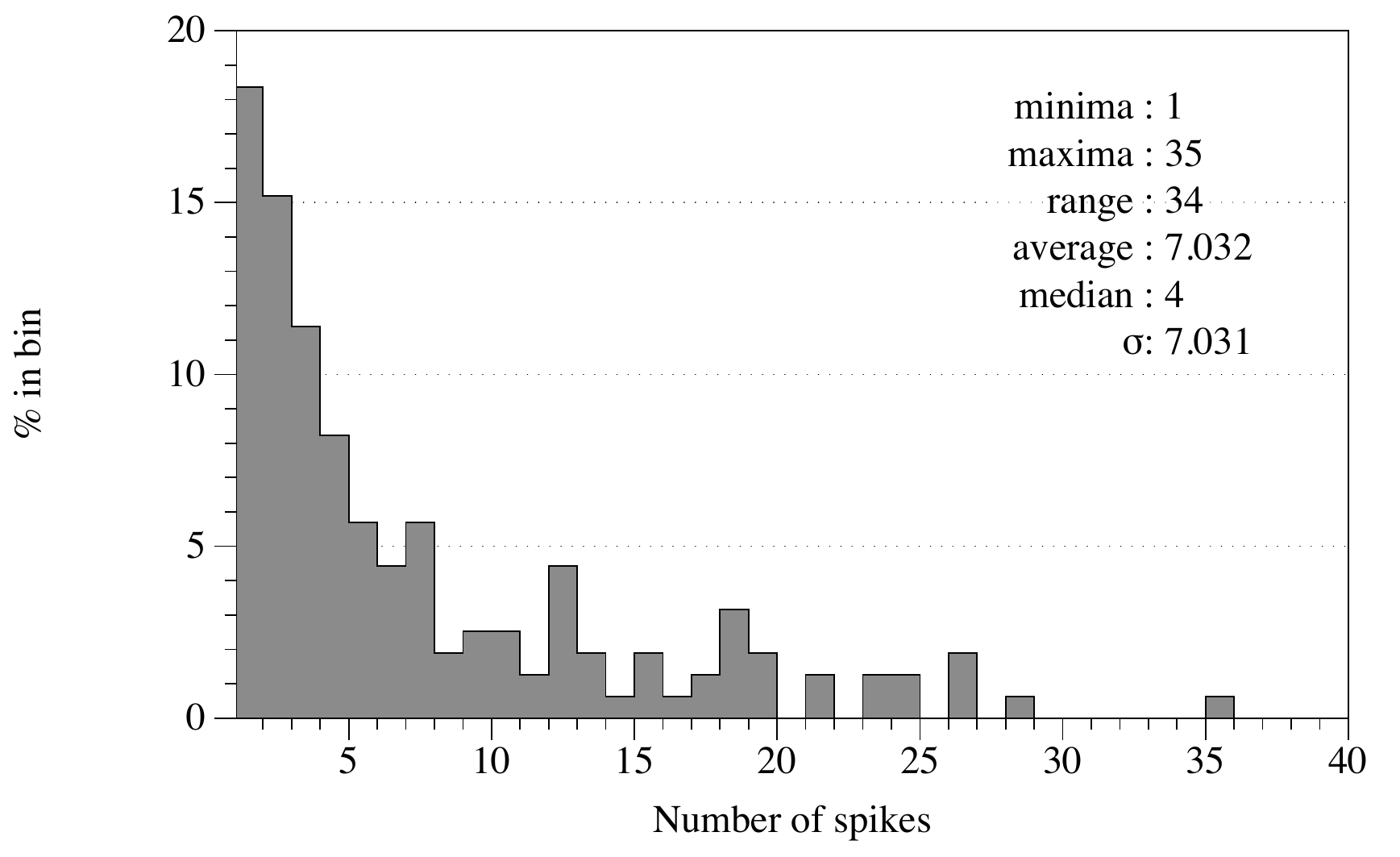}}
    \caption{Distribution of a number of spikes in trains, i.e. of the fungal words' lengths, of (ab)~\emph{C. militaris}, (cd)~\emph{F. velutipes}, (ef)~\emph{S. commune}, (gh)~\emph{O. nidiformis} for the train separation thresholds $a$ (acef) and $2\cdot a$ (bdfh), where $a$ is a species specific average interval between two consequent spikes, see Tab.~\ref{tab:spiking}. }
    \label{fig:trainsdistributions}
\end{figure}

\begin{figure}[!tbp]
    \centering
\subfigure[]{\includegraphics[width=0.45\textwidth]{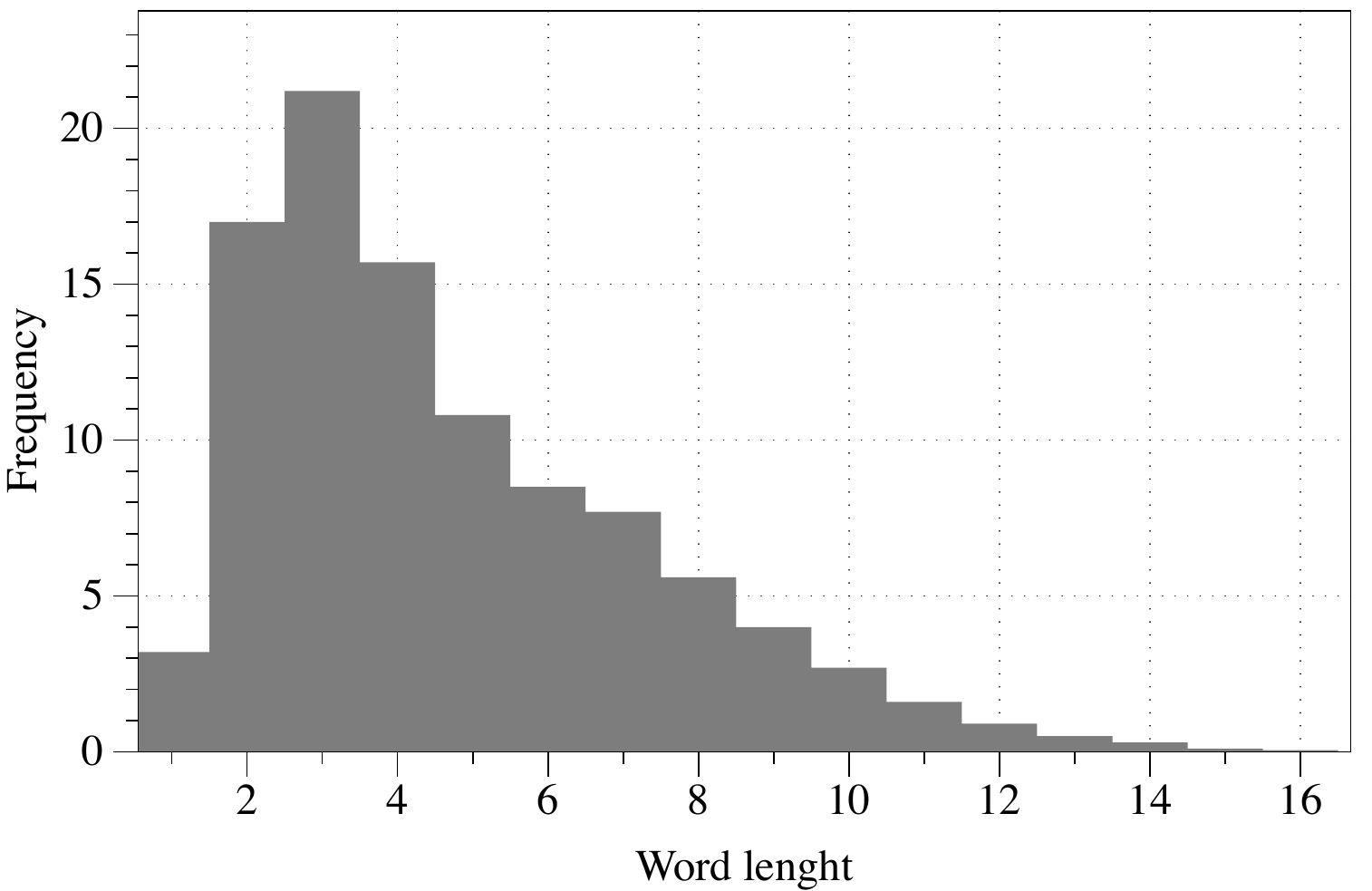}}
\subfigure[]{\includegraphics[width=0.45\textwidth]{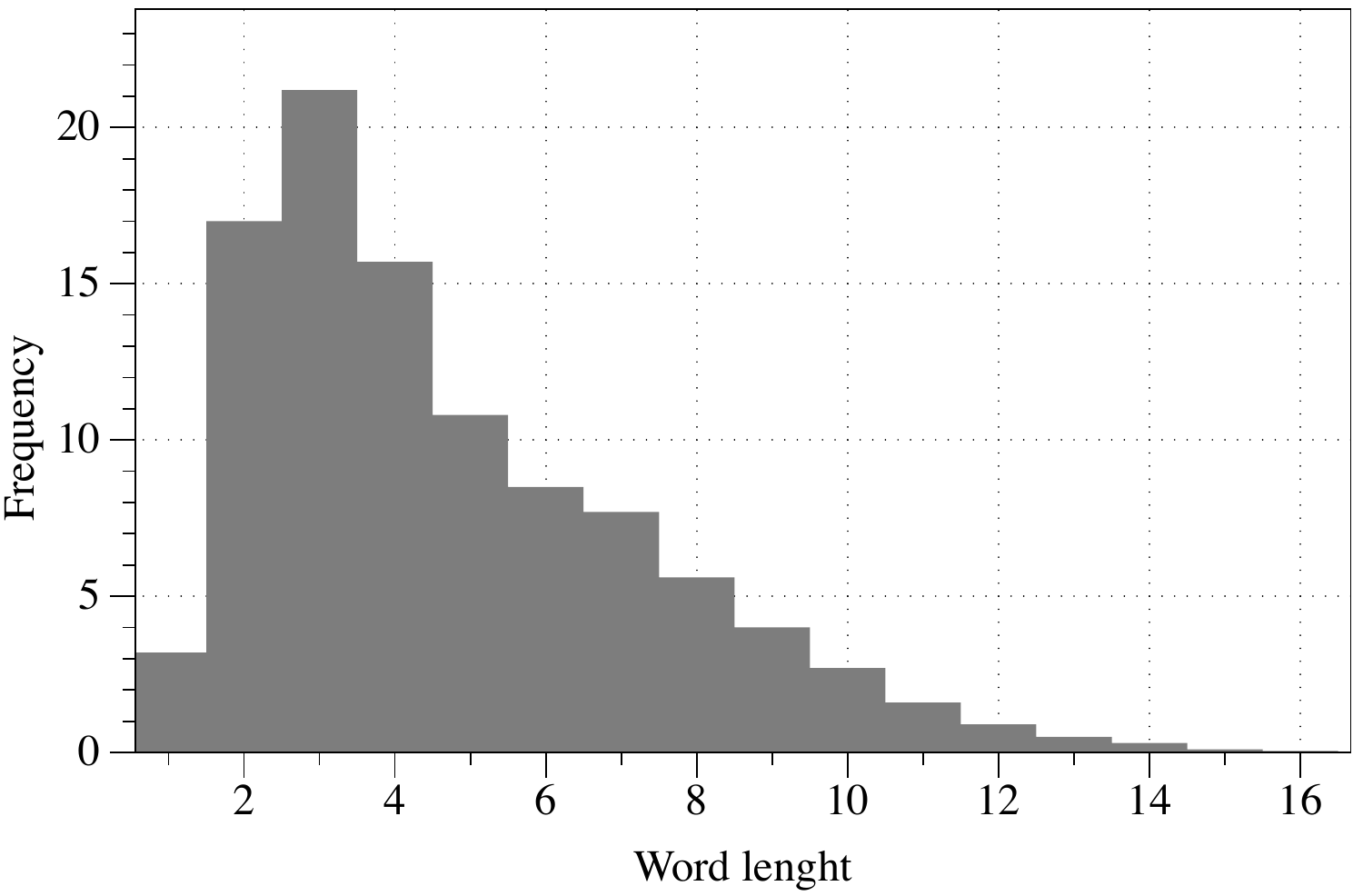}}
    \caption{Word length frequencies in (a)~English and (b)~Swedish, data are taken from Tab. 1 in~\cite{sigurd2004word}.
    }
    \label{fig:EnglishSwedishDistributions}
\end{figure}

\begin{table}[!tbp]
    \centering
    {\footnotesize 
    \begin{tabular}{l|l|l|l|l}
                      & $l_1$  & $l_2$ \\ \hline
\emph{C. militaris}   &  4.7 & 8.9  \\
\emph{F. velutipes}   &  3.6 & 7.4  \\
\emph{S. commune}     &  4.4  & 8.5  \\
\emph{O. nidiformis}  &  3.3  & 7  \\ \hline \hline
                      &  $m$   & &  & \\ \hline
English language      &  4.8    \\
Russian language      &  6 \\
Greek language        &  4.45 
    \end{tabular}
    }
\caption{Average word lengths in fungal and human languages. $l_1$ is an average word length in the spike grouping using $\theta=a$ and  $l_2$ using $\theta=2 \cdot a$, $m$ is an average word length of 1950+ Russian and English language approximated from the evolutionary plots in~\cite{bochkarev2015average} and average word length in Greek language approximated from Hellenic National Corpus~\cite{hatzigeorgiu2001word}.}
\label{tab:wordelength}
\end{table}

To convert the binary sequences representing spikes into sentences of the speculative fungal language, we must split the strings into words.  We assumed that if a distance between consequent spikes is not more than $\theta$ the spikes belong to the same word. To define $\theta$ we adopted analogies from English language. An average vowel duration in English (albeit subject to cultural and dialect variations) is 300~ms, minimum 70~ms and maximum 400~ms~\cite{house1961vowel}, with average post-word onset of c. 300~ms~\cite{weber1999functional}. We explored two options of the separation the spike trains into words: $\theta=a(s)$ and $\theta=2\cdot a(s)$, where $a(s)$ is an average interval between two subsequent spikes recorded in species $s \in \{\text{\emph{C. militaris}, \emph{F. velutipes}, \emph{S. commune},\emph{O. nidiformis}}\}$.  Distributions of fungal word lengths, measured in a number of spikes in $\theta$-separated trains of spikes are shown in Fig.~\ref{fig:trainsdistributions}.  The distributions follow predictive values 
$f_{exp}=\beta \cdot 0.73 \cdot l^c$, where $l$ is a length of a word, and $a$ varies from 20 to 26, and $b$ varies from 0.6 to 0.8, similarly to frequencies of word lengths in English and Swedish (Tab.`~\ref{fig:EnglishSwedishDistributions} in ~\cite{sigurd2004word}). As detailed in Tab.~\ref{tab:wordelength}, average word length in fungi, when spikes grouped with $\theta=a$ are the same range as average word lengths of human languages. For example, average number of spikes in train of \emph{C. militaris} is 4.7 and average word length in English language is 4.8. Average word length of  \emph{S. commune} is 4.4 and average word length in Greek language is 4.45.

 \begin{figure}[!tbp]
     \centering
\subfigure[]{\includegraphics[width=0.24\textwidth]{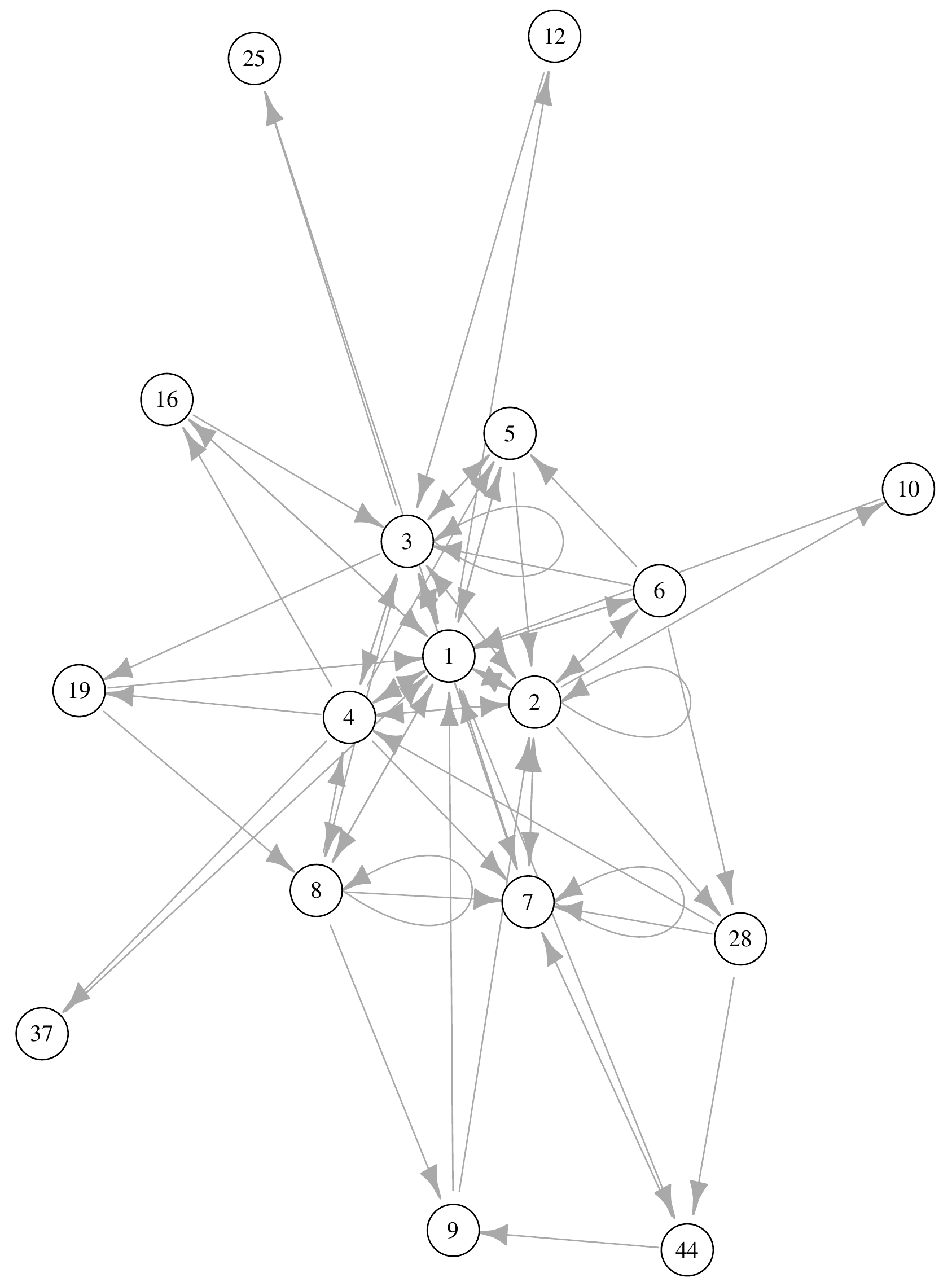}}
\subfigure[]{\includegraphics[width=0.24\textwidth]{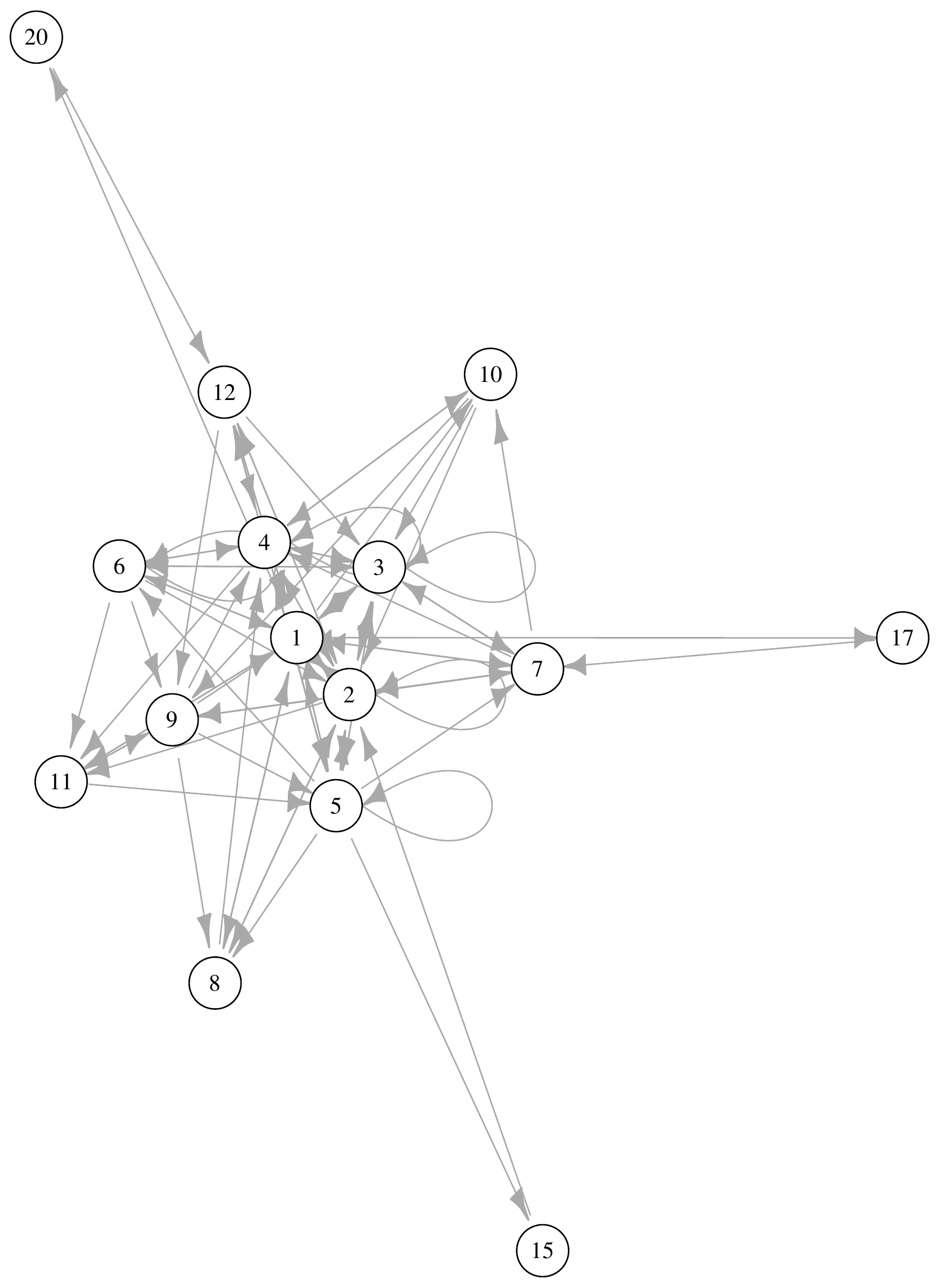}}
\subfigure[]{\includegraphics[width=0.24\textwidth]{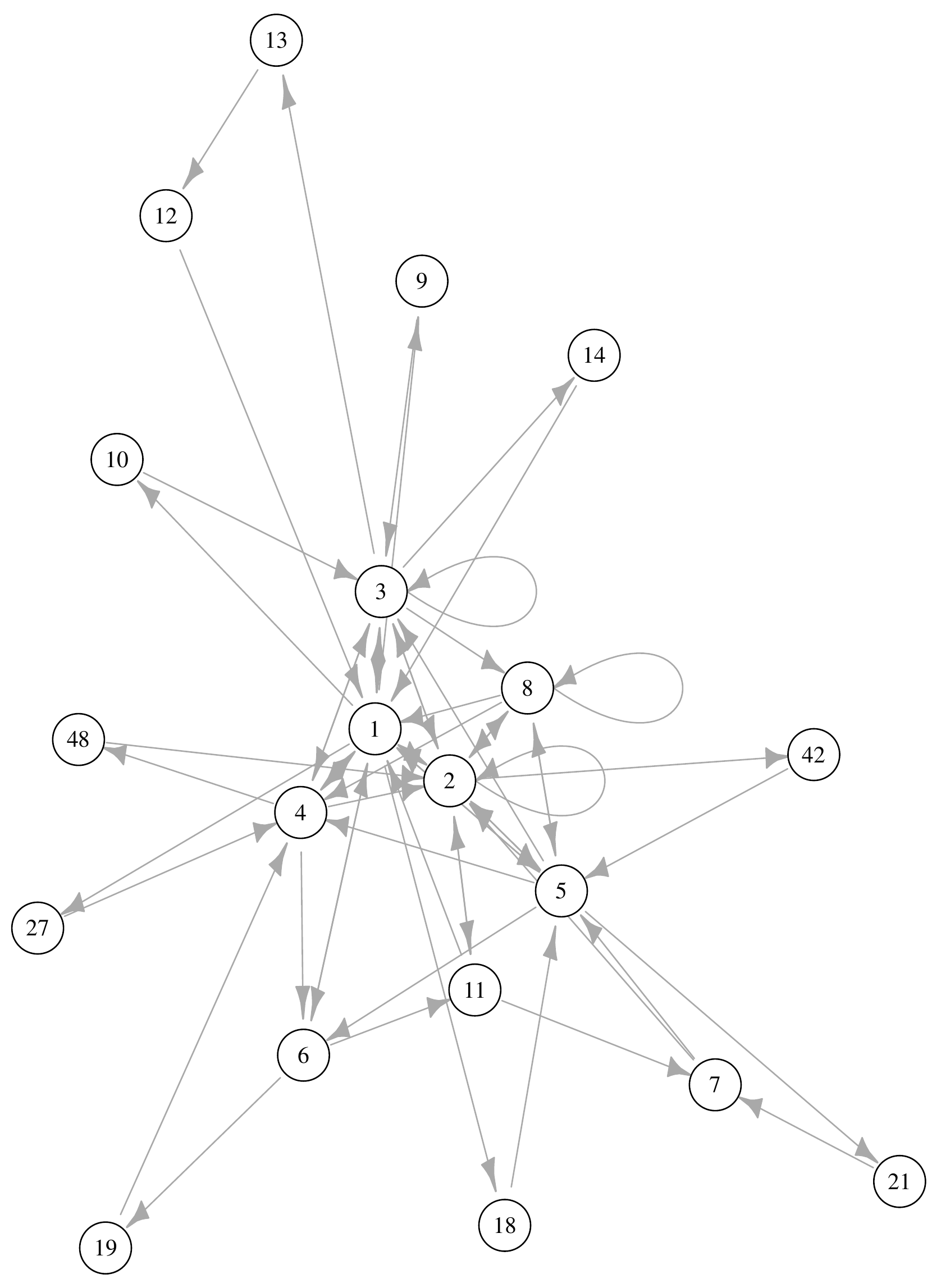}}
\subfigure[]{\includegraphics[width=0.24\textwidth]{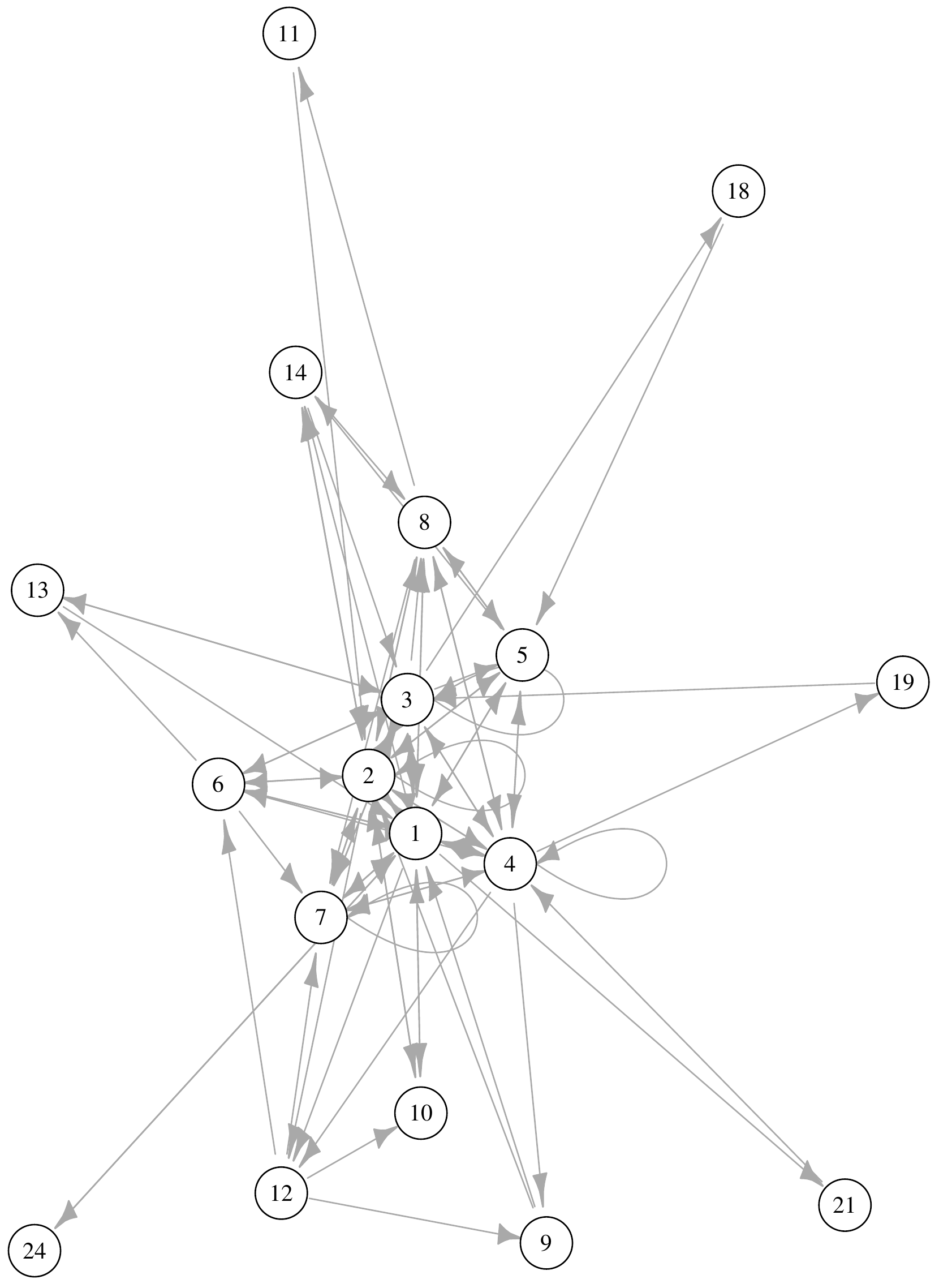}}
\subfigure[]{\includegraphics[width=0.24\textwidth]{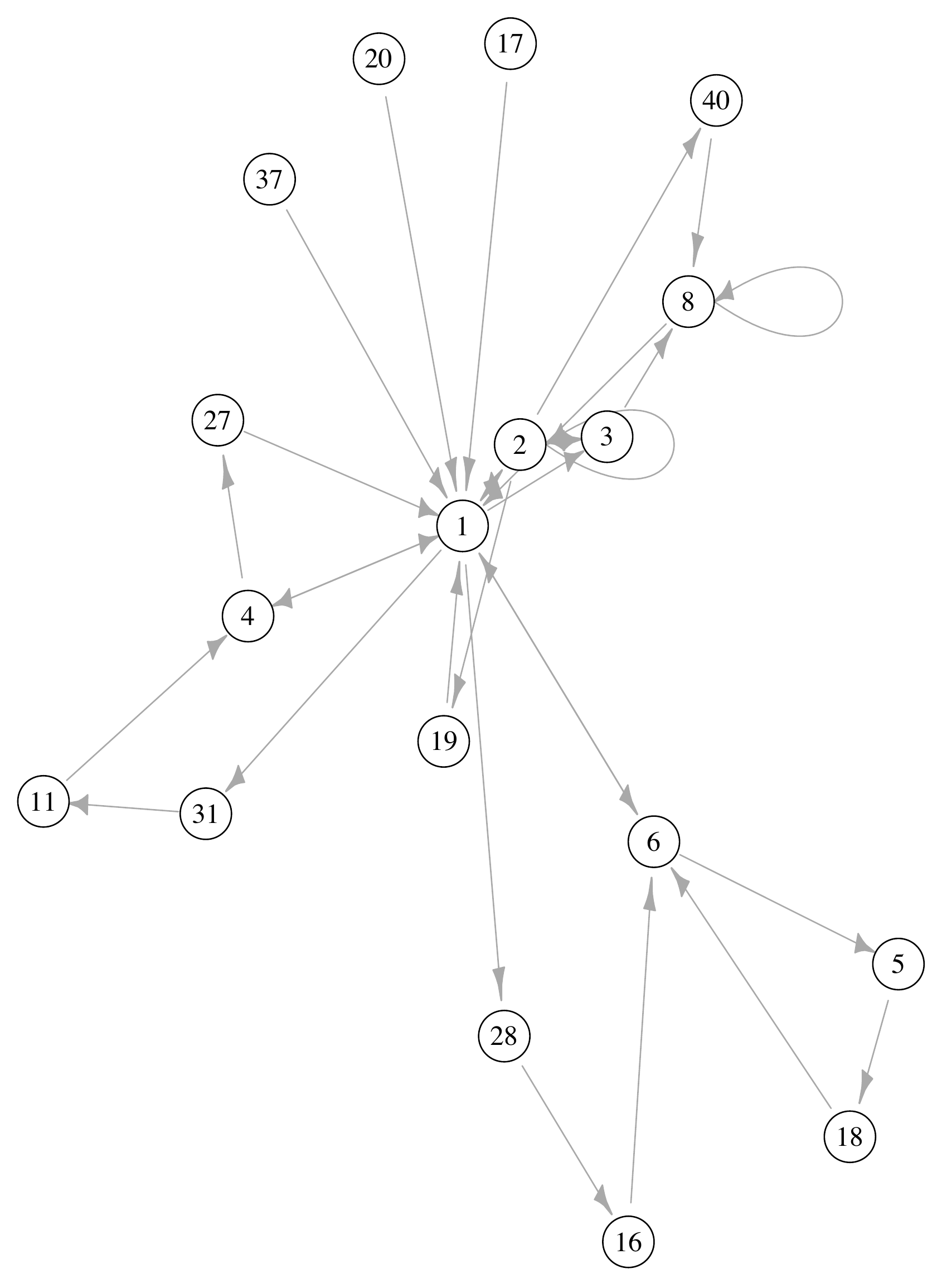}}
\subfigure[]{\includegraphics[width=0.24\textwidth]{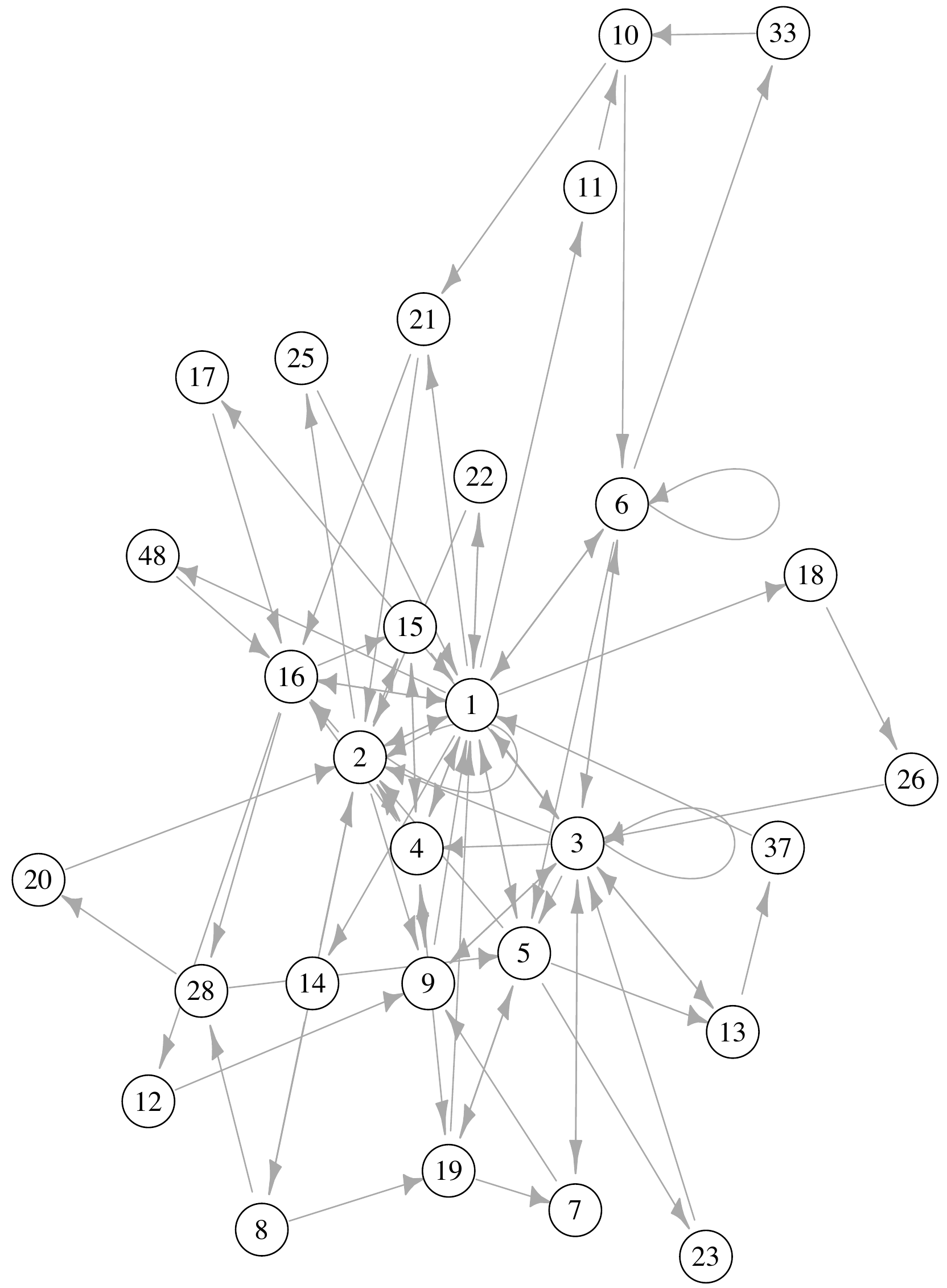}}
\subfigure[]{\includegraphics[width=0.24\textwidth]{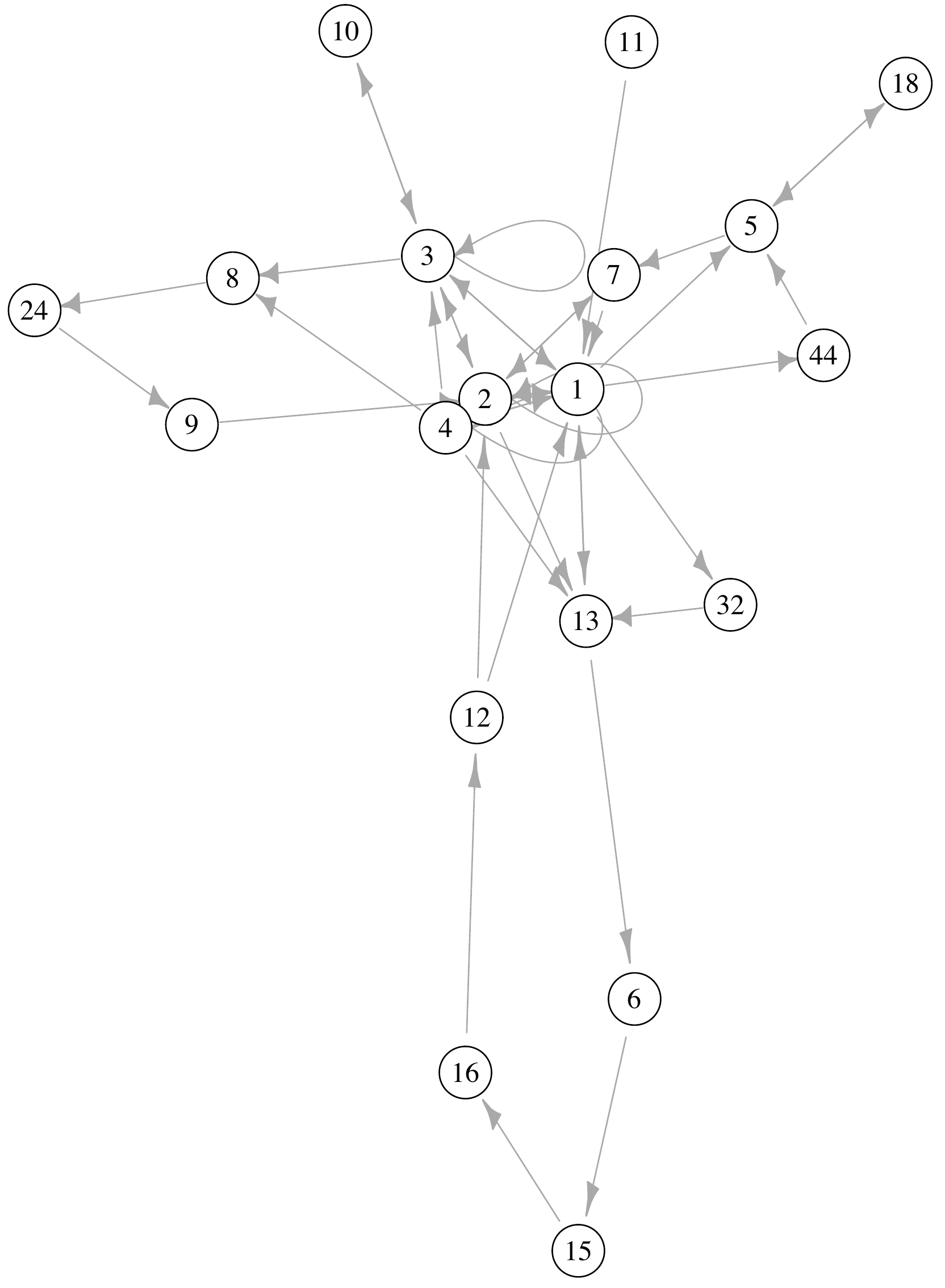}}
\subfigure[]{\includegraphics[width=0.24\textwidth]{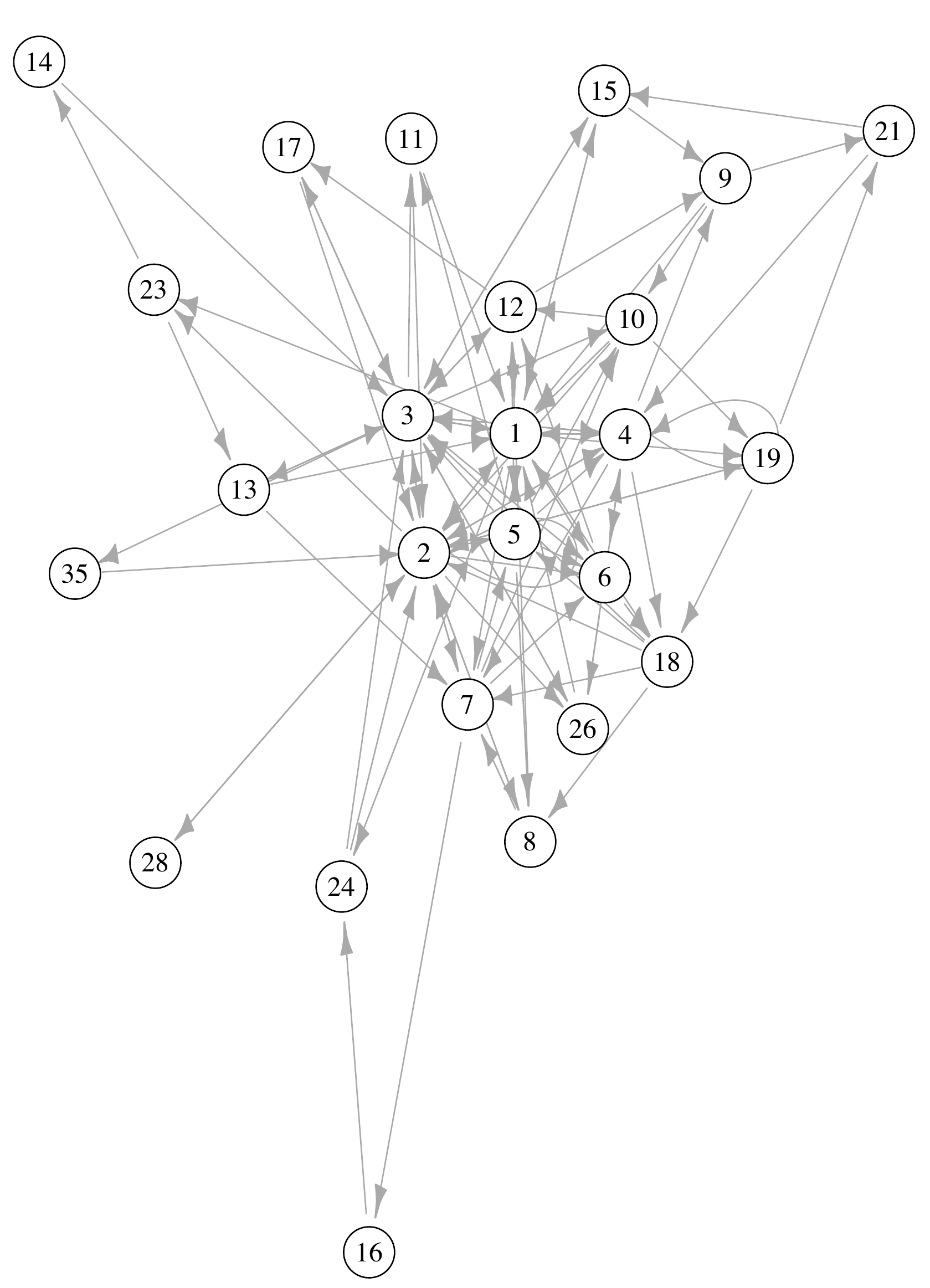}}
     \caption{State transition graphs of fungal spiking machines, where spikes have been grouped using $\theta=a$ (a--d) and $\theta=2\cdot a$ (e--h). (ae)~\emph{C. militaris}, (bf)~\emph{F. velutipes}, (cg)~\emph{S. commune}, (dh)~\emph{O. nidiformis}
     }
     \label{fig:spikingmachine}
 \end{figure}
 
  \begin{figure}[!tbp]
     \centering
\subfigure[]{\includegraphics[width=0.24\textwidth]{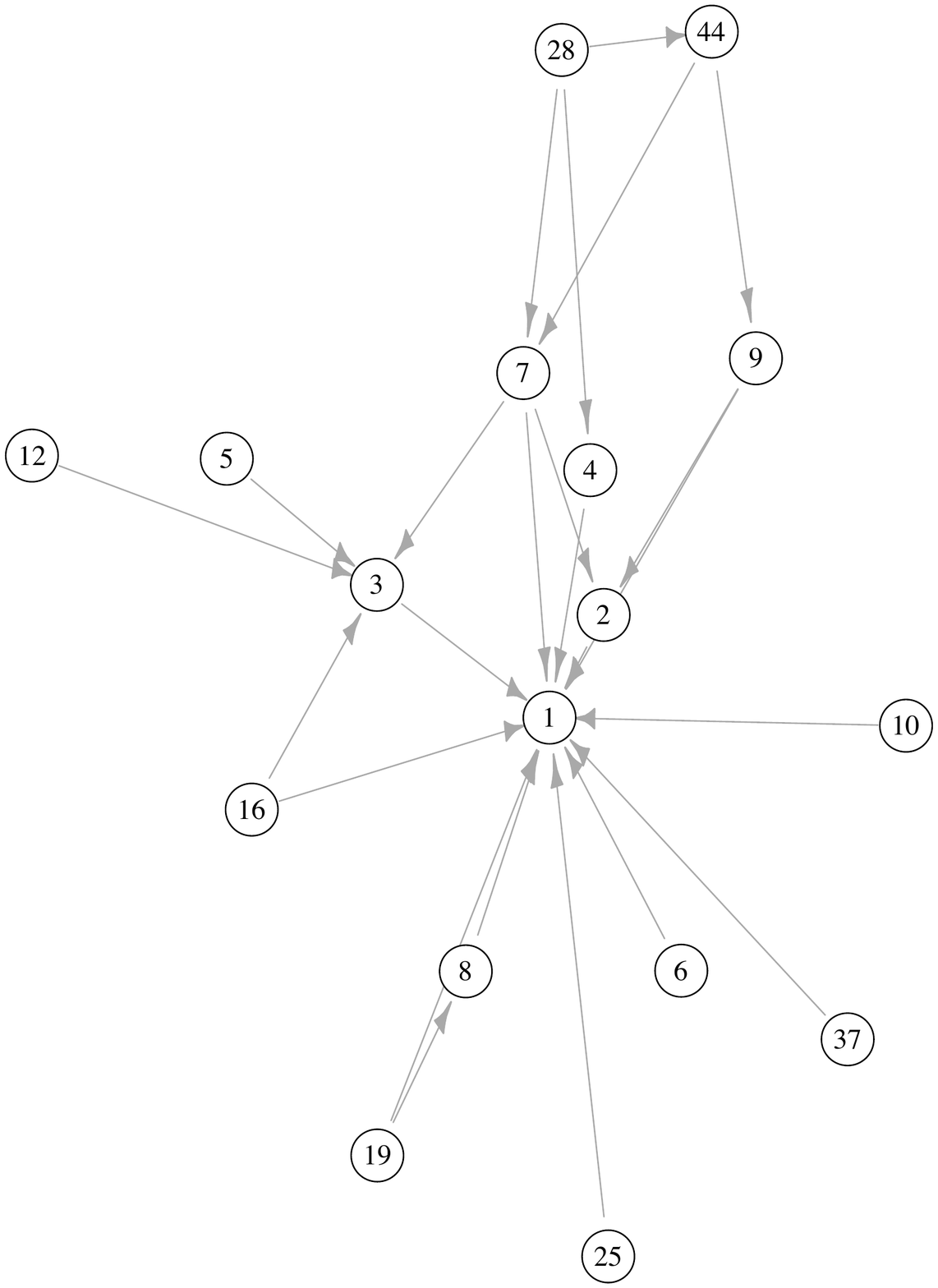}}
\subfigure[]{\includegraphics[width=0.24\textwidth]{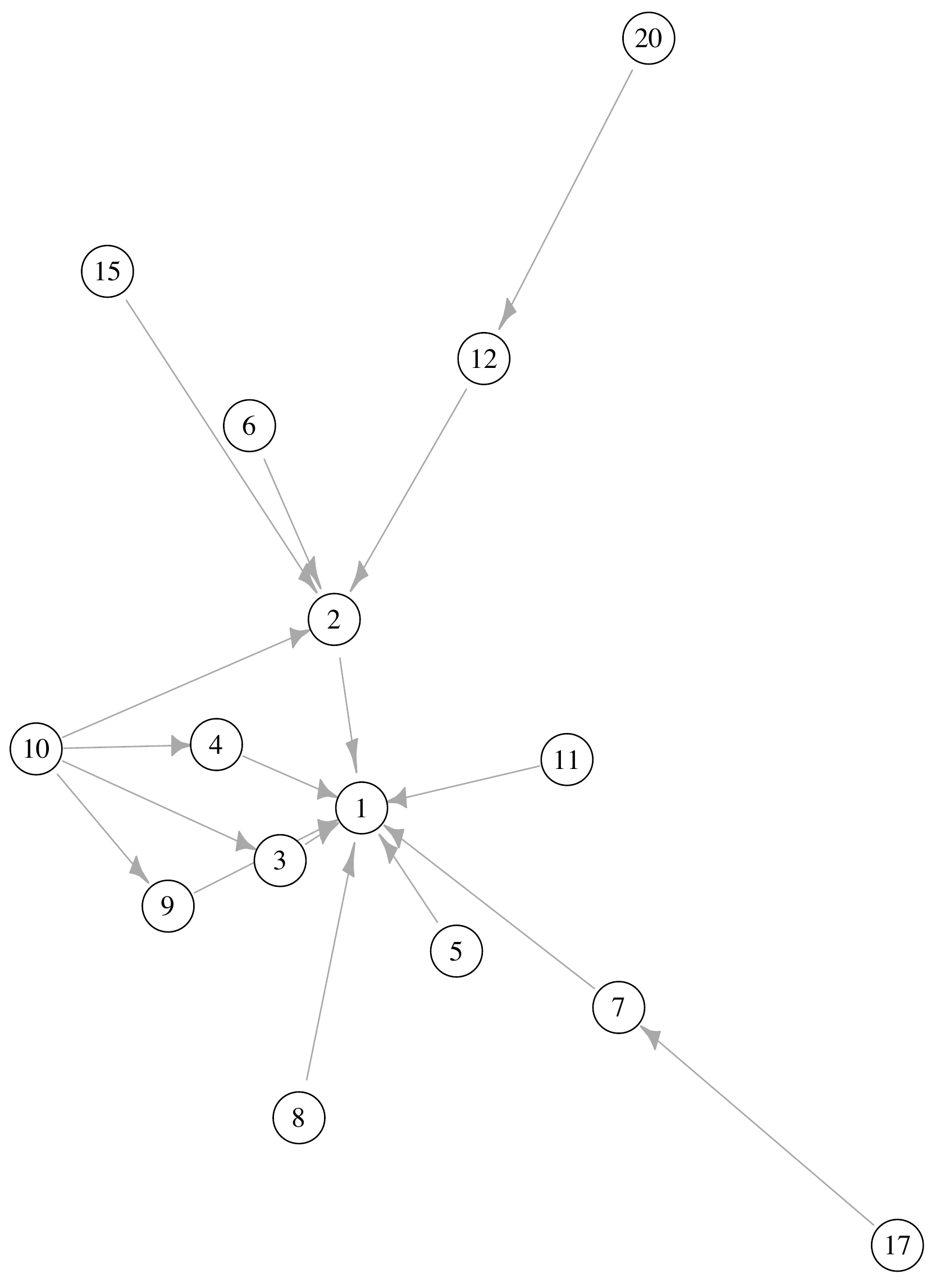}}
\subfigure[]{\includegraphics[width=0.24\textwidth]{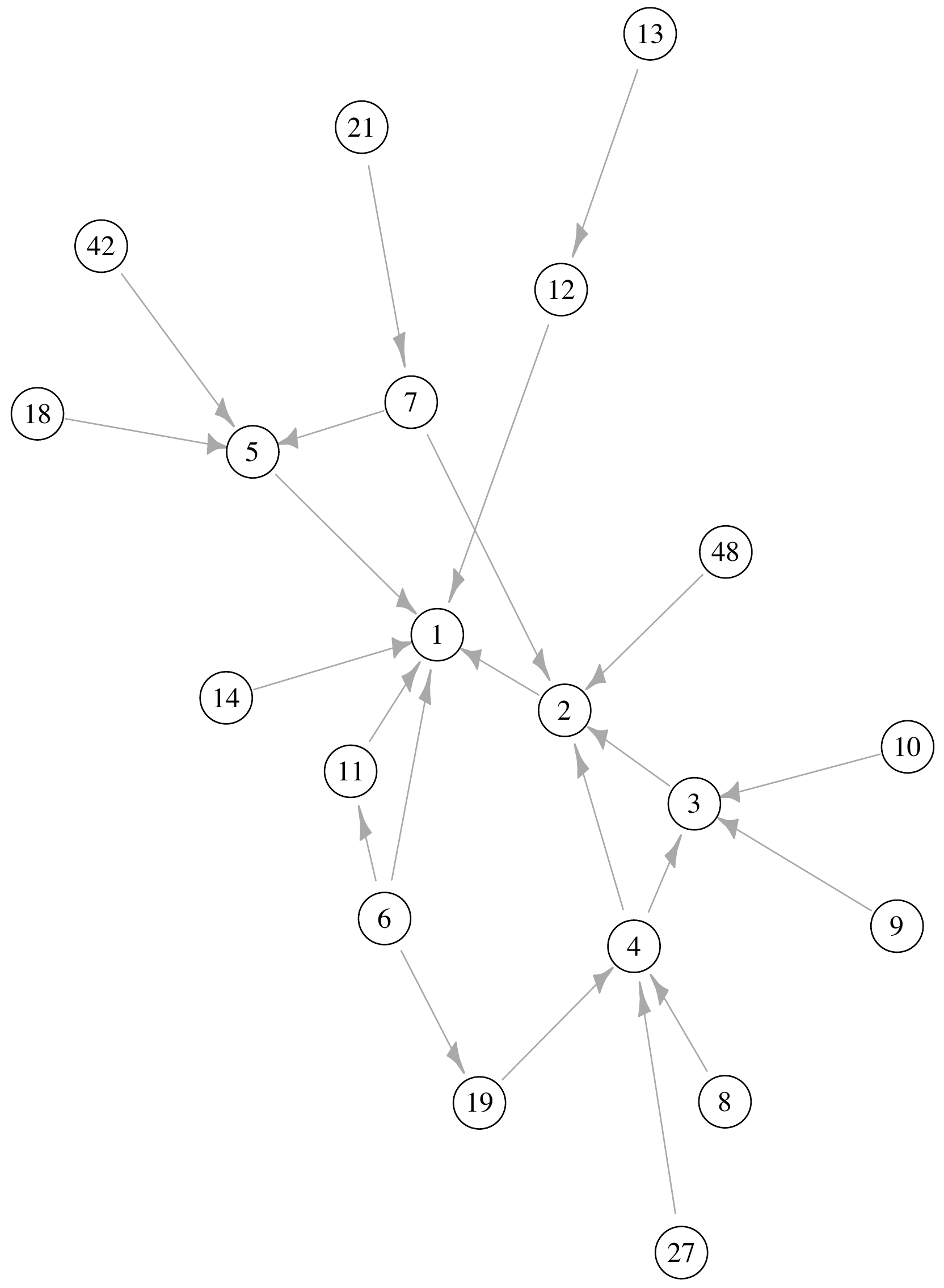}}
\subfigure[]{\includegraphics[width=0.24\textwidth]{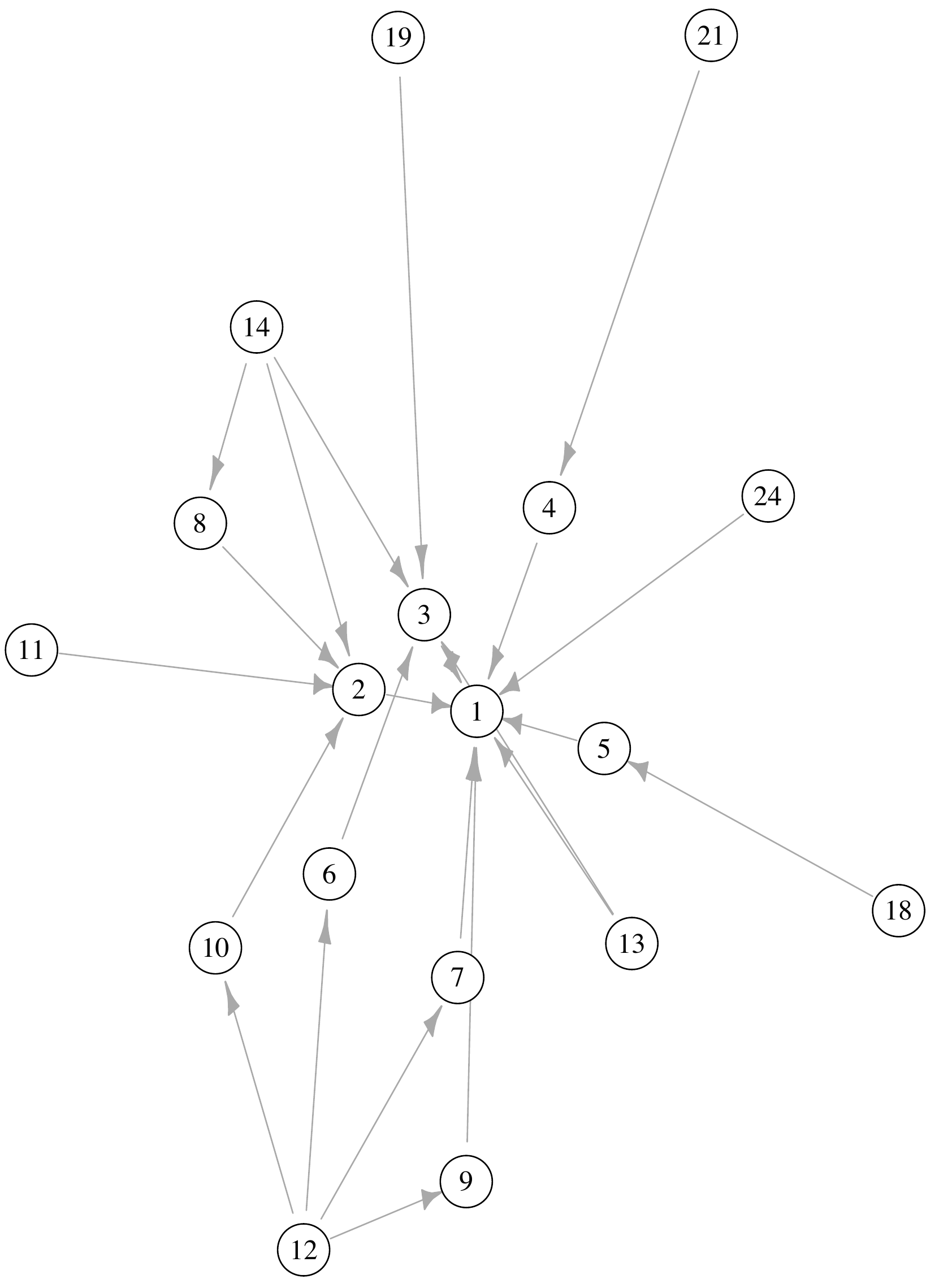}}
\subfigure[]{\includegraphics[width=0.24\textwidth]{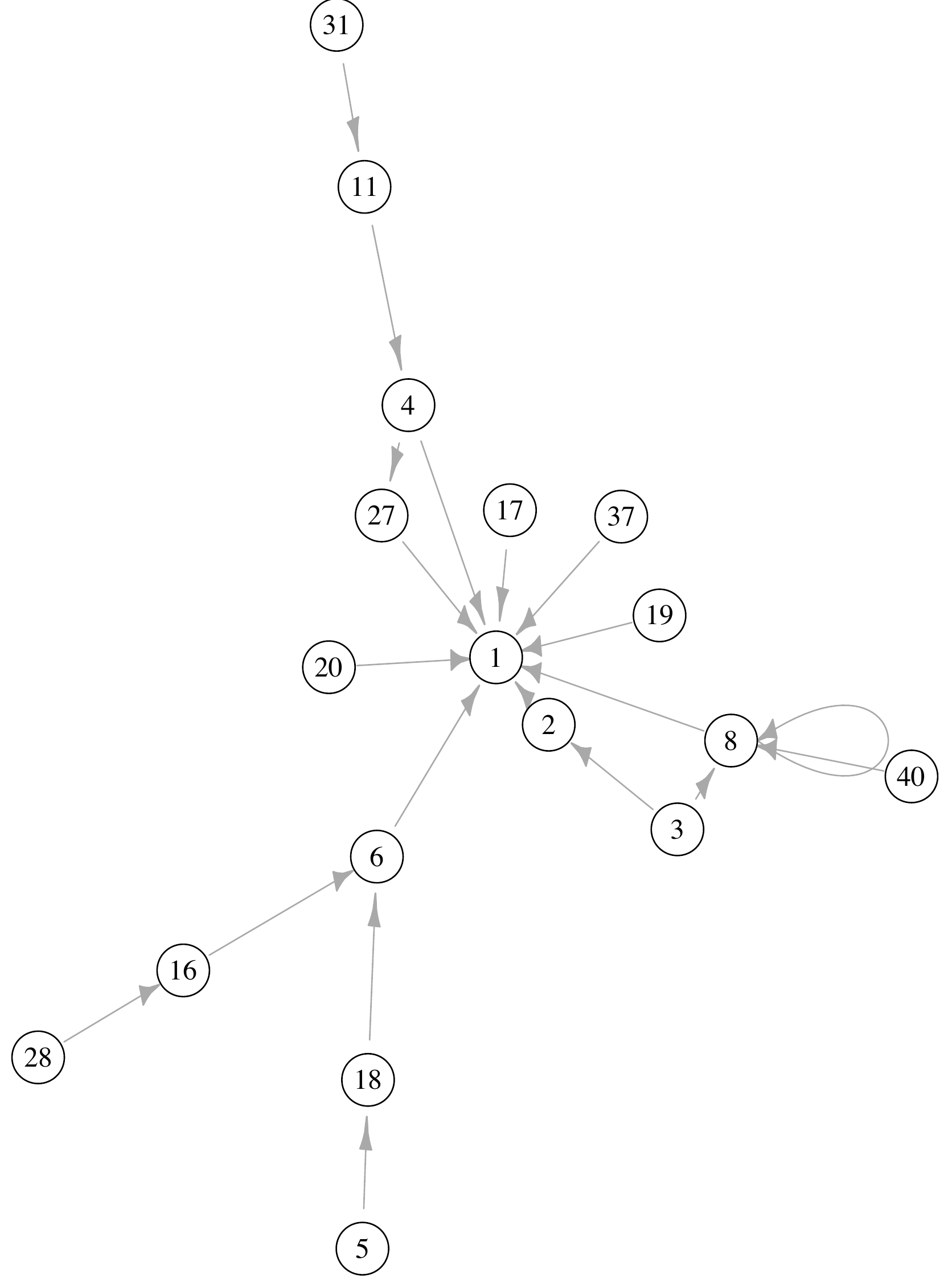}}
\subfigure[]{\includegraphics[width=0.24\textwidth]{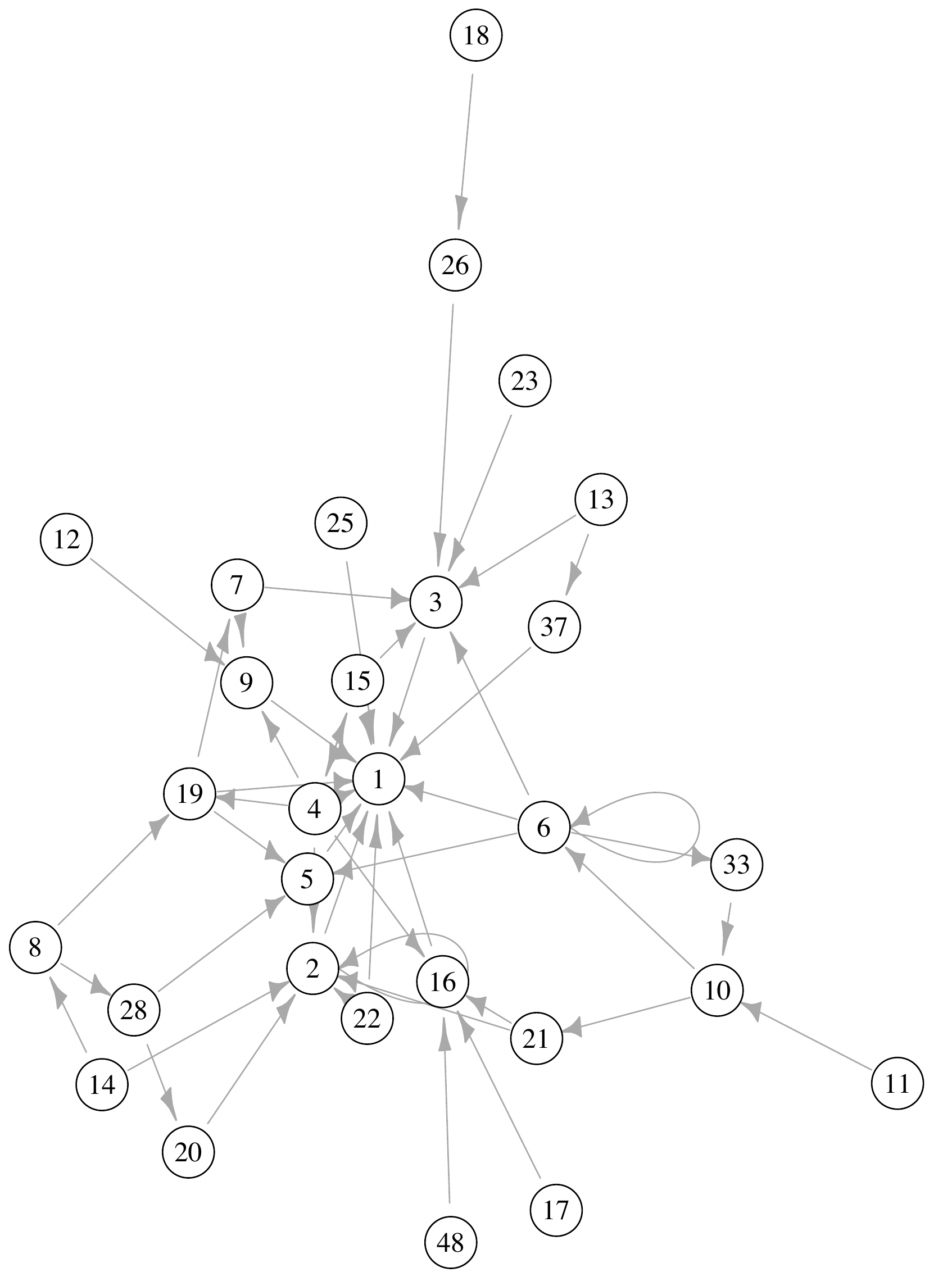}}
\subfigure[]{\includegraphics[width=0.24\textwidth]{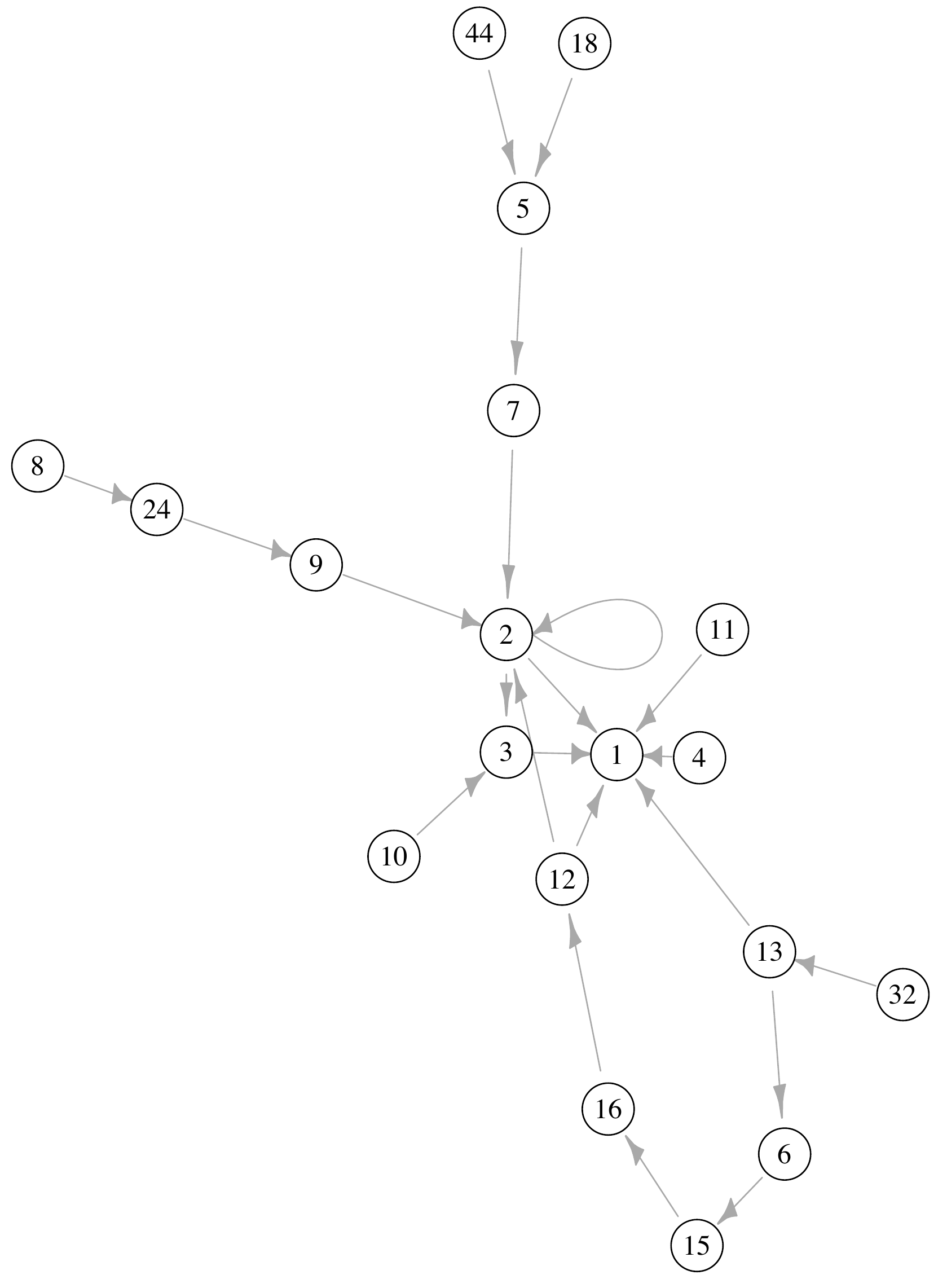}}
\subfigure[]{\includegraphics[width=0.24\textwidth]{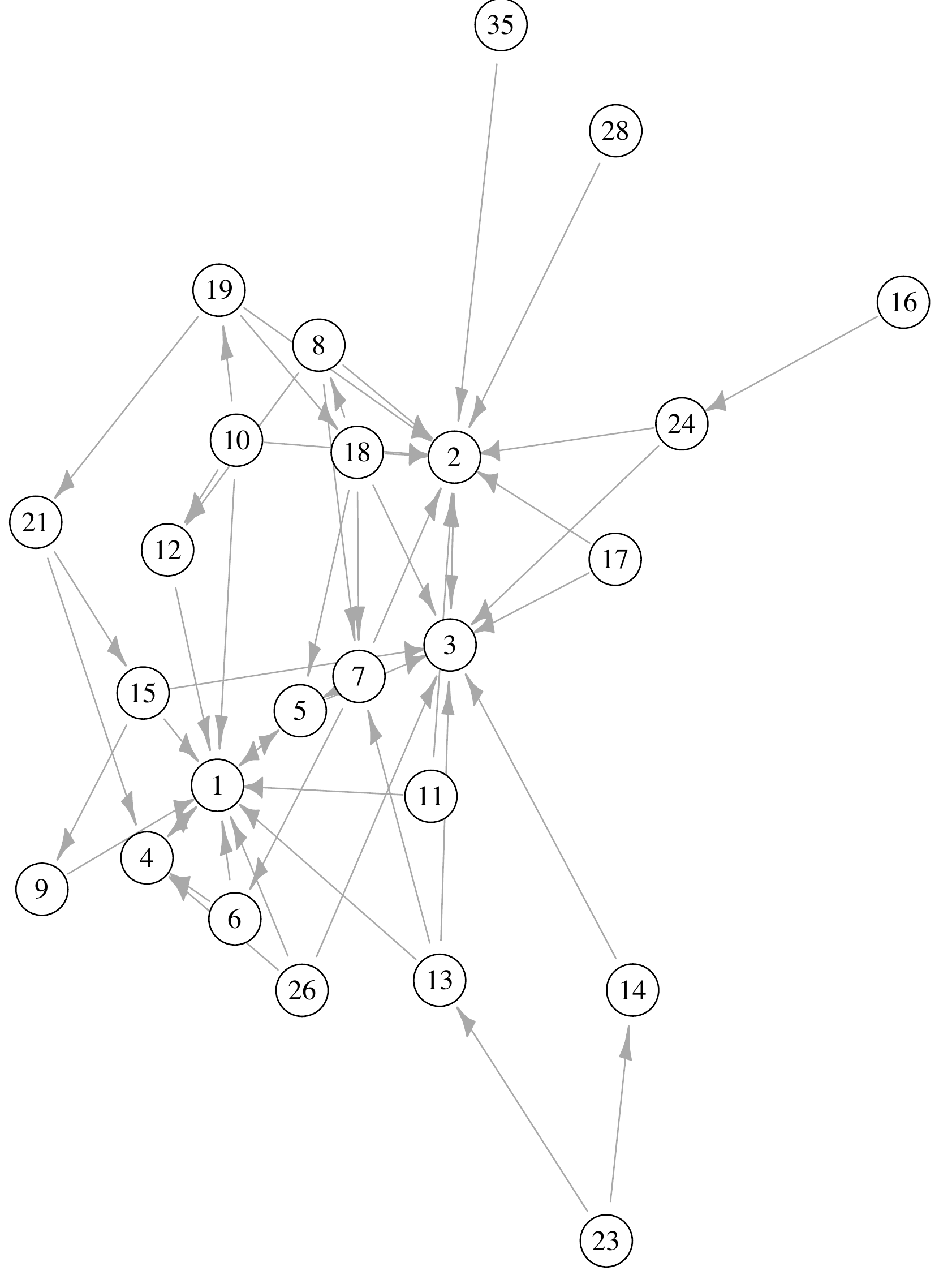}}
     \caption{Filtered state transition graphs of fungal spiking machines, where spikes have been grouped using $\theta=a$ (a--d) and $\theta=2\cdot a$ (e--h). (ae)~\emph{C. militaris}, (bf)~\emph{F. velutipes}, (cg)~\emph{S. commune}, (dh)~\emph{O. nidiformis}. The transitions were filtered in such manner that for each state $i$ we select state $j$ such that the weight $w(i,j)$ is maximal over $w(i,z)$, where $z \in {\bf S}$, ${\bf S}$ is a set of states. 
     }
     \label{fig:spikingmachineFiltered}
 \end{figure}
 
 To uncover syntax of the fungal language we should estimate what is most likely order of the words in fungal sentences. We do this via characterisation of global transition graphs of fungal spiking machines. A fungal spiking machine is a finite state machine. It takes states from ${\mathbf S} \in {\mathbf N}$ and updates its states according to probabilistic transitions:
${\mathbf S} \times [0,1] \rightarrow {\mathbf S}$, being in a state $s^t \in {\mathbf S}$ at time $t+1$ the automaton takes state  $s^{t+1} \in {\mathbf S}$ with probability $p(s^t,s^{t+1}) \in [0,1]$. The probabilities of the state transitions are estimated from the sentences of the fungal language.

The state transition graphs of the fungal spiking machines are shown in Fig.~\ref{fig:spikingmachine} for full dictionary case and in Fig.~\ref{fig:spikingmachineFiltered} for the filtered states sets when states over 9 are removed. 

\begin{table}[!tbp]
    \centering
    {\footnotesize 
    \begin{tabular}{l|l|l}
                      & $\theta=a$          & $\theta=2 \cdot a$ \\ \hline
\emph{C. militaris}   &  $1, \ldots, 8$     & $1, \ldots, 3$,  \\
\emph{F. velutipes}   &  $1, \ldots, 8$, 9  & 1, 2, 4, 15, 16  \\
\emph{S. commune}     &  $1, \ldots, 4$, 8  & $1, \ldots, 4$, 7  \\
\emph{O. nidiformis}  &  $1, \ldots, 5$     & $1, \ldots, 5$, 10, 12  \\  \hline
    \end{tabular}
    }
\caption{Attractive cores in the probabilistic state spaces of fungal spiking machines.}
\label{tab:attractivemeasures}
\end{table}

The probabilistic state transition graphs shown in Fig.~\ref{fig:spikingmachine} are drawn using physical model spring-based Kamada-Kawai algorithm~\cite{kamada1989algorithm}. Thus we can clearly see cores of the state space as clusters of closely packed states. The cores act as attractive measures in the probabilistic state space. The attractive measures are listed in Tab.~\ref{tab:attractivemeasures}. The membership of the cores well matches distribution of spike trains lengths (Fig.~\ref{fig:trainsdistributions}). 

A leaf, or Garden-of-Eden, state is a state which has no predecessors. \emph{C. militaris} probabilistic fungal spiking machine has leaves `25' and `37' in case of in grouping $\theta=a$ (Fig.~\ref{fig:spikingmachine}a) and `20', `17' and `37' in case of in grouping $\theta=2 \cdot a$ (Fig.~\ref{fig:spikingmachine}e). All other probabilistic fungal machines do not have leaves apart of \emph{S. commune} which has one leaf `11' in case of in grouping $\theta=2 \cdot a$ (Fig.~\ref{fig:spikingmachine}g).

An absorbing state of a finite state machine is a state in which the machine remains forever once it takes this state. All spiking fungal machines, derived in grouping $\theta=a$, have the only absorbing state `1' (Fig.~\ref{fig:spikingmachineFiltered}a--d). They have no cycles in the state space.  There are between 8, \emph{F. velutipes} (Fig.~\ref{fig:spikingmachineFiltered}c) and \emph{O. nidiformis} (Fig.~\ref{fig:spikingmachineFiltered}g), and 11 leaves, \emph{S. commune}) (Fig.~\ref{fig:spikingmachineFiltered}e). A maximal length of a transient period, measured in a maximal number of transitions required to reach the absorbing state from a leaf state varies from 3 (\emph{F. velutipes}) to 11 (\emph{S. commune}).

State transition graphs get more complicated, as we evidence further, when grouping $\theta=2 \cdot a$ is used (Fig.~\ref{fig:spikingmachineFiltered}h). Fungal spiking machine \emph{O. nidiformis} has one absorbing state, `1' (Fig.~\ref{fig:spikingmachineFiltered}h). Fungal spiking machines \emph{S. commune} (Fig.~\ref{fig:spikingmachineFiltered}g) and \emph{C. militaris} (Fig.~\ref{fig:spikingmachineFiltered}e) have two absorbing state each, `1' and `2' and `1' and `8', respectively. The highest number of absorbing states is found in the state transition graph of the \emph{F. velutipes} spiking machine (Fig.~\ref{fig:spikingmachineFiltered}f). They are `1', `6' and `2'. A number of leaves varies from 7, \emph{S. commune}, to 9, \emph{O. nidiformis} and \emph{C. militaris}, to 12,\emph{F. velutipes}. Only \emph{O. nidiformis} spiking machine has cycles in each state transition graph (Fig.~\ref{fig:spikingmachineFiltered}h). The cycles are $1 \longleftrightarrow 5$ and $2 \longleftrightarrow 3$.

\begin{table}[!tbp]
    \centering
    {\footnotesize 
\subfigure[]{
    \begin{tabular}{p{4.7cm}|l|l|l|l}
 &  \emph{C. militaris} & \emph{F. velutipes} &  \emph{S. commune} &  \emph{O. nidiformis}\\ \hline
Algorithmic complexity, bits  &	1211 & 1052 & 981 & 1243 \\
Algorithmic complexity normalised  &	6.51 & 3.94 & 7.55 & 3.67 \\
Logical depth, steps &	4321 &  4957 & 3702 & 5425 \\
Logical depth normalised &	23 &  19 & 28 & 16 \\
Shannon entropy, bits &	2.4  & 2.3 & 2.3 & 2.3\\
Second order entropy, bits & 3.8 &  3.7 & 3.7 & 3.7  \\
LZ complexity , bits & 1153 & 1495 &  910 & 1763\\
LZ complexity (normalised), bits &	6.2 & 5.6 &  7 & 5.2\\
Input string length & 186 & 267 & 130 & 339 \\
    \end{tabular}
    }
\subfigure[]{
    \begin{tabular}{p{4.7cm}|l|l|l|l}
 &  \emph{C. militaris} & \emph{F. velutipes} &  \emph{S. commune} &  \emph{O. nidiformis}\\ \hline
Algorithmic complexity, bits  &	1047 & 1295 & 980  & 1393 \\
Algorithmic complexity normalised  &	10.57 & 10.04 & 14.4  & 8.82 \\
Logical depth, steps &	2860  &  4147 &  3046 &  4731\\
Logical depth normalised &	29  &  32 &  45 &  30\\
Shannon entropy, bits &	  2.5 &  2.5 &  2.5 & 2.6 \\
Second order entropy, bits &  4 &  4.2  & 4.2 &  4.3 \\
LZ complexity, bits &	 594 &  993 &  666 & 1232 \\
LZ complexity normalised &	 6 &  7.7 &  9.8 & 7.8 \\
Input string length & 99 & 129 & 68 & 158 \\
    \end{tabular}
    }
\subfigure[]{
    \begin{tabular}{p{4.7cm}|l|l|l|l}
 &  \emph{C. militaris} & \emph{F. velutipes} &  \emph{S. commune} &  \emph{O. nidiformis}\\ \hline
Algorithmic complexity, bits  &	    679 & 976 & 466  & 1276 \\
Algorithmic complexity normalised & 3.96 & 3.97 & 4.05  & 4 \\
Shannon entropy, bits &	           2.5  &  2.6 & 2.5  & 2.5 \\
Second order entropy, bits &       4.7  &  5  & 4.5   & 4.9 \\
LZ complexity, bits &  735  & 1009  & 563  & 1208  \\
LZ complexity normalised & 4.3  &  4.1 & 4.9   &  3.8 \\
Input string length & 171 & 246 & 115 & 319 \\
    \end{tabular}
    }
    }
    \caption{
Block Decomposition Method (BDM) algorithmic complexity estimation,
BDM logical depth estimation, Shannon entropy, 
Second order entropy, LZ complexity. The measures are estimated using  The Online Algorithmic Complexity Calculator (\url{https://complexitycalculator.com/index.html}) block size 12, alphabet size 256. Spike trains are extracted with (a)~$\theta=a$ and (b)~$\theta=2\cdot a$, where $a$ is an average interval between two consequent spikes, see Tab.~\ref{tab:spiking}. We also provide values of the LZ complexity and algorithmic complexity normalised by input string lengths.
In table (c) we provide data on the strings of train powers (in number of spikes) calculated with $\theta=a$ and then filtered so value over 9 are removed and the complexity is estimated in alphabet of 9 symbols.
}
    \label{tab:complexity}
\end{table}

To study complexity of the fungal language algorithmic complexity~\cite{zenil2020review}, Shannon entropy~\cite{lin1991divergence} and Liv-Zempel complexity~\cite{ziv1977universal,ziv1978compression} of the fungal words (sequences of spike trains lengths) are estimated using The Online Algorithmic Complexity Calculator\footnote{\url{https://complexitycalculator.com/index.html}}~\cite{zenil2020review,zenil2018decomposition,gauvrit2014algorithmic,delahaye2012numerical} in Tab.~\ref{tab:complexity}. Shannon entropy of the strings recorded is not shown to be species specific, it is 2.3 for most species but 2.4 for~\emph{C. militaris} in case of $\theta=a$ grouping and 2.5 for most species but 2.6 for \emph{O. nidiformis} in case of $\theta=2 \cdot a$. The same can be said about second order entropy (Tab.~\ref{tab:complexity}). \emph{O. nidiformis} shows highest values of algorithmic complexity for both cases of spike trains separation (Tab.~\ref{tab:complexity}ab) and filtered sentences (where only words with up to 9 spikes are left) (Tab.~\ref{tab:complexity}c). In other of decreasing algorithmic complexity we then have ~\emph{C. militaris}, \emph{F. velutipes}  and \emph{S. commune}. 

The hierarchy of algorithmic complexity changes when we normalise the complexity values dividing them by the string lengths. For the case $\theta=a$ the hierarchy of descending complexity will be 
 \emph{S. commune} (7.55), ~\emph{C. militaris} (6.51), \emph{F. velutipes} (3.94), \emph{O. nidiformis} (3.67) (Tab.~\ref{tab:complexity}a). Note that in this case a normalised algorithmic complexity of \emph{S. commune} is nearly twice higher than that of \emph{O. nidiformis}. For the case  $\theta=2\cdot a$  \emph{S. commune}  still has the highest normalised algorithmic complexity amongst the four species studied (Tab.~\ref{tab:complexity}b). Complexities of  \emph{C. militaris} and \emph{F. velutipes} are almost the same, and the complexity of \emph{O. nidiformis} is the lowest. When we consider filtered sentences of fungal electrical activity, where words with over 9 spikes are removed, we get nearly equal values of the algorithmic complexity, ranging from 3.96 to 4.05 (Tab.~\ref{tab:complexity}c).
 LZ complexity hierarchy is the same for all three cases --- $\theta=a$  (Tab.~\ref{tab:complexity}a), $\theta=2 \cdot a$  (Tab.~\ref{tab:complexity}b) and filtered sentences  (Tab.~\ref{tab:complexity}c): \emph{S. commune}, \emph{C. militaris}, \emph{F. velutipes}, \emph{O. nidiformis}. To summarise, in most conditions, \emph{S. commune} is an uncontested champion in complexity of the sentences generated. It is followed by  \emph{C. militaris}.  Two other species studied \emph{F. velutipes} and \emph{O. nidiformis} occupy lower levels of the complexity hierarchy.

\section{Discussion}
\label{discussion}

We recorded extracellular electrical activity of four species of fungi. We found evidences of the spike trains propagating along the mycelium network. We speculated that fungal electrical activity is a manifestation of the information communicated between distant parts of the fungal colonies. We assumed that types of characters used to code the information in electrical communication of fungi are trains of spikes. We therefore attempted to uncover key linguistic phenomena of the proposed fungal language. We found that distributions of lengths of spike trains, measured in a number of spikes, follows the distribution of word lengths in human languages.  We found that size of fungal lexicon can be up to 50 words, however the core lexicon of most frequently used words does not exceed 15-20 words. Species \emph{S. commune} and  \emph{O. nidiformis} have largest lexicon while species \emph{C. militaris} and \emph{F. velutipes} have less extensive one. Depending on the threshold of spikes grouping into words, average word length varies from 3.3 (\emph{O. nidiformis}) to 8.9 (\emph{C. militaris}). A fungal word length averaged over four species and two methods of spike grouping is 5.97 which is of the same range as an average word length in some human languages, e.g. 4.8 in English and 6 in Russian. To characterise a syntax of the fungal language we analysed state transition graphs of the probabilistic fungal spiking machines. We found that attractive measures, or communication cores, of the fungal machines are composed of the words up to ten spikes long with longer words appearing less often. We analysed complexity of the fungal language and found that species \emph{S. commune} generates most complex, amongst four species studied, sentences. The species \emph{C. militaris} is slightly below \emph{S. commune} in the hierarchy of complexity and \emph{F. velutipes} and \emph{O. nidiformis} occupy lower levels of the hierarchy. Future research should go into three following directions: study of inter-species variations, interpretation of a fungal grammar and re-consideration of the coding type.  First, we should increase a number of fungi species studied to uncover if there is a significant variations in the language syntax between the species. Second, we should try to uncover grammatical constructions, if any in the fungal language, and to attempt to semantically interpret syntax of the fungal sentences. Third, and more like the most important direction of future research, would to make a thorough and detailed classification of fungal words, derived from the train of spikes. Right now we classed the word based solely on a number of spikes in the corresponding trains. This is indeed quite a primitive classification aka to interpreting binary words only by sums of their bits and not exact configurations of 1s and 0s. Said that we should not expect quick results, we still did not decipher language or cats and dogs despite living with them for centuries but research into electrical communication of fungi is in its pure infant stage.

\section*{Acknowledgement}

This project has received funding from the European Union's Horizon 2020 research and innovation programme FET OPEN ``Challenging current thinking'' under grant agreement No 858132. 


\end{document}